\documentclass[11pt,a4paper]{article}
\usepackage[T1]{fontenc}
\usepackage[utf8]{inputenc}
\usepackage{lmodern}
\usepackage{mathptmx}
\usepackage[margin=25mm]{geometry}
\usepackage{amsmath,amssymb,bm,booktabs,array,longtable,multirow}
\usepackage{graphicx,xcolor,tikz}
\usetikzlibrary{arrows.meta,positioning,fit}
\usepackage{caption,placeins,microtype,float}
\usepackage[numbers,sort&compress]{natbib}
\usepackage[colorlinks=true,linkcolor=blue!50!black,citecolor=blue!50!black,urlcolor=blue!50!black]{hyperref}
\definecolor{flowblue}{HTML}{17456C}
\newcommand{\Rey}{\mathrm{Re}}

\newcommand{\vct}[1]{\boldsymbol{#1}}
\newif\ifstandalone
\standalonefalse
\title{Regional accuracy of physics-informed neural networks in deep lid-driven cavities}
\author{Ehsan Roohi\thanks{Corresponding author. Email: \href{mailto:roohie@umass.edu}{roohie@umass.edu}.}\,\textsuperscript{a}\quad and\quad Christopher J. McDevitt\,\textsuperscript{b}\\[5pt]
\small \textsuperscript{a}Department of Mechanical and Industrial Engineering,\\ \small University of Massachusetts Amherst,\\ \small 160 Governors Drive, Amherst, MA 01003, USA\\
\small \textsuperscript{b}Department of Materials Science and Engineering, Nuclear Engineering Program,\\ \small University of Florida, P.O. Box 116400, Gainesville, FL 32611, USA}

\date{}
\begin{document}
\maketitle
\begin{abstract}
A deep lid-driven cavity has strong motion beneath the lid and a stack of much weaker eddies below it, so agreement in bulk velocity is an incomplete test of a computed flow. We examine data-free physics-informed neural networks (PINNs) for this steady two-dimensional flow and compare velocity, pressure and vorticity region by region with spectral/$hp$ and steady finite-volume solutions that use the same smoothed lid. With an established hard-constrained stream-function formulation, the PINNs reproduce the eddy stack of a cavity five widths deep for Reynolds numbers from 100 to 1000, but an error of several per cent remains, almost all of it within a tenth of a width of the two lid corners, and changes to the training move it by at most a factor of two. The cause is the polynomial factor that imposes the wall conditions. Near the lid it is proportional to the distance from the lid divided by the depth, so in a deep cavity it is small exactly where the flow is most demanding. Trained in the same way, this factor is as accurate as any other in a square cavity, degrades with Reynolds number at depth two, and fails at $\Rey=1000$ at depth five and at $\Rey=250$ at depth ten. Factors that are local to the walls, with the wall distances as additional network inputs, remove the limit: at depth five they reduce the velocity, pressure and vorticity errors by about two orders of magnitude, and they remain accurate at every depth tested. In the cavity ten widths deep the speed falls by about sixteen orders of magnitude from the lid to the floor. Below the upper one or two eddies, every eddy is about $1.40$ widths long and about 360 times weaker than the one above it, at every Reynolds number from 100 to 1000, as Moffatt's analysis of Stokes eddies in a channel predicts; the Reynolds number changes only the upper eddies and so shifts the whole stack. The PINN follows this decay to the floor. The continuation reaches $\Rey=2250$, and a residual test on held-out points identifies the failures beyond it without a reference solution.
\end{abstract}
\noindent\textbf{Keywords:} physics-informed neural networks; deep cavity; lid-driven flow; Reynolds-number continuation; regional collocation; weak recirculation; spectral/$hp$ elements.

\section{Introduction}
A deep lid-driven cavity tests more than the reproduction of its dominant vortex. Motion imposed at the lid drives a sequence of much weaker recirculating vortices below it, while the strongest velocity gradients remain close to the moving wall and its junctions with stationary sidewalls. A numerical approximation can therefore agree in bulk velocity while differing in local shear or in weak motion at depth. This study asks how consistently a data-free physics-informed neural network (PINN) reproduces velocity, pressure and sampled rotational structure across these regions.

The physical configuration is established. Burggraf connected viscous cavity eddies with the rotational core that develops as Reynolds number increases \cite{burggraf1966separated}; Pan and Acrivos examined rectangular cavities, including creeping-flow calculations at depths up to five widths \cite{pan1967rectangular}. Moffatt showed that Stokes flow near a sharp corner contains an infinite sequence of ever weaker eddies, and that the same happens between two parallel walls driven from one end, where each eddy is about 1.4 widths long and about 360 times weaker than the one above it \cite{moffatt1964eddies}; Pan and Acrivos observed such eddies in their experiments \cite{pan1967rectangular}. Cheng and Hung studied $0.1\leq D/W\leq7$ and $0.01\leq\Rey\leq5000$ \cite{cheng2006deep}. Zhuo et al. included $D/W=5$ at $\Rey=100$, 500 and 1000 among broader ranges \cite{zhuo2013deep}. The present work does not claim the first simulation of this geometry or Reynolds number. Its subject is the regional accuracy of a neural solution procedure for this established flow.

Classical square-cavity benchmarks also show why the boundary-value problem and assessment protocol matter. Multigrid, continuation and spectral calculations provide accurate reference solutions \cite{ghia1982multigrid,schreiber1983cavity,botella1998benchmark,erturk2005cavity}, while steady and time-dependent studies distinguish solution accuracy from physical stability \cite{bruneau2006revisited,shankar2000review}. In particular, regularizing the moving lid changes the problem at the wall junctions. Here the PINN and reference solve the same Gaussian-regularized-lid problem; comparisons with conventional discontinuous-lid studies are qualitative. A reported steady field or maximum studied Reynolds number is not a measured instability threshold. Table~\ref{tab:deep-literature} records these distinctions for the deep-cavity literature.

\begin{table}[htbp]\centering\small
\caption{Deep-cavity literature: a maximum studied Reynolds number is not a transition threshold. Depth is normalized by lid width.}\label{tab:deep-literature}
\begin{tabular}{p{2.8cm}p{4.1cm}p{6.3cm}}\toprule
Study & Verified geometry and Reynolds scope & Flow regime and relevance\\\midrule
Pan and Acrivos \cite{pan1967rectangular} & Creeping-flow calculations through $D/W=5$; experiments at $\Rey=20$--4000 & Steady-flow structure; 4000 is an experimental range limit, not a measured Hopf threshold.\\
Cheng and Hung \cite{cheng2006deep} & $D/W=0.1$--7; $\Rey=0.01$--5000 & Steady vortex structures, including deep vortices; the reported range is not an onset criterion.\\
Zhuo et al., Part I \cite{zhuo2013deep} & $D/W=1.5$, 3, 5, 7; $\Rey=100$, 500, 1000, 3200, 5000 & Steady-flow positions, sizes and strengths; directly relevant geometric precedent.\\
Zhuo et al., Part II \cite{zhuo2013bifurcation} & First Hopf bifurcations for $D/W=1$--5 & Periodic and quasiperiodic states above geometry-dependent critical values; not a universal turbulence threshold.\\
Lin et al. \cite{lin2013transition} & Transition study at $D/W=1$--3 & Documents depth-dependent Hopf onset; also discusses earlier unsteady cases at $\Rey=7500$ for $D/W=1.5$ and 4. These are not depth-five thresholds.\\\bottomrule
\end{tabular}
\end{table}

Neural trial functions can enforce boundary conditions by construction \cite{lagaris1998neural}; modern PINNs train through differential-equation residuals \cite{raissi2019pinn,karniadakis2021piml,cuomo2022review}. For incompressible flow, variable choice, loss weighting and transfer affect optimization and differentiation \cite{jin2021nsfnets}. Gradient imbalance, conditioning and unequal convergence rates complicate the relation between a small training loss and field accuracy \cite{wang2021gradient,wang2022ntk,krishnapriyan2021failure,rathore2024landscape}. A deep cavity makes this distinction physically concrete: an objective dominated by strong upper motion need not adequately constrain weak lower recirculation. Hard constraints, localized representations and adaptive sampling are different responses to these difficulties \cite{sukumar2022boundary,jagtap2020conservative,moseley2023fbpinn,wu2023sampling,wang2024das2}; the regional sampling used here is fixed in advance and is not adaptive.

McDevitt et al.\ \cite{mcdevitt2024cavity} examined square and shallower trapezoidal cavities; the present study extends the assessment to rectangular cavities with $D/W=1$, 2, 5 and 10.
For the square cavity, PINNs now reach high Reynolds numbers. The stream-function formulation of \cite{mcdevitt2024cavity} was trained without data up to $\Rey=5000$ and with sparse simulation data up to about $2\times10^4$. Methods that couple automatic differentiation with numerical differences between neighbouring points \cite{chiu2022canpinn}, or replace the convection term by a finite-volume discretization \cite{wei2025ffvpinn}, have given data-free square-cavity solutions up to $\Rey=10^4$, and training schedules that activate collocation points progressively from the boundary have been tested on the square cavity \cite{cestola2026flowaware}. A deep cavity is difficult for a different reason: the range of velocity scales between the lid and the floor, which grows exponentially with depth.

Table~\ref{tab:position} places these extensions alongside the reported geometry, training-data requirements and assessment scope of related cavity PINN studies. It distinguishes the present deep-cavity, region-resolved comparison from earlier applications; it does not rank solvers whose boundary conditions and evaluation protocols differ.

\begin{table}[htbp]
\centering\small
\caption{Scope of related cavity PINN studies and the extensions
examined in the present work.}
\label{tab:position}
\begin{tabular}{p{2.7cm}p{4.5cm}p{6cm}}
\toprule
Study & Reported scope & Distinction of the present work\\
\midrule

McDevitt et al.\ \citep{mcdevitt2024cavity}
&
Square cavities: data-free cases through $\Rey=5000$ and
sparse-data-assisted cases through $\Rey=21000$;
parametric trapezoidal cavities with $0.5\leq D/W\leq1$.
Velocity, streamfunction and pressure fields are examined.
&
Extends the assessment to rectangular cavities with $D/W$ up to 10.
Measures velocity, pressure and vorticity errors region by region
against matched spectral/$hp$ and finite-volume references,
including the weak floor eddies.
\\

Fowler et al.\ \citep{fowler2024thermal}
&
Two-dimensional natural convection in a thermal cavity,
with Rayleigh number included as a network input.
&
Addresses mechanically driven recirculation rather than
buoyancy-driven flow. The fixed-depth Reynolds-number series
examines regional agreement across the strong upper circulation
and weak lower vortices.
\\

Takao and Ii \citep{takao2026finetuning}
&
Rectangular cavities with $D/W=0.5$--2 at $\Rey=100$;
coordinate transformation and fine-tuning using interior
velocity observations.
&
Examines the deeper geometry $D/W=5$ through $\Rey=1000$,
without supervised interior solution values in training.
Transferred initialization is obtained from earlier PINN states.
\\

Wang et al.\ \citep{wang2024das2}
&
Data-free square-cavity calculations through $\Rey=1000$,
using adaptive sampling.
&
Examines deep-cavity flow using prescribed regional collocation,
rather than adaptive sampling, and evaluates workflow changes
through separate upper, interior and floor metrics.
\\

Chiu et al.\ \citep{chiu2022canpinn}; Wei et al.\ \citep{wei2025ffvpinn}
&
Square cavity; coupled automatic and numerical differentiation, and a simplified finite-volume convection term; data-free solutions up to $\Rey=10^4$.
&
Keeps a mesh-free formulation with automatic differentiation only; addresses the depth, not the Reynolds number of a square cavity.
\\

Cestola et al.\ \citep{cestola2026flowaware}
&
Square cavity at $\Rey=100$ among seven benchmarks; collocation points activated progressively from the boundary.
&
Changes the representation of the wall conditions rather than the training schedule.
\\

Present work
&
Data-free rectangular cavities at $D/W=1$, 2, 5 and 10; $\Rey=100$--1000 at every depth and up to 2250 at $D/W=5$.
&
Compares velocity, pressure and vorticity region by region; identifies the boundary factor as the source of the lid-corner error and replaces it by wall-local factors; checks the decay of the eddy stack against Stokes theory down to the floor.
\\

\bottomrule
\end{tabular}
\end{table}

Recent cavity-specific reviews and reliability studies emphasize that formulation, boundary treatment and evaluation metrics prevent simple rankings across implementations \cite{sheikholeslami2026review,sluzalec2026reliable}. We address this issue with matched lid data and independent Nektar++ spectral/$hp$ fields \cite{cantwell2015nektar,moxey2020nektar}. Velocity, pressure after a specified gauge alignment, and vorticity computed from both sampled velocity fields are assessed separately. The last quantity tests sampled derivative agreement, rather than the native automatic-differentiation vorticity of the network.

The paper has two parts. In the first we assess the established formulation at $D/W=5$ and $\Rey=100$, 250, 500 and 1000, region by region and for velocity, pressure and vorticity, together with four training workflows at $\Rey=1000$ that differ in regional collocation, initialization, residual scaling and depth envelope. The PINNs reproduce the eddy stack and the weak floor flow, but every workflow leaves an error of several per cent that sits almost entirely at the two lid corners, and no change to the training removes it. This raises the question of the second part: what limits the accuracy at the lid, and how does the limit depend on the Reynolds number and the depth? We show that the cause is the boundary factor that imposes the wall conditions, test factors that are local to the walls with identical training, separate their ingredients, vary the depth from one to ten widths, and extend the Reynolds number. Because the time-stepped reference turned out to be unconverged near the floor of the deepest cases, we add steady finite-volume references and explain why a solution integrated in time from rest is slow to reach the lowest eddies. In the cavity ten widths deep the two references and the PINN then let us follow the eddy stack over sixteen orders of magnitude in speed and compare it with Moffatt's Stokes theory.

Other neural approaches to partial differential equations include the deep Galerkin and Deep Ritz methods \cite{sirignano2018dgm,eyu2017deepritz}, variational domain decomposition \cite{kharazmi2021hpvpinn} and the DeepXDE library \cite{lu2021deepxde}; gradient-enhanced losses, self-adaptive weights and re-initialization are other training strategies \cite{yu2022gradient,mcclenny2023adaptive,lee2025reinitialization}. We use none of them. Pal et al.\ also impose the boundary conditions of the Navier--Stokes equations exactly \cite{pal2025hardbc}, and De Ryck et al.\ give error estimates for Navier--Stokes PINNs \cite{deryck2024error}.

Section~\ref{sec:problem} specifies the problem. Section~\ref{sec:formulation} describes the inherited representation and its training, and then the wall-local factors. Section~\ref{sec:cfd} describes the two reference solvers and the assessment protocol. Section~\ref{sec:results} gives the results of the established formulation, and Section~\ref{sec:local} those of the wall-local factors, including the depth study and higher Reynolds numbers. The discussion and conclusions follow; loss histories are given in an appendix.

\section{Physical problem and cavity campaign}\label{sec:problem}
$\Rey$ denotes the Reynolds number, and all variables are nondimensional unless stated otherwise. In the method name spectral/$hp$, $h$ is the element size and $p$ the polynomial degree; they are unrelated to the height $h$ and the pressure $p$.
\subsection{Governing equations and matched lid}
Consider a two-dimensional rectangular cavity of width $W$ and depth $D$, with $0\leq x\leq W$, $1-D\leq y\leq1$ and the lid at $y=1$. In the figures, $h=y+D-1$ is the height above the floor. Lengths are scaled by $W$ and velocities by the lid velocity scale $U$, so that $\Rey=UW/\nu$; here $U=W=1$. The steady incompressible equations are
\begin{align}
 u_x+v_y&=0,\\
 R_x&=u u_x+v u_y+p_x-\Rey^{-1}(u_{xx}+u_{yy})=0,\label{eq:rx}\\
 R_y&=u v_x+v v_y+p_y-\Rey^{-1}(v_{xx}+v_{yy})=0.\label{eq:ry}
\end{align}
Here $u$ and $v$ are the horizontal and vertical velocity components, $p$ is pressure, $\nu$ is the kinematic viscosity, and $R_x$ and $R_y$ are the steady momentum residuals. Subscripts denote differentiation, for example $u_x=\partial u/\partial x$ and $u_{xx}=\partial^2u/\partial x^2$. In Eq.~\eqref{eq:lid}, $U_{\rm lid}$ is the nondimensional lid speed and $\delta_x$ its horizontal smoothing length.
Pressure is defined up to an additive constant. The sidewalls and floor are stationary, and the lid speed falls smoothly to zero at its ends:
\begin{equation}
 u(x,1)=U_{\rm lid}(x)=\bigl[1-e^{-x^2/\delta_x^2}\bigr]\bigl[1-e^{-(x-W)^2/\delta_x^2}\bigr],\qquad v(x,1)=0,\quad \delta_x=\sqrt{0.001}\,W.\label{eq:lid}
\end{equation}
The smoothing in Eq.~\eqref{eq:lid} removes the velocity discontinuity at the lid ends. The PINN and CFD calculations use the same profile, so they solve the same boundary-value problem. A broader transition would ease the approximation near the corners, but it would change the problem, and we do not use one. Standard cavity results, such as the high-Reynolds-number solutions of Erturk et al.\ \citep{erturk2005cavity}, use other lid conditions; they give context but are not compared quantitatively, and neither are earlier finite-volume calculations with different lid profiles.

\subsubsection{Nondimensionalization, pressure and vorticity}
An asterisk identifies a dimensional quantity; $t$ denotes time, $p_0$ an arbitrary constant reference pressure, and $\boldsymbol{u}=(u,v)$ the velocity vector. The gradient is $\nabla=(\partial_x,\partial_y)$ and the Laplacian is $\nabla^2=\partial_{xx}+\partial_{yy}$. In the component sum in Eq.~\eqref{eq:pressure_poisson_expanded}, $x_1=x$, $x_2=y$, $u_1=u$, $u_2=v$, and both indices $i,j$ run from 1 to 2. The streamfunction $\psi$ in Eq.~\eqref{eq:vorticity_expanded} satisfies $u=\psi_y$ and $v=-\psi_x$.
For a fluid of constant density $\rho$ and kinematic viscosity $\nu$, the dimensional variables are scaled as $x=x^*/W$, $y=y^*/W$, $t=t^*U/W$, $\boldsymbol{u}=\boldsymbol{u}^*/U$ and $p=(p^*-p_0)/(\rho U^2)$. The Reynolds number is therefore based on the cavity width, not its depth, and varying $D/W$ at fixed $\Rey$ changes only the geometry. Since $h=y+D-1$ is a translation, derivatives with respect to $h$ and $y$ are the same.

The CFD calculation integrates the unsteady equations
\begin{equation}
 \partial_t\boldsymbol{u}+\boldsymbol{u}\cdot\nabla\boldsymbol{u}
 =-\nabla p+\Rey^{-1}\nabla^2\boldsymbol{u},\qquad
 \nabla\cdot\boldsymbol{u}=0.\label{eq:unsteady_expanded}
\end{equation}
The PINN seeks a steady field, for which the time derivative vanishes; there is no body force. The lid drives the flow through tangential motion, viscous stress carries this motion into the interior, and it is dissipated at the stationary walls. In a deeper cavity the upper circulation drives progressively weaker motion below it, and the gradients of this weak motion must still be resolved.

In the neural formulation pressure is a separate network output, not reconstructed from velocity after training. Taking the divergence of Eq.~\eqref{eq:unsteady_expanded} and using continuity gives
\begin{equation}
 \nabla^2p=-\sum_{i,j}\frac{\partial u_i}{\partial x_j}
                         \frac{\partial u_j}{\partial x_i}
 =-\left[u_x^2+2u_yv_x+v_y^2\right].\label{eq:pressure_poisson_expanded}
\end{equation}
This relation links pressure curvature to velocity gradients. It is not added to the PINN loss, because the momentum equations already couple pressure to velocity. Since an additive constant leaves both equations unchanged, we remove the mean PINN-minus-CFD pressure difference over the comparison region before comparing fields. This aligns the two gauges, which the solvers fix differently, without changing gradients or extrema.

The out-of-plane vorticity, positive for counterclockwise rotation, is
\begin{equation}
 \omega=v_x-u_y=-\nabla^2\psi.\label{eq:vorticity_expanded}
\end{equation}
Taking the curl of the momentum equations eliminates pressure and yields
\begin{equation}
 \partial_t\omega+u\omega_x+v\omega_y
 =\Rey^{-1}(\omega_{xx}+\omega_{yy}).\label{eq:vort_transport_expanded}
\end{equation}
In the steady cavity, vorticity is generated at the walls, carried by convection and spread into the interior by diffusion. Two fields with similar streamlines can therefore have different shear layers or corner gradients. Pressure reflects the momentum balance across the cavity; vorticity, a derivative of velocity, is more sensitive to local resolution. In the sampled-field comparisons both velocity fields are differentiated with the same operator on the same grid, so neural automatic differentiation is not mixed with a different CFD derivative.

\subsection{Cavity cases}
Table~\ref{tab:campaign} lists the cases. Velocity, pressure and vorticity of the established formulation are compared with CFD at $D/W=5$ and four Reynolds numbers, with detailed comparisons of the training workflows at $\Rey=1000$.
\begin{table}[htbp]\centering\small
\caption{Cases reported. The region-by-region assessment of the established formulation is at $D/W=5$; the matched comparison of the polynomial and wall-local factors covers $D/W=1$, 2, 5 and 10.}\label{tab:campaign}
\begin{tabular}{p{2cm}p{3cm}p{8.2cm}}\toprule
$\Rey$ & $D/W$ & Role\\\midrule
100, 250, 500 & 1, 2, 5, 10 & Continuation in Reynolds number; established formulation at $D/W=5$ (Section~\ref{sec:results}); polynomial and wall-local factors at every depth (Section~\ref{sec:depth}).\\
1000 & 1, 2, 5, 10 & Principal case: training workflows of the established formulation at $D/W=5$, and the matched depth comparison.\\
1500, 2000, 2250 & 5 & Higher Reynolds numbers of the wall-local formulation (Section~\ref{sec:highRe}).\\\bottomrule
\end{tabular}
\end{table}

\begin{figure}[!htbp]\centering
\includegraphics[width=.8\linewidth]{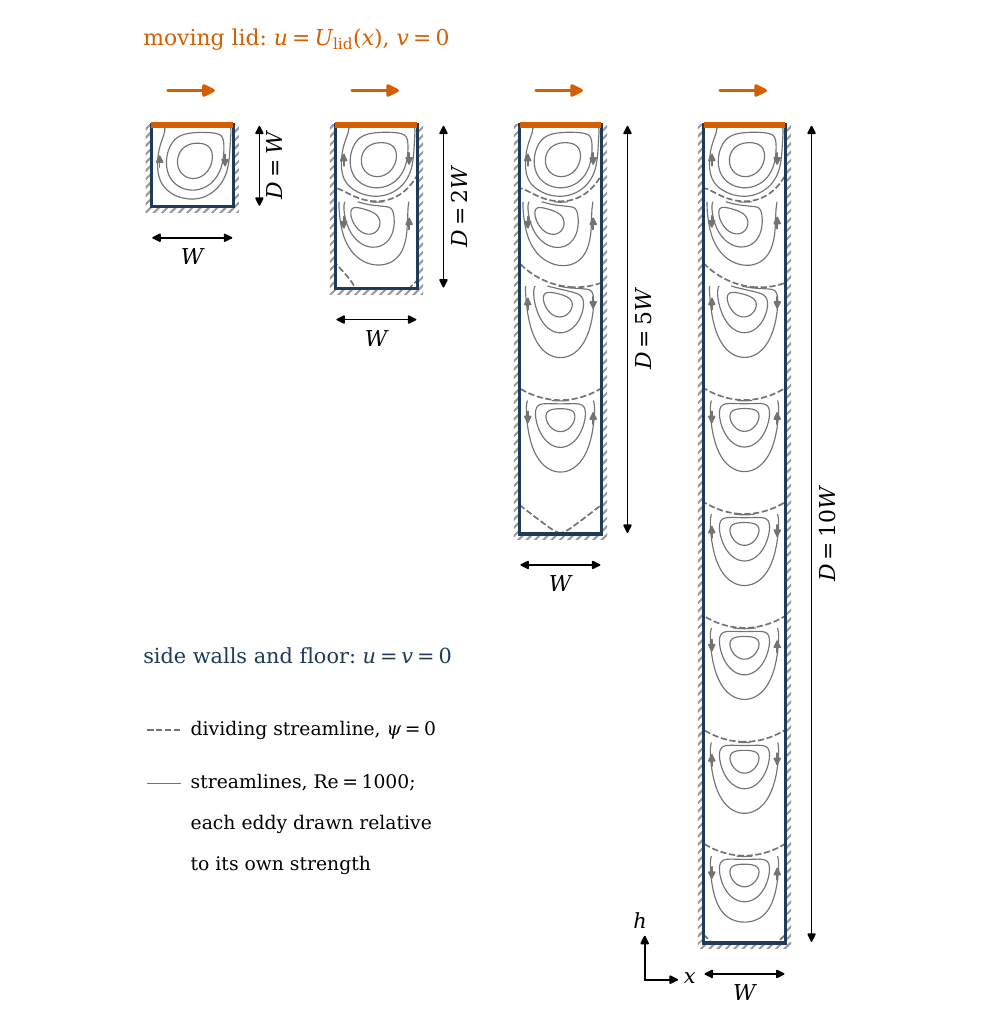}
\caption{The four cavities studied, drawn to scale with their lids aligned: width $W$ and depth $D=W$, $2W$, $5W$ and $10W$, with height $h$ measured from the floor. The lid moves to the right with the smoothed profile $U_{\rm lid}(x)$ of Eq.~(\ref{eq:lid}); the side walls and floor are at rest. Inside each cavity, thin lines are streamlines of the wall-local PINN at $\Rey=1000$; each eddy is contoured relative to its own strength, since the strengths span up to sixteen orders of magnitude. Dashed lines are the dividing streamlines $\psi=0$ and the arrows give the sense of rotation.}\label{fig:cavity-geometry}
\end{figure}

Figure~\ref{fig:cavity-geometry} shows the four cavities to scale, with the eddies that form in each at $\Rey=1000$. The lid turns the primary eddy clockwise; its lower flank drags the fluid beneath it the other way, and the same coupling repeats down to the floor, so the eddies alternate in sense. In a deep cavity the upper flow dominates both the velocity field and an unweighted momentum residual, and the weak lower eddies are easily neglected.

How weak are they? Far below the lid the flow is slow enough to be governed by the Stokes equations, and between two parallel walls a distance $W$ apart the stream function then has solutions $\psi\propto f(x)\,e^{-\lambda h/W}$. Moffatt \citep{moffatt1964eddies} showed that the slowest-decaying one has $\lambda=2(2.1061+1.1254\,\mathrm{i})$, so each eddy is $\pi W/(2\times1.1254)\approx1.40W$ long and $e^{2.1061\pi/1.1254}\approx357$ times weaker than the eddy above it. The speed then falls by a factor of ten every $0.55W$, or by about 1.83 decades per width. A cavity ten widths deep should therefore span some sixteen decades of speed from the lid to the floor, and Section~\ref{sec:lowRe} shows that it does.

\section{Neural formulation and training}\label{sec:formulation}
\subsection{Hard-constrained streamfunction--pressure representation}
The trainable weights and biases form the parameter vector $\boldsymbol{\theta}$. The scalar outputs $N_\psi$ and $N_p$ are the neural streamfunction correction and pressure. The lifting $\psi_{\rm lid}$ supplies the prescribed wall motion, $B$ is a boundary-vanishing factor, and $E$ is the depth envelope. The local coordinates $\xi$ and $\eta$ range from zero to one across the width and depth. The vertical lifting length is $\delta_y$. Depth below the lid is $d$; $d_0$ is the envelope transition depth, $\kappa$ its decay rate, and $\beta$ the inverse length controlling the smoothness of the softplus function $s_\beta$. Its argument $z$ is a scalar, and $\sigma$ denotes the logistic function.
Two network outputs $N_\psi(x,y;\vct\theta)$ and $N_p(x,y;\vct\theta)$ define
\begin{align}
 \psi&=\psi_{\rm lid}+B^2 E(y)N_\psi, &p&=E(y)N_p,\label{eq:ansatz}\\
 u&=\partial_y\psi, &v&=-\partial_x\psi,\label{eq:uv}\\
 B&=16\xi(1-\xi)\eta(1-\eta),\quad
 \xi=x/W, &\eta&=[y-(1-D)]/D.\label{eq:b}
\end{align}
The analytic lifting is
\begin{equation}
 \psi_{\rm lid}=(y-1)\eta^2
 \bigl[1-e^{-(x-W)^2/\delta_x^2}\bigr]
 \bigl[1-e^{-x^2/\delta_x^2}\bigr]
 e^{-(1-y)^2/\delta_y^2},\qquad \delta_y=0.1W.\label{eq:lifting}
\end{equation}
The factor $B^2$ makes the neural correction and its first derivatives vanish on the walls. The lifting~\eqref{eq:lifting} gives the lid motion and satisfies the conditions on the stationary walls, and Eq.~\eqref{eq:uv} satisfies continuity identically. Neither continuity nor the velocity boundary conditions therefore need penalty terms. This removes one source of competition in the optimization, but does not by itself make the momentum equations accurately solved.

The depth envelope inherited from the deep-cavity baseline is
\begin{align}
 E(d)&=\exp\{-\kappa[s_\beta(d-d_0)-s_\beta(-d_0)]\},\quad d=1-y,\label{eq:e}\\
 s_\beta(z)&=\beta^{-1}\log(1+e^{\beta z}),\quad
 \beta=10/W,\quad d_0=0.25W,\quad \kappa=4.212/W.
\end{align}
The value $\kappa=4.212/W$ is motivated by the real decay exponent in the parallel-wall Stokes-flow limit discussed by Moffatt \citep{moffatt1964eddies}. It sets the scale of the representation; it is not fitted to the CFD. Section~\ref{sec:lowRe} shows that it is in fact the decay rate of the lower eddies at every Reynolds number studied. In Eq.~\eqref{eq:ansatz}, $E(y)$ means $E(d)$ at $d=1-y$. The envelope gives a smooth depth scale but prescribes neither the number nor the location of the vortices. Its derivatives enter the velocity, which is the derivative of the complete streamfunction, and the softplus must be evaluated stably because the residual involves high derivatives of $\psi$.

\subsubsection{Differentiation of the transformed fields}
The network maps the coordinates $(x,y)$ to two scalars; it does not predict the velocity components directly. With $A=B^2$, differentiating the full representation gives
\begin{align}
 u&=\psi_{{\rm lid},y}+A_yEN_\psi+AE_yN_\psi+AE\partial_yN_\psi,\\
 v&=-\psi_{{\rm lid},x}-A_xEN_\psi-AE\partial_xN_\psi,\label{eq:product_velocity_expanded}
\end{align}
because $E$ depends only on depth. The product-rule terms are part of the physical field, so changing the envelope changes the velocity even with the same network parameters. The lifting and the first derivatives of $B$ and $E$ are evaluated analytically, and the network derivatives by automatic differentiation. The momentum residuals need derivatives of the streamfunction up to third order and of the pressure output up to first order.

For the smooth envelope in Eq.~\eqref{eq:e}, its derivative is
\begin{equation}
 E_y=\kappa\,\sigma\!\left(\beta[d-d_0]\right)E,
 \qquad \sigma(z)=\frac{1}{1+e^{-z}}.\label{eq:envelope_derivative_expanded}
\end{equation}
The sign is positive because $d=1-y$: the envelope grows toward the lid. Subtracting $s_\beta(-d_0)$ gives $E(0)=1$. Below the transition the envelope carries the depth dependence and the network a slowly varying correction. The softplus is evaluated as a log-add-exp expression, which avoids overflow and keeps the higher derivatives smooth near the transition.

On each wall $B=0$ and $\nabla(B^2)=2B\nabla B=0$, so the neural correction adds no normal or tangential velocity there, whatever the weights. At the lid $\partial_y\psi_{\rm lid}$ gives the regularized lid speed, and on the stationary walls its first derivatives vanish. Since $u_x+v_y=\psi_{yx}-\psi_{xy}=0$ everywhere, the optimizer has only the momentum balance to fit. The streamfunction lifting, depth envelope and optimizer implementation were supplied by C.~J.~McDevitt; the regional training distributions and the assessment of the deep cavity at higher Reynolds numbers are the extensions examined here.

\subsection{Residual scaling and corner treatment}
In Eq.~\eqref{eq:objective}, $J$ is the empirical loss, $N$ the number of collocation points, and $i$ their index. The quantities $M_i$, $E_i$, $R_{x,i}$ and $R_{y,i}$ are the mask, envelope and momentum residuals evaluated at point $i$. The mask $M$ uses the corner length $r_m$ defined in Eq.~\eqref{eq:mask}; the notation $MR$ below means the product of the mask and either momentum residual.
The training objective is
\begin{equation}
 J(\vct\theta)=\frac{1}{N}\sum_{i=1}^{N}
 \left[\left(\frac{M_iR_{x,i}}{E_i}\right)^2+
 \left(\frac{M_iR_{y,i}}{E_i}\right)^2\right],\label{eq:objective}
\end{equation}
with the inherited upper-corner mask
\begin{equation}
 M=\left[1-e^{-[x^2+(1-y)^2]/r_m^2}\right]
   \left[1-e^{-[(x-W)^2+(1-y)^2]/r_m^2}\right],\qquad r_m=0.01W.\label{eq:mask}
\end{equation}
Dividing by $E$ compensates for the decay of the physical residuals with depth; it scales the residual and does not modify Eqs.~\eqref{eq:rx}--\eqref{eq:ry}. The mask vanishes at the lid corners and is close to unity away from them. In addition, collocation points in the immediate corner neighborhoods are omitted, as described below. Because the mask weights the residual at the remaining points, a small weighted loss does not imply local accuracy there. The field comparisons below use the unweighted fields and include the near-lid discrepancies.

The objective contains no interior velocity, pressure or vorticity data. Transferred parameters come from earlier PINN training, not from fitting CFD, and the CFD solution is used only for comparison. The method is therefore data-free, unlike cavity methods that use interior observations.

Two variants of the scaling are tested. The ``soft residual'' variant replaces $MR/E$ by $MR/\sqrt{E}$ and keeps the output envelope. The ``half envelope'' variant uses $\kappa=2.106/W$ in both the output and the residual scale. The first changes only the emphasis of the residual; the second also changes the representation. The half-envelope case has an additional $\Rey=500$ reconditioning phase before the $\Rey=1000$ phase, so it does not isolate one factor.

\subsection{Regional collocation}
Uniform sampling allocates points by area rather than by local difficulty. The balanced sampler used here draws 50\% of its points uniformly over the cavity, 30\% from the upper one-width strip, 15\% from the lower one-width strip, and 5\% from logarithmically distributed neighborhoods of the lower corners. Upper points within a radius $0.02W$ of the lid corners are excluded. Here $r$ is distance from the selected lower corner, and the angle is measured into its interior quadrant. The logarithmic component uses $r/W\in[10^{-4},0.25]$, angles in $[0.01,\pi/2-0.01]$, and equally likely left and right corners.

All points have equal weight in Eq.~\eqref{eq:objective}, so the mixture deliberately weights the regions differently; it is neither an unbiased estimate of the area integral nor residual-adaptive. Nonuniform and residual-adaptive sampling are established approaches \citep{wu2023sampling}; here the allocation is fixed in advance for the deep cavity.

\subsection{Architecture, optimization and transfer}
The network uses six hidden tanh layers of width 64, with 21,122 trainable parameters. Weights are initialized from a Xavier-normal distribution and biases to zero, unless transferred from another model. All calculations use double precision.

Figure~\ref{fig:pinn-training-workflow} summarizes the procedure: the constrained representation, the regional residual, a SOAP phase with changing batches and a self-scaled Broyden phase on fixed points. The final stage compares the fields with CFD, which plays no part in training.

\begin{figure}[!htbp]
\centering
\begingroup
\definecolor{pinnNavy}{HTML}{183B56}
\definecolor{pinnBlue}{HTML}{EDF4FA}
\definecolor{pinnTeal}{HTML}{EAF5F1}
\definecolor{pinnGold}{HTML}{FFF5DE}
\pgfdeclarelayer{pinnpanel}
\pgfdeclarelayer{pinncard}
\pgfsetlayers{pinnpanel,pinncard,main}
\begin{tikzpicture}[
  font=\small,
  txt/.style={inner sep=0pt,outer sep=0pt,anchor=north west,
    align=left,text=black},
  card/.style={draw=pinnNavy!45,line width=.6pt,fill=white,
    rounded corners=1.6mm,inner xsep=3mm,inner ysep=2.8mm},
  panel/.style={fill=#1,draw=none,rounded corners=3mm,
    inner xsep=6mm,inner ysep=4.5mm},
  inset/.style={fill=#1,draw=none,rounded corners=1.2mm,
    inner xsep=2.6mm,inner ysep=2.4mm,text=black},
  flow/.style={-{Latex[length=2.3mm,width=1.7mm]},draw=pinnNavy,
    line width=.8pt},
  iterate/.style={flow,rounded corners=2.2mm}
]

\node[txt,text width=43mm] (n1) at (3mm,0mm) {
  \textcolor{pinnNavy}{\textbf{1\quad Specify the problem and initialization}}\\[1mm]
  Set $\mathrm{Re}$ and $D/W$; prescribe the matched lid
  (Eq.~\eqref{eq:lid}), residual variant and source PINN initialization.
};
\node[txt,text width=82mm] (n2) at (61mm,0mm) {
  \textcolor{pinnNavy}{\textbf{2\quad Construct the constrained neural fields}}\\[1mm]
  $(x,y)\mapsto(N_\psi,N_p)\mapsto(\psi,p)$
  (Eq.~\eqref{eq:ansatz}); recover $u=\psi_y$ and $v=-\psi_x$
  (Eq.~\eqref{eq:uv}).\\
  Boundary factor $B$, lid lifting $\psi_{\rm lid}$ and depth envelope $E$:
  Eqs.~\eqref{eq:b}, \eqref{eq:lifting} and \eqref{eq:e}.
};
\begin{pgfonlayer}{pinncard}
  \node[card,fit=(n1) (n1.west |- n2.south)] (c1) {};
  \node[card,fit=(n2) (n2.east |- n1.south)] (c2) {};
\end{pgfonlayer}

\node[txt,text width=140mm] (n3h) at ([xshift=3mm,yshift=-11.8mm]c1.south west) {
  \textcolor{pinnNavy}{\textbf{3\quad Sample points and assemble the physics loss}}
};
\node[txt,text width=72mm] (n3) at ([yshift=-2mm]n3h.south west) {
  Draw the regional mixture (Eq.~\eqref{eq:mixture_expanded}).
  Differentiate the complete fields and evaluate $R_x,R_y$
  (Eqs.~\eqref{eq:rx} and \eqref{eq:ry}).\\
  Apply corner mask $M$ (Eq.~\eqref{eq:mask}) and envelope scaling;
  accumulate globally normalized loss and gradient:
};
\node[inset=pinnBlue,anchor=east] (n3eq) at (n3.east -| n3h.east) {$\begin{gathered}
    J(\boldsymbol{\theta})=\frac{1}{N}\sum_{i=1}^{N}
    \left[\left(\frac{M_iR_{x,i}}{E_i}\right)^2+
          \left(\frac{M_iR_{y,i}}{E_i}\right)^2\right]\\[1mm]
    \text{(Eq.~\eqref{eq:objective}).}
  \end{gathered}$};
\begin{pgfonlayer}{pinncard}
  \node[card,fit=(n3h) (n3) (n3eq)] (c3) {};
\end{pgfonlayer}

\node[txt,text width=39mm] (n4) at ([xshift=3mm,yshift=-16.8mm]c3.south west) {
  \textcolor{pinnNavy}{\textbf{4\quad SOAP optimization: 10,000 updates}}\\[1mm]
  Minimize $J$ in Eq.~\eqref{eq:objective}; learning rate $0.002$.
  Redraw the regional batch every 50 steps.
  Recompute the loss and gradient at each update.
};
\node[txt,text width=34mm] (n5) at ([xshift=56mm,yshift=-16.8mm]c3.south west) {
  \textcolor{pinnNavy}{\textbf{5\quad Freeze the collocation set}}\\[1mm]
  Fix $N=131072$ points and retain the same objective
  (Eq.~\eqref{eq:objective}) throughout the quasi-Newton phase.
};
\node[txt,text width=39mm] (n6) at ([xshift=104mm,yshift=-16.8mm]c3.south west) {
  \textcolor{pinnNavy}{\textbf{6\quad Self-scaled Broyden: 20,000 updates}}\\[1mm]
  Compute the search direction and parameter update
  (Eq.~\eqref{eq:ssb_update}); perform line-search evaluations
  and curvature updates while minimizing the same $J$.
};
\begin{pgfonlayer}{pinncard}
  \node[card,fit=(n4) (n4.west |- n5.south) (n4.west |- n6.south)] (c4) {};
  \node[card,fit=(n5) (n5.west |- n4.south) (n5.west |- n6.south)] (c5) {};
  \node[card,fit=(n6) (n6.east |- n4.south) (n6.east |- n5.south)] (c6) {};
\end{pgfonlayer}

\node[txt,text width=140mm] (n7h) at ([xshift=3mm,yshift=-16.8mm]c4.south west) {
  \textcolor{pinnNavy}{\textbf{7\quad Independent field assessment}}
};
\node[txt,text width=88mm] (n7) at ([yshift=-2mm]n7h.south west) {
  Evaluate $u,v,p,\psi,\omega$ on common physical coordinates;
  use Eqs.~\eqref{eq:uv} and \eqref{eq:vorticity_expanded}.\\
  Compare with matched CFD fields after pressure-gauge alignment.
  Report bulk and regional discrepancies
  (velocity metric: Eq.~\eqref{eq:error}).
};
\node[inset=pinnGold,anchor=east,text width=40mm,align=center] (n7cfd)
  at (n7.east -| n7h.east) {
  \textbf{CFD fields are used for assessment only, not for training.}
};
\begin{pgfonlayer}{pinncard}
  \node[card,fit=(n7h) (n7) (n7cfd)] (c7) {};
\end{pgfonlayer}

\draw[flow] (c1.east) -- (c2.west);
\draw[flow] (c2.south) -- (c2.south |- c3.north);
\draw[flow] (c3.south -| c4.north) -- (c4.north);
\draw[flow] (c4.east) -- (c5.west);
\draw[flow] (c5.east) -- (c6.west);
\draw[flow] (c6.south) -- (c6.south |- c7.north);
\draw[iterate] ([yshift=-8mm]c4.west) -- ++(-4.5mm,0) |- ([yshift=8mm]c4.west);
\draw[iterate] ([yshift=-8mm]c6.east) -- ++(4.5mm,0) |- ([yshift=8mm]c6.east);

\begin{pgfonlayer}{pinnpanel}
  \node[panel=pinnBlue,fit=(c1) (c2) (c3)] {};
  \node[panel=pinnTeal,fit=(c4) (c5) (c6)] {};
  \node[panel=pinnGold,fit=(c7)] {};
\end{pgfonlayer}
\end{tikzpicture}
\endgroup
\caption{PINN training and independent assessment. The two-output representation enforces incompressibility and wall velocities by construction. SOAP uses periodically redrawn regional batches, followed by self-scaled Broyden optimization on a fixed point set. Equation references identify the definitions used at each stage. The iteration counts are those of the final training phase; the initialization and any earlier reconditioning differ between workflows.}
\label{fig:pinn-training-workflow}
\end{figure}

\subsubsection{Optimization on a regional objective}
The iteration index is $k$, $\boldsymbol{g}_k=\nabla_{\boldsymbol{\theta}}J(\boldsymbol{\theta}_k)$ is the loss gradient, $H_k$ approximates the inverse Hessian, $\boldsymbol{d}_k$ is the search direction, and $\alpha_k$ is its line-search step length. The constants $c_1$ and $c_2$ control sufficient decrease and curvature. The update vectors are $\boldsymbol{s}_k=\boldsymbol{\theta}_{k+1}-\boldsymbol{\theta}_k$ and $\boldsymbol{y}_k=\boldsymbol{g}_{k+1}-\boldsymbol{g}_k$; this bold $\boldsymbol{y}_k$ is not the spatial coordinate $y$.
The symbol $q$ denotes the sampling probability density. Its components $q_\Omega$, $q_{\rm upper}$, $q_{\rm floor}$ and $q_{\rm corner}$ correspond to the full domain $\Omega$, upper strip, floor strip and lower-corner neighborhoods, respectively. SOAP denotes the Shampoo-based optimizer with Adam-like updates \cite{vyas2024soap}; SSB denotes self-scaled Broyden optimization in the tables. Optimizer studies motivate examining such staged training separately from physical accuracy \cite{urban2025optimization,kiyani2025optimizer}.
The regional sampler can be written as the mixture
\begin{equation}
 q=0.50q_{\Omega}+0.30q_{\rm upper}
                  +0.15q_{\rm floor}+0.05q_{\rm corner}.\label{eq:mixture_expanded}
\end{equation}
The components overlap: a point in the floor strip can also come from the full-domain component, so the proportions allocate training effort rather than partition the domain. After the upper-corner disks are rejected, every point has equal weight, and no importance weights are applied; raising the floor and corner probabilities therefore raises their weight in the objective.

All four detailed $\Rey=1000$ workflows use the balanced mixture in their final phase, and each final history completes 10,000 SOAP and 20,000 SSB steps. They differ in their starting fields. The control and soft-residual cases start from the same $\Rey=500$ field. The refined case starts from an earlier $\Rey=1000$ model trained with 50\% full-domain, 30\% floor-strip and 20\% lower-corner points. The half-envelope case has an additional 10,000-SOAP/20,000-SSB reconditioning phase at $\Rey=500$, because changing $E$ changes the initial velocity. Two links in these histories could not be recovered from the archive: whether the precursor of the refined initializer (identified by a hash of its configuration) ran 20,000 or 30,000 steps, and the history of the extra half-envelope phase. The total training effort therefore differs between the workflows and is not known exactly. Only control and soft residual start from the same parameters; the comparison of control and refined measures the combined effect of initialization and regional sampling, not the benefit of sampling alone.

Training first applies SOAP for 10,000 steps with learning rate $0.002$, drawing a new regional batch every 50 steps. SOAP makes adaptive updates in a preconditioned coordinate system and gives the initial reduction of the weighted residual. Its batches contain 131,072 points for the four detailed workflows and 262,144 points for the original $\Rey=250$ and 500 fields. The following 20,000 self-scaled Broyden steps use a fixed set of $N=131072$ points in all cases. With gradient $\boldsymbol{g}_k$ and inverse-Hessian approximation $H_k$, the search direction and step are
\begin{equation}
 \boldsymbol{d}_k=-H_k\boldsymbol{g}_k,\qquad
 \boldsymbol{\theta}_{k+1}
 =\boldsymbol{\theta}_k+\alpha_k\boldsymbol{d}_k.
 \label{eq:ssb_update}
\end{equation}
The line search applies sufficient-decrease and curvature tests with $c_1=10^{-4}$ and $c_2=0.9$. The differences $\boldsymbol{s}_k$ and $\boldsymbol{y}_k$ update the self-scaled curvature approximation, which is stored as vector generators and small coefficient blocks rather than as a dense matrix. All generators of the phase are kept, rather than a short rolling history. Because the points are fixed, successive gradients refer to the same objective.

To limit memory, the loss is evaluated in chunks of 65,536 points; each chunk's residual sum is divided by the global $N$, so the full-batch objective and gradient are unchanged. Training uses seed 12345 and the held-out points seed 20264. A change of Reynolds number or residual definition starts a new optimizer phase from the transferred network; a resumed phase keeps both the network and the optimizer state.

\subsection{Why the boundary factor limits the accuracy near the lid}\label{sec:cause}
\begin{figure}[!htbp]\centering
\includegraphics[width=\linewidth]{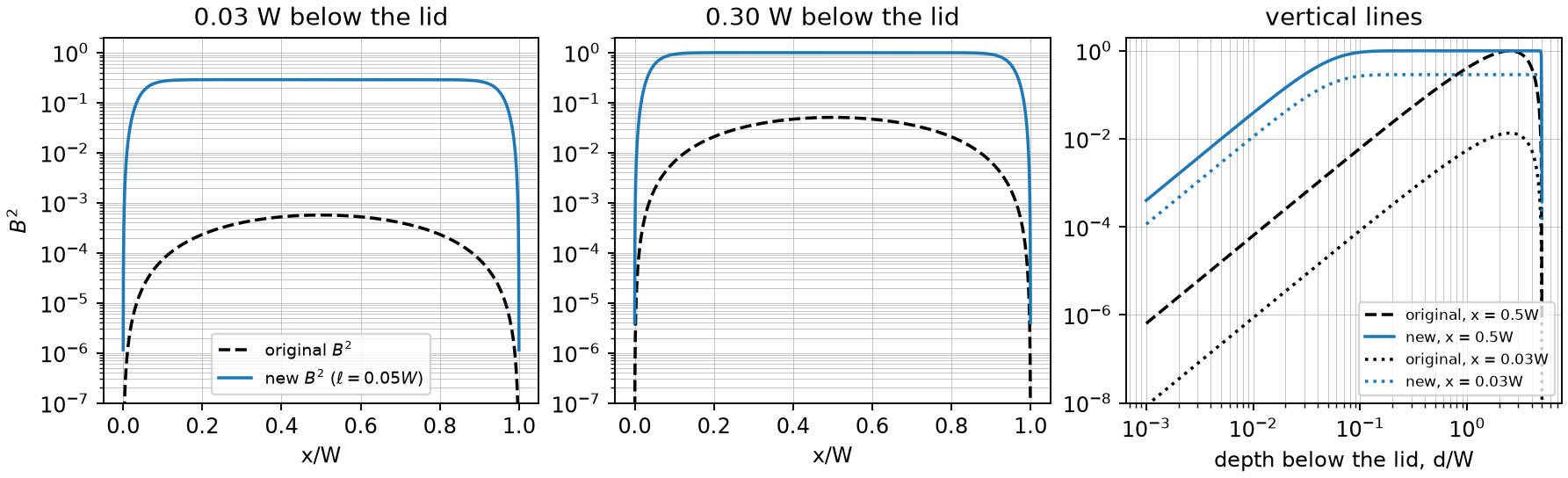}
\caption{Square of the boundary factor, which multiplies the network output in Eq.~(\ref{eq:ansatz}), for $D=5W$. Left and centre: along horizontal lines $0.03W$ and $0.3W$ below the lid. Right: along vertical lines through the centre and $0.03W$ from the left wall. The polynomial factor (\ref{eq:b}) is small everywhere in the lid layer; the wall-local factor (\ref{eq:Btanh}) is of order one beyond about $0.1W$ from every wall.}
\label{fig:bfactor}
\end{figure}

The results of Section~\ref{sec:results} show that, whatever the training workflow, almost all of the remaining vorticity error of the established formulation lies within a tenth of a width of the two lid corners; a recent study of neural pressure surrogates in a three-dimensional cavity finds the same concentration of error at the lid edges \cite{zhou2026where}. The representation (\ref{eq:ansatz}) suggests why. The boundary factor (\ref{eq:b}) is the simplest function that vanishes on the four walls, but its shape is set by the whole cavity. At a distance $d=1-y$ below the lid, $1-\eta=d/D$, so for $d\ll D$
\begin{equation}
B^2\approx 256\,\xi^2(1-\xi)^2\,\frac{d^2}{D^2}.
\label{eq:Bnear}
\end{equation}
Three hundredths of a width from both the lid and a side wall of a cavity five widths deep, $B^2$ is below $10^{-5}$ (Fig.~\ref{fig:bfactor}), while in the middle of the upper eddy it is of order $10^{-1}$. The lid shear layer and the corner flow are where $\psi$ departs most from $\psi_{\rm lid}$, so there the network output must be very large compared with the correction it is meant to produce, and it must change by several orders of magnitude over a few hundredths of a width. Networks with smooth activations learn such sharp, large-amplitude features slowly and inaccurately \cite{wang2022ntk}. Neither regional collocation nor residual scaling changes this, which is consistent with the limited effect of the workflow changes.

Equation~(\ref{eq:Bnear}) makes two predictions. The difficulty should grow with the Reynolds number, because as the lid layer thins to about $\Rey^{-1/2}W$ the correction concentrates where $B^2$ is smallest. And it should grow with depth, because near the lid $B^2$ scales as $D^{-2}$ although nothing changes in the flow there. Sections~\ref{sec:matched} and \ref{sec:depth} test both.

\subsection{Wall-local boundary factors}\label{sec:method}
Any function that vanishes linearly on each wall can replace (\ref{eq:b}). What matters, according to Eq.~(\ref{eq:Bnear}), is that the factor reaches order one within a short distance of every wall, whatever the size of the domain. We test two such factors. The first is a product of hyperbolic tangents of the four wall distances,
\begin{equation}
B_{\tanh}=\tanh\frac{x}{\ell}\,\tanh\frac{W-x}{\ell}\,\tanh\frac{1-y}{\ell}\,\tanh\frac{y-1+D}{\ell},\qquad \ell=0.05W.
\label{eq:Btanh}
\end{equation}
Close to a wall each factor equals the scaled distance to that wall, and a few $\ell$ away it approaches one. We chose $\ell$ once, comparable with the lid smoothing length and with the boundary-layer thickness at $\Rey=1000$, and did not tune it. The second is the approximate distance function (ADF) of Sukumar and Srivastava \cite{sukumar2022boundary}, built from the four wall distances with the R-conjunction $a\wedge b=a+b-\sqrt{a^2+b^2}$,
\begin{equation}
B_{\mathrm{ADF}}=\frac{\phi(x,y)}{\phi(W/2,\,1-D/2)},\qquad \phi=\bigl[x\wedge(W-x)\bigr]\wedge\bigl[(1-y)\wedge(y-1+D)\bigr].
\label{eq:Badf}
\end{equation}
It behaves like the distance to the nearest wall and keeps growing away from the walls instead of saturating. Both factors and their normal derivatives vanish on every wall, so the wall conditions remain exact, and near the lid corners both are orders of magnitude larger than the polynomial factor. Everything else in Eqs.~(\ref{eq:ansatz})--(\ref{eq:e}) is unchanged.

With the wall-local factors the network also receives the wall distances as inputs: besides $x$ and $y$, the functions $\tanh(x/s)$ and $\tanh((W-x)/s)$ for two short scales $s$, and $\tanh((1-y)/s)$ for four scales from $0.03W$ to $W$. These inputs let the first layer resolve the lid smoothing and the shear layers without very large weights, similar to Fourier features \cite{tancik2020fourier,wang2021eigenvector} but with scales attached to the walls. We call $B_{\tanh}$ with these inputs the wall-local formulation. The corner mask (\ref{eq:mask}) is removed, and only points within $10^{-3}W$ of the lid corners are excluded from the collocation set.

To separate the representation from the training history, every formulation compared in Section~\ref{sec:local}, including the polynomial factor, is trained in exactly the same way (Table~\ref{tab:setup}). Each is trained at $\Rey=100$ from a random start and continued step by step to higher Reynolds numbers, each time from the previous weights with a new optimizer. On one NVIDIA B200 GPU each step takes one to three hours. A steady finite-volume solution (Section~\ref{sec:openfoam}) on one processor core takes a few hours at $D/W=5$ and about a day at $D/W=10$ to reach the residuals given there. We do not claim a speed advantage for the PINN; the aim is to understand and remove an accuracy limit of a mesh-free formulation.

\begin{table}[htbp]\centering\small
\caption{Network, sampling and training settings for the matched comparisons of Section~\ref{sec:local}, identical for all formulations.}
\label{tab:setup}
\begin{tabular}{lp{0.72\linewidth}}\toprule
Network & 6 hidden layers of width 64, $\tanh$; 21{,}634 parameters (10 inputs) or 21{,}122 (2 inputs)\\
Precision & float64\\
Collocation points & 131{,}072 per batch: 45\% uniform, 25\% in the upper width, 10\% in the lowest width, 5\% log-uniform around the floor corners, 15\% log-uniform around the lid corners\\
Optimizer & 10{,}000 SOAP steps (new batch every 50 steps), then self-scaled Broyden on a fixed batch: 10{,}000 steps for $\Rey\le500$, 20{,}000--30{,}000 above\\
Continuation & $\Rey=100\to250\to500\to1000$ at every depth; $\to1500\to2000\to2250$ at $D/W=5$\\
Held-out test & 131{,}072 uniform points, $0.1W$ around the lid corners excluded\\
Seeds & one seed for all runs; a second seed for the wall-local formulation at $D/W=5$\\\bottomrule
\end{tabular}
\end{table}

A small training loss does not show that the solution is right. As a test that needs no reference solution, we evaluate the momentum residual on the independent held-out points and divide it by the training loss. A ratio much larger than one means that the network fits its collocation points but not the equations between them. For every run reported as converged in Section~\ref{sec:local}, the ratio is of order one.

\section{Independent CFD and evaluation protocol}\label{sec:cfd}
\subsection{\texorpdfstring{Spectral/$hp$ reference and sensitivity cases}{Spectral/hp reference and sensitivity cases}}
In Eq.~\eqref{eq:hp_expansion_expanded}, $f_h$ is a discrete scalar field on element $\Omega_e$, $e$ is the element index, $\widehat f^{\,e}_{ij}$ are its expansion coefficients, $\phi_i$ and $\phi_j$ are basis functions, and $P$ is polynomial degree. The indices $i,j$ enumerate modes from zero to $P$. Here the subscript $h$ denotes a discrete field, not height above the floor.
\subsubsection{\texorpdfstring{Spectral/$hp$ discretization and velocity correction}{Spectral/hp discretization and velocity correction}}
Superscripts $n$ and $n+1$ identify successive time levels separated by time step $\Delta t$. The vector $\boldsymbol{A}$ is the explicit multistep combination defined below and $\mathcal N$ is the convective operator. In Eq.~\eqref{eq:weak_helmholtz_expanded}, $u_h$ denotes one discrete velocity component, $F_h$ its assembled right-hand-side forcing, and $w$ a test function. The inner product is $(a,b)=\int_\Omega ab\,d\Omega$, with the dot product understood for vector arguments. In the velocity solve, $F_h$ represents the corresponding component of $\boldsymbol{A}/\Delta t-\nabla p^{n+1}$, including the boundary lifting.
The reference calculations use the incompressible Navier--Stokes solver in Nektar++ 5.9.0, with a continuous Galerkin projection on structured quadrilateral elements. Within an element, each field is expanded in a tensor
product of modified polynomial basis functions
\cite{karniadakis2005spectral},\begin{equation}
 f_h\big|_{\Omega_e}=\sum_{i=0}^{P}\sum_{j=0}^{P}
 \widehat f^{\,e}_{ij}\phi_i(\zeta_1)\phi_j(\zeta_2),\label{eq:hp_expansion_expanded}
\end{equation}
where $(\zeta_1,\zeta_2)$ are element coordinates. Each of $u,v,p$ uses $P+1$ modes in each direction: seven at order six and nine at order eight. The base mesh has $32\times160$ elements and the refined mesh $48\times240$, both with the same element aspect ratio at $D/W=5$. The nonlinear terms are dealiased.

We solve the nonlinear unsteady
equations with the velocity-correction scheme
\cite{karniadakis1991splitting}
and second-order implicit--explicit time integration, which treats advection explicitly and diffusion implicitly. In second-order multistep notation, let
\begin{equation}
 \boldsymbol{A}=2\boldsymbol{u}^n-\tfrac12\boldsymbol{u}^{n-1}
 -\Delta t\left[2\mathcal{N}(\boldsymbol{u}^n)
                     -\mathcal{N}(\boldsymbol{u}^{n-1})\right],
 \quad \mathcal{N}(\boldsymbol{u})=\boldsymbol{u}\cdot\nabla\boldsymbol{u}.
\end{equation}
After startup, each step solves a pressure Poisson problem and one Helmholtz problem per velocity component:
\begin{align}
 \nabla^2p^{n+1}&=\frac{\nabla\cdot\boldsymbol{A}}{\Delta t},\\
 \left(\frac{3}{2\Delta t}-\Rey^{-1}\nabla^2\right)
 \boldsymbol{u}^{n+1}&=\frac{\boldsymbol{A}}{\Delta t}-\nabla p^{n+1}.
 \label{eq:velocity_correction_expanded}
\end{align}
Pressure and velocity are thus obtained from separate elliptic problems rather than one coupled system. The weak Helmholtz problem is
\begin{equation}
 \frac{3}{2\Delta t}(w,u_h)+\Rey^{-1}(\nabla w,\nabla u_h)
 =(w,F_h),\label{eq:weak_helmholtz_expanded}
\end{equation}
for test functions $w$ that vanish on velocity Dirichlet boundaries, with the boundary lifting included in the assembly; the pressure problem is assembled in the same way.

Both velocity components satisfy Dirichlet conditions: zero on the stationary walls and the regularized lid speed on the lid. Pressure uses Nektar++'s high-order boundary condition, \texttt{USERDEFINEDTYPE="H"} in the Extensible Markup Language (XML) session file. The accompanying \texttt{VALUE="0"} does not set the normal pressure derivative to zero; the solver constructs the condition consistent with velocity correction. No coupled Stokes solver, substepping or selective-frequency damping is used. Because the CFD is time-dependent and the PINN steady, each comparison states the physical time of the CFD snapshot.

The base calculation uses $\Delta t=2\times10^{-4}$. Polynomial order, mesh and time step are refined one at a time, with the same lid and viscosity, and the fields are compared at the same physical time, so that discretization effects are not confused with the slow evolution of the flow. We compare the sensitivity fields at $t\simeq170$; the later base field at $t\simeq317$ is used for the PINN comparisons. The numerical framework and splitting method follow \cite{cantwell2015nektar,moxey2020nektar,karniadakis1991splitting,karniadakis2005spectral}.

The fields are sampled on a common $201\times1001$ grid with $\Delta x=\Delta h=0.005W$. Values duplicated at element boundaries are averaged; they did not differ. Bilinear resampling between the different node locations adds an interpolation error, so we report a sampled vector norm, not a spectral-element quadrature norm or a Richardson-extrapolated estimate.

Refining the polynomial order or the mesh changes the interior velocity by less than a tenth of a per cent, and halving the time step changes it more than a hundred times less (Table~\ref{tab:cfd}). The change between $t\simeq160$ and 170 is also small, so the flow evolves slowly at this time.
\begin{table}[htbp]\centering\small
\caption{Common-time CFD sensitivity at $\Rey=1000$, $D/W=5$, $t\simeq170$. Interior excludes a $0.02W$ wall band. Differences are relative sampled vector norms against the base field, except the last row, on the $201\times1001$ comparison grid. Floor uses $0.02<x/W<0.98$ and $0.02<h/W<0.4$; its normalization uses the base velocity RMS on those same samples. The last regional entry is unavailable because the $t\simeq160$ source field was not recovered.}\label{tab:cfd}
\begin{tabular}{lrrrrr}\toprule
Case & Order & Elements & $\Delta t$ & Interior (\%) & Floor (\%)\\\midrule
Base & 6 & $32\times160$ & $2\times10^{-4}$ & --- & ---\\
Polynomial refinement & 8 & $32\times160$ & $2\times10^{-4}$ & 0.054 & 0.0045\\
Mesh refinement & 6 & $48\times240$ & $2\times10^{-4}$ & 0.065 & 0.074\\
Time-step halving & 6 & $32\times160$ & $10^{-4}$ & 0.0003 & 0.0001\\
$t=160\rightarrow170$, same $\Delta t/2$ mesh & 6 & $32\times160$ & $10^{-4}$ & 0.003 & \textemdash\\\bottomrule
\end{tabular}
\end{table}

The floor velocity RMS of the reference at this time is $3.1\times10^{-8}U$. Pressure offsets, pressure-shape differences and vorticity differences for the interior, upper-left and floor regions are given in the accompanying data. The profile overlays use a separate $321\times1601$ display grid.

\begin{figure}[!htbp]
\centering
\includegraphics[width=\textwidth]{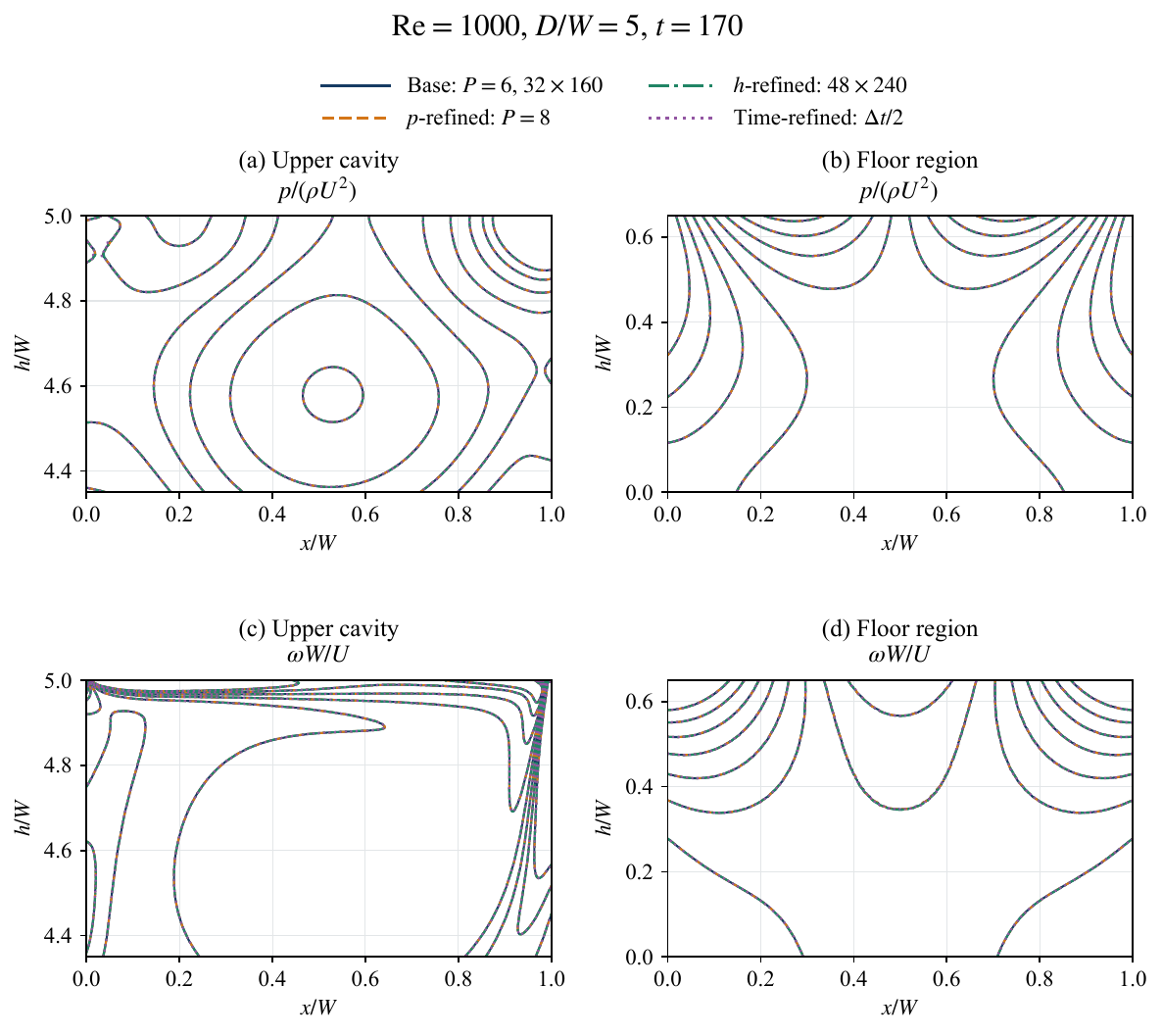}
\caption{Discretization sensitivity of pressure (top) and vorticity (bottom) at $\mathrm{Re}=1000$, $D/W=5$ and the matched physical time $t=170$. Left: upper cavity; right: floor region. The base calculation ($P=6$, $32\times160$ elements, $\Delta t=2\times10^{-4}$) is compared with polynomial refinement ($P=8$), mesh refinement ($48\times240$ elements) and time-step refinement ($\Delta t=10^{-4}$), varied separately. Identical contour levels are used for all four calculations in each panel. Pressure is referenced to the mean in each displayed region. All fields are sampled on a common physical grid; vorticity is evaluated using the same differentiation operator.}
\label{fig:cfd-pressure-vorticity-sensitivity}
\end{figure}

In Fig.~\ref{fig:cfd-pressure-vorticity-sensitivity} the pressure and vorticity contours of the four calculations are almost indistinguishable, both near the lid and near the floor.

\subsection{Steady finite-volume reference}\label{sec:openfoam}
For the depth study and the higher Reynolds numbers we use a second, independent reference: a steady solution with OpenFOAM (v2506), obtained with the SIMPLE algorithm, second-order linear-upwind convection and central diffusion, on $160\times[600+100(D/W-2)]$ cells graded towards the walls ($160\times320$ for the square cavity),
 with the lid profile (\ref{eq:lid}) imposed through the same Gaussian expression. The iterations are continued until the normalized residuals are below $10^{-12}$. For $D/W=2$ and 10 and for $\Rey=2000$, new spectral/$hp$ solutions were computed with the same session settings on $32\times(32D/W)$ elements. The two solvers are weak in different places. The steady solution involves no time integration, so it gives the weak lower eddies reliably, but its second-order discretization error is larger than that of the spectral solution in the strong upper flow. At $D/W=5$ and $\Rey=1000$ the wall-local PINN of Section~\ref{sec:local} differs from the spectral solution by less than a tenth of a per cent in interior velocity and from the steady solution by about a quarter of a per cent, so the steady solution carries an error of this order there.

\subsection{Why a time-stepped solution is slow to converge near the floor}\label{sec:nektar}
The two solvers agree in the upper part of the cavity and, in the deeper and faster cases, disagree near the floor. The reason is the time integration rather than the spatial resolution of the spectral solution. Each eddy is driven by viscous stresses from the eddy above it. The time for momentum to diffuse across one eddy is of order $W^2/\nu=\Rey\,W/U$, and the motion reaches an eddy only after the eddies above it have formed, so the spin-up time of the lower eddies grows with both the Reynolds number and the depth. At $\Rey=2000$ the spectral solution at $D/W=5$ was integrated to $t=400\,W/U$, a fifth of a single viscous time; at $D/W=10$ and $\Rey=1000$ it was integrated to $t=480\,W/U$, against a diffusion time of order $(DW)^2/\nu$ over the whole depth. In both cases the lowest eddies had not developed. At $\Rey=2000$ the floor eddy of the spectral solution was still growing between $t=320$ and $400$ towards the steady values (Section~\ref{sec:highRe}). At $D/W=10$ the spectral solution below mid-depth was a nearly linear profile of about $10^{-8}U$, while the steady solution continues to decay eddy by eddy (Section~\ref{sec:depth}). In a test at $\Rey=100$, tightening the tolerance of the iterative linear solvers by several orders of magnitude did not change this profile. Splitting schemes of the velocity-correction type also carry a time-step-dependent error near the walls \cite{guermond2006overview}; we have not separated the two effects, but at this level either would affect only velocities below about $10^{-8}U$. Integrating to a steady state from rest is the usual procedure for cavity benchmarks \cite{patil2006deep,zhuo2013deep}; in a deep cavity it requires integration times that are impractical, and a steady solver is the better reference for the weakest eddies. The same consideration applies to the floor comparisons of Section~\ref{sec:results}, which use spectral solutions at $t=120$--$317\,W/U$.

A steady solver has its own limit in the deepest cavity. In double precision, velocities below about $10^{-15}U$ are lost to round-off, because they are computed as small differences between order-one quantities. For $D/W=10$ we therefore also solve the lowest four widths separately. The steady equations are unchanged by $u\to Su$, $\nu\to S\nu$, $p\to S^2p$, so we solve this sub-domain with $S=10^8$, on the same cells, with the velocity on its upper boundary taken from the full solution, where it is still accurate to six digits. The flow is the same, but its smallest velocities are now well above round-off.

\subsection{Field and regional comparison}
Table~\ref{tab:assessment-provenance} lists the neural fields and the reference used in each comparison.
\begin{table}[htbp]\centering\small
\caption{Neural fields and references used in each comparison. The two neural assessments use different sampling grids and floor masks. The Reynolds-number series uses the control, not the refined field, at $\Rey=1000$.}\label{tab:assessment-provenance}
\begin{tabular}{p{3cm}p{5.6cm}p{5cm}}\toprule
Assessment & Neural field and preparation & Reference and evaluation\\
\midrule
Common Reynolds series & Original envelope fields at $\Rey=100$, 250 and 500; balanced control at 1000 initialized from 500 & $121\times605$ grid; times in Table~\ref{tab:re-sources}; floor $0.02<h/W<0.65$\\
Detailed control and soft residual & Same specified $\Rey=500$ source; same final balanced sampler; residual definitions differ & $161\times801$ grid; CFD $t\simeq317$; floor $0.02<h/W<0.4$\\
Detailed refined workflow & Prior floor-refined $\Rey=1000$ source, then balanced final sampling & Same detailed grid, reference, gauge and masks\\
Detailed half envelope & Changed envelope and additional $\Rey=500$ reconditioning before final phase & Same detailed grid, reference, gauge and masks\\
CFD sensitivity & No neural training; order, mesh and time-step variants & Matched $t\simeq170$; $201\times1001$ sampled grid\\
\bottomrule
\end{tabular}
\end{table}

The detailed $\Rey=1000$ comparisons use a $161\times801$ grid and the CFD field at $t\simeq317$; the Reynolds-number series uses a $121\times605$ grid and the reference times of Table~\ref{tab:re-sources}. Numbers from the two should not be mixed. Let $\mathcal R$ be a set of $N_{\mathcal R}$ grid points, indexed by $i$. The superscript CFD denotes the reference, and fields without superscript are the PINN prediction. We report dimensional root-mean-square (RMS) discrepancies,
\begin{equation}
\begin{aligned}
 E_{\boldsymbol u}(\mathcal R)&=U\left[\frac{1}{N_{\mathcal R}}\sum_{i\in\mathcal R}\bigl((u_i-u_i^{\rm CFD})^2+(v_i-v_i^{\rm CFD})^2\bigr)\right]^{1/2},\\
 E_p(\mathcal R)&=\rho U^2\left[\frac{1}{N_{\mathcal R}}\sum_{i\in\mathcal R}(p_{g,i}-p_{g,i}^{\rm CFD})^2\right]^{1/2},\\
 E_\omega(\mathcal R)&=\frac{U}{W}\left[\frac{1}{N_{\mathcal R}}\sum_{i\in\mathcal R}(\omega_i-\omega_i^{\rm CFD})^2\right]^{1/2}.
\end{aligned}\label{eq:error}
\end{equation}
The means are taken over the grid points of the stated region, and the velocity RMS includes both components without a factor of one half. The tables list $E_{\boldsymbol u}/U$, $E_p/(\rho U^2)$ and $E_\omega W/U$, which are absolute differences on the lid scales. Because the lower flow is weak, these absolute differences are small there even when the relative error is not. We therefore also give profiles, which show the local amplitudes, and ratios to the reference RMS of the same region (Table~\ref{tab:matched-velocity-nrmse} and Appendix~\ref{app:matched-diagnostics}).

The interior is $0.02<x/W<0.98$, $0.02<h/W<4.98$; it excludes a band of width $0.02W$ along every wall, and with it the corners. Within it, the upper-left region is $x/W<0.25$, $h/W>4.75$ and the floor region $h/W<0.4$. These evaluation regions are unrelated to the training mask, and the discrepancies are unweighted. For pressure, each field's mean over the interior is subtracted, giving $p_g$, and the same gauge is used in the upper-left and floor regions. Diagnostics labeled `shape' also subtract the regional mean of the difference.

For a region $\mathcal R$, write $\langle f\rangle_{\mathcal R}=N_{\mathcal R}^{-1}\sum_{i\in\mathcal R}f_i$. The matched-reference diagnostics are
\begin{equation}
 \mathrm{NRMSE}_{\boldsymbol u}=\frac{E_{\boldsymbol u}/U}{\langle (u^{\rm CFD})^2+(v^{\rm CFD})^2\rangle_{\mathcal R}^{1/2}},\qquad
 \mathrm{NRMSE}_{\omega}=\frac{E_{\omega}W/U}{\langle(\omega^{\rm CFD})^2\rangle_{\mathcal R}^{1/2}}.
 \label{eq:matched-nrmse}
\end{equation}
Numerator and denominator use the same points and uniform weights, so the ratios are regional, not pointwise, and never divide by a near-zero local velocity. The pressure NRMSE is normalized by the RMS of the gauge-aligned reference pressure, and the shape NRMSE by that of the regionally centered reference; where this denominator is tiny, it is reported with the ratio.

Cases are ranked by their discrepancy from CFD, not by training loss or held-out residual, which measure different quantities. Streamlines are integrated from the sampled velocities with the same seeds for every field, and are supplemented by horizontal profiles of velocity, pressure and vorticity at $h/W=4.90$, 4.95 and 4.98. Streamlines show the geometry of the flow but measure neither its circulation nor the area enclosed by a separatrix. Locating a stagnation point would require solving for zeros of the velocity and classifying its Jacobian: in this incompressible flow a positive determinant indicates an elliptic point and a negative one a saddle.
\FloatBarrier

\subsection{Comparison-grid and interpolation sensitivity}\label{sec:grid-sensitivity}
How much does the comparison grid matter? The same $\Rey=1000$ control field and CFD snapshot were sampled bilinearly on both grids. Going from $121\times605$ to $161\times801$, with the same masks and gauge, changes the interior RMS discrepancy by about 10\% in velocity, 5\% in pressure and 20\% in vorticity (Table~\ref{tab:retained-grid-sensitivity}). Vorticity changes most, since both the sampled locations and its finite-difference operator change with the grid. The pressure here is resampled directly from the native CFD snapshot; the Reynolds-series table is unchanged.

The $321\times1601$ row resamples the $161\times801$ neural field and so measures only the postprocessing; the network has not yet been evaluated directly on finer grids.

\begin{table}[htbp]\centering\small
\caption{Comparison-grid sensitivity using the same control arrays and the same $t\simeq317$ CFD source, original bilinear sampling, interior mask, gauge, and uniform sample weights. Vorticity is recomputed from each sampled velocity grid. The $321\times1601$ row samples the $161\times801$ neural-field interpolant; it is not direct network refinement or evidence of CFD/neural convergence.}\label{tab:retained-grid-sensitivity}
\begin{tabular}{lrrr}\toprule
Grid & $E_{\boldsymbol u}/U$ & $E_p/(\rho U^2)$ & $E_\omega W/U$\\\midrule
$121\times605$ & $6.542\times10^{-3}$ & $2.665\times10^{-3}$ & $6.649\times10^{-1}$\\
$161\times801$ & $7.181\times10^{-3}$ & $2.793\times10^{-3}$ & $8.009\times10^{-1}$\\
$321\times1601$ & $7.053\times10^{-3}$ & $2.855\times10^{-3}$ & $8.660\times10^{-1}$\\
\bottomrule\end{tabular}\end{table}
\FloatBarrier

\section{Results of the established formulation}\label{sec:results}
We first compare the four Reynolds numbers on a common grid, and then the training workflows at $\Rey=1000$.
\subsection{A common assessment at Reynolds numbers 100, 250, 500 and 1000}\label{sec:reseries}
The four $D/W=5$ cases are processed in the same way. At $\Rey=100$, 250 and 500 the neural fields come from the original envelope networks; at $\Rey=1000$ we use the balanced control, which starts from the $\Rey=500$ field and uses the regional sampling. The training therefore changes along with the Reynolds number. The reference fields are Nektar++ solutions with the regularized lid~\eqref{eq:lid} (Table~\ref{tab:re-sources}).

All velocity and pressure fields are sampled on the same $121\times605$ grid, with $0.002\leq x/W\leq0.998$ and $0.002\leq h/W\leq4.998$. The $\Rey=100$ arrays are already on this grid; at $\Rey=250$ and 500 the networks are evaluated on it in double precision, and the $\Rey=1000$ neural arrays and all CFD fields are sampled bilinearly. One mean is removed from each pressure field over $0.02<x/W<0.98$, $0.02<h/W<4.98$, and $\omega=\partial_xv-\partial_hu$ is computed from every sampled velocity with the same second-order difference operator. The vorticity discrepancy therefore includes interpolation and differencing errors. Because this series uses a coarser grid and a taller floor region ($0.02<h/W<0.65$) than the detailed comparison, its discrepancies for the same control field differ from the detailed ones (Table~\ref{tab:retained-grid-sensitivity}).

\begin{table}[htbp]\centering\small
\caption{Sources for the common $D/W=5$ Reynolds-number assessment. Time is in units of $W/U$. The saved CFD times differ across the series.}\label{tab:re-sources}
\begin{tabular}{rrp{90mm}}\toprule
$\Rey$ & CFD time & Neural field\\\midrule
100 & 120 & Original depth-envelope comparison arrays\\
250 & 160 & Original formulation, completed 20,000-step SSB weights\\
500 & 160 & Original formulation, completed 20,000-step SSB weights\\
1000 & 317 & Balanced control, initialized from the completed $\Rey=500$ field\\\bottomrule
\end{tabular}
\end{table}

\begin{figure}[H]\centering
\includegraphics[width=\linewidth]{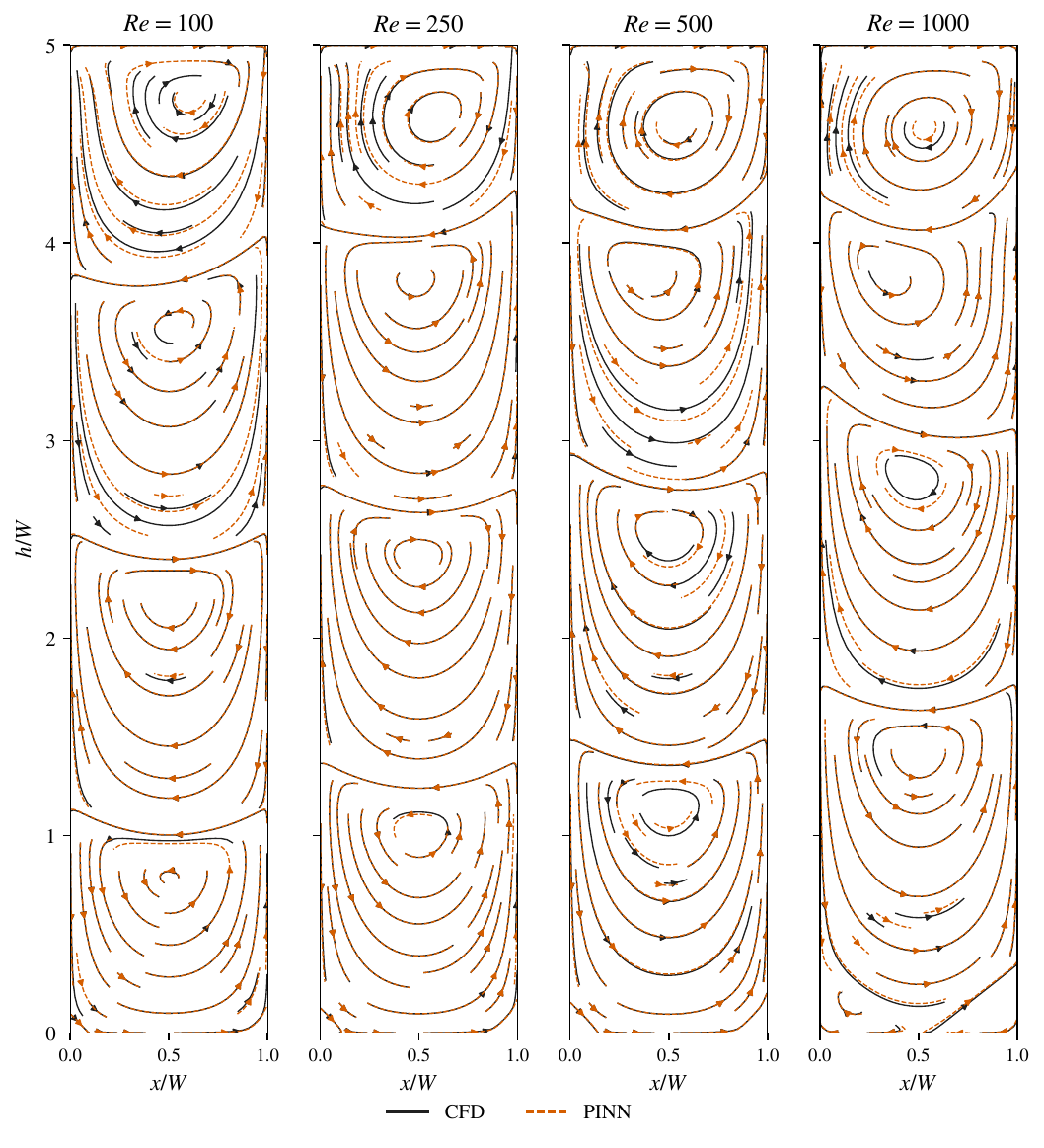}
\caption{Full-depth streamline comparison at $D/W=5$, $\Rey=100$, 250, 500 and 1000, from left to right. Black solid integral curves are CFD, and orange dashed curves are PINN. Both use identical integration and automatic-seeding parameters on the frozen $121\times605$ velocity grid. CFD times are $tU/W=120$, 160, 160 and 317, respectively. Streamline density is not a measure of velocity magnitude.}\label{fig:re-streamlines}
\end{figure}

\begin{figure}[H]\centering
\includegraphics[width=\linewidth]{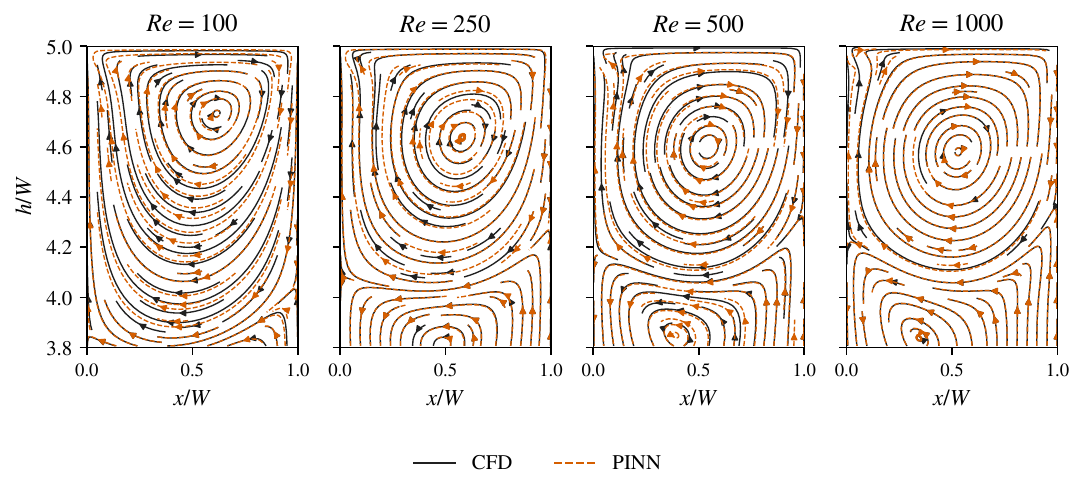}
\caption{Upper-cavity streamline enlargement, $3.8\leq h/W\leq5$, for $\Rey=100$, 250, 500 and 1000 at $D/W=5$. Colors, line styles and reference times are those of Fig.~\ref{fig:re-streamlines}. The enlarged view resolves changes in the primary circulation and differences between the CFD and neural fields.}\label{fig:re-streamlines-upper}
\end{figure}

Figures~\ref{fig:re-streamlines} and \ref{fig:re-streamlines-upper} show the streamlines. Four prominent vortices of alternating rotation are visible over the depth. As the Reynolds number increases, the intermediate and lower vortex centers move upward and the primary vortex becomes confined to the upper cavity, changing from a vertically elongated circulation at $\Rey=100$ to a rounder core at $\Rey=500$ and 1000. The PINN reproduces the same stack, with local differences near the core and along the dividing streamlines. The streamlines show only geometry: the lower vortices look as prominent as the upper ones but are far weaker, as the depth amplitudes below show.

Why does the primary vortex become rounder? In Eq.~\eqref{eq:vort_transport_expanded} a larger Reynolds number weakens diffusion relative to advection, so near the lid the circulation itself redistributes vorticity, and a round core can coexist with sharper gradients at the walls. The lower vortices have a much smaller local Reynolds number and are coupled to the upper flow only through weak viscous stresses. Their upward shift in Fig.~\ref{fig:re-streamlines} is a rearrangement of the stack, not a sign that they too become dominated by inertia.

\FloatBarrier

\begin{figure}[H]\centering
\includegraphics[width=\linewidth]{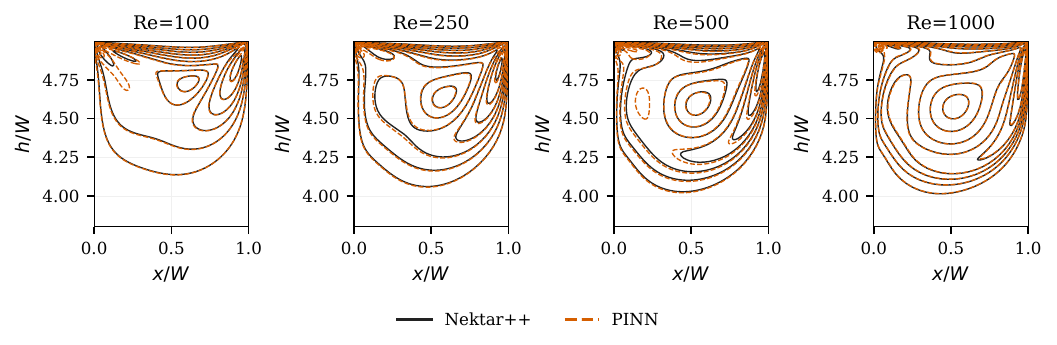}
\caption{Speed contours over $3.8\leq h/W\leq4.998$ at $D/W=5$, with $\Rey=100$, 250, 500 and 1000 from left to right. Black solid lines are Nektar++ and orange dashed lines are PINN. The contour index $j$ enumerates the stated levels. Shared levels are $0.0856j$, $j=1,\ldots,9$, in units of $U$. The blank lower portion is omitted.}\label{fig:re-speed-full}
\end{figure}

\begin{figure}[H]\centering
\includegraphics[width=\linewidth]{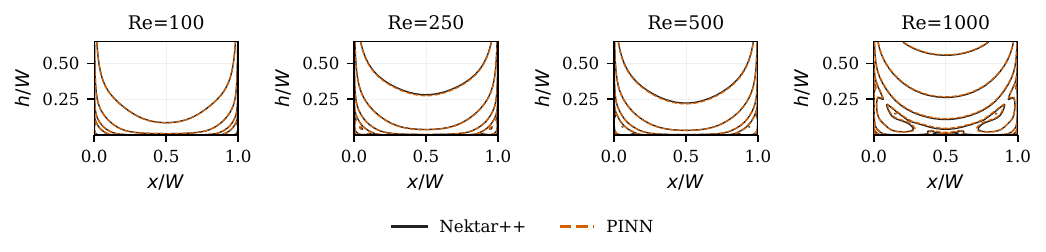}
\caption{Floor speed enlargement at $D/W=5$, $\Rey=100$, 250, 500 and 1000. The upper row is omitted because it repeats Fig.~\ref{fig:re-speed-full}. Line styles are those of that figure. Levels are $1.48\times10^{-12}\,10^{5j/6}$, $j=0,\ldots,6$, in units of $U$.}\label{fig:re-speed-zoom}
\end{figure}

The speed contours of Figs.~\ref{fig:re-speed-full} and \ref{fig:re-speed-zoom} show the amplitudes that the streamlines do not. CFD and PINN share the contour levels in each panel, and the floor levels are the same at all four Reynolds numbers, so the floor panels show the weak lower motion on its own scale.

\FloatBarrier

\begin{figure}[H]\centering
\includegraphics[width=\linewidth]{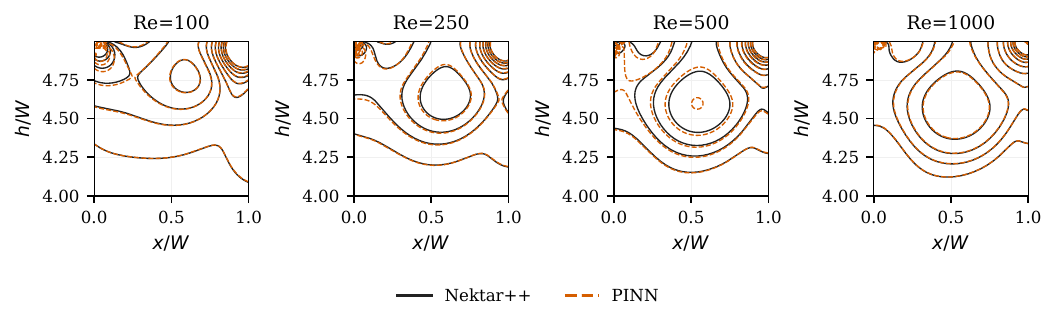}
\caption{Pressure contours over $4\leq h/W\leq4.998$ at $D/W=5$, $\Rey=100$, 250, 500 and 1000, in units of $\rho U^2$ after interior mean alignment. Black solid lines are CFD; orange dashed lines are PINN. Shared levels are $0.0277j$, $j=-5,\ldots,5$.}\label{fig:re-p-full}
\end{figure}

\begin{figure}[H]\centering
\includegraphics[width=\linewidth]{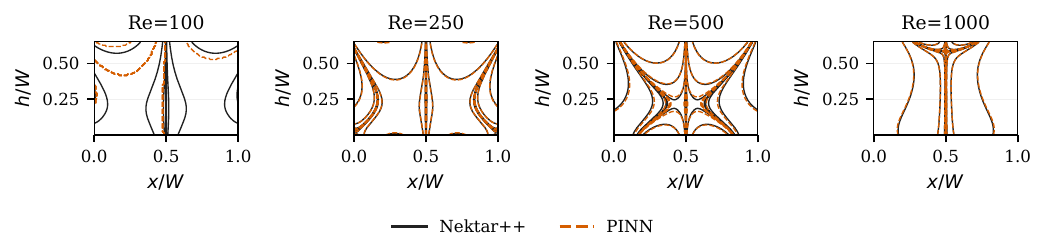}
\caption{Floor pressure-contour enlargement at $D/W=5$, $\Rey=100$, 250, 500 and 1000 with the mean of each displayed strip removed. The repeated upper row is omitted. Levels are $\pm1.11\times10^{-14}\,10^{5j/6}$, $j=0,\ldots,6$, shared across the four cases in units of $\rho U^2$.}\label{fig:re-p-zoom}
\end{figure}

In the pressure contours of Figs.~\ref{fig:re-p-full} and \ref{fig:re-p-zoom}, pressure gradients turn the fast fluid around the main vortex and near the sidewalls. In the floor view the mean of each strip is removed to show its weak variation; this is for display only, and all pressure errors and profiles use the single interior gauge.

\FloatBarrier

\begin{figure}[H]\centering
\includegraphics[width=\linewidth]{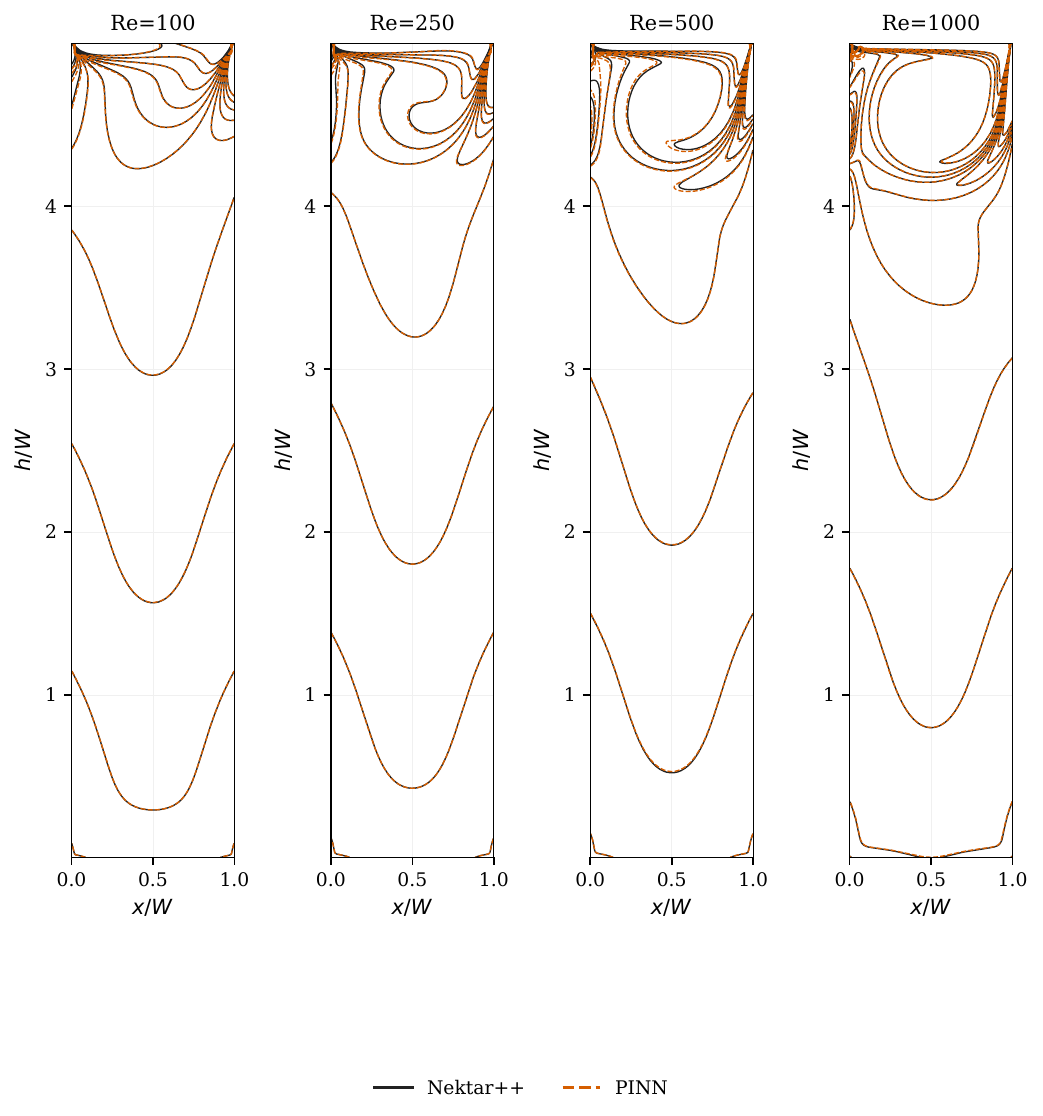}
\caption{Full-depth sampled vorticity at $D/W=5$, $\Rey=100$, 250, 500 and 1000 in units of $U/W$. The same velocity-difference operator is used for all cases. Black solid contours denote CFD and orange dashed contours PINN. Shared levels are $1.25j$, $j=-5,\ldots,5$.}\label{fig:re-omega-full}
\end{figure}

\begin{figure}[H]\centering
\includegraphics[width=\linewidth]{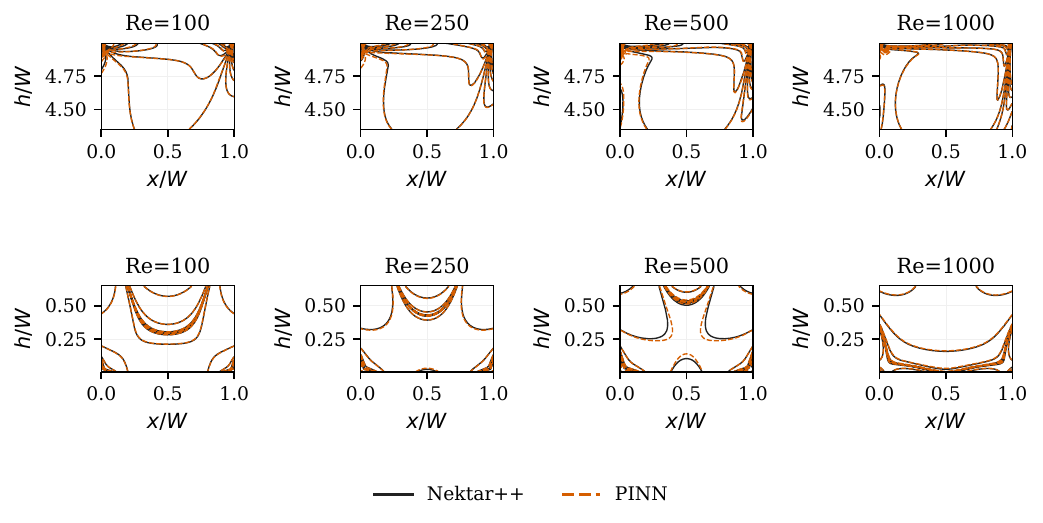}
\caption{Upper and floor vorticity-contour enlargements for $\Rey=100$, 250, 500 and 1000. Upper levels are $3.93j$, $j=-5,\ldots,5$; floor levels are $\pm5.34\times10^{-12}\,10^{5j/6}$, $j=0,\ldots,6$, in units of $U/W$. Shared levels permit comparison of shape and amplitude across Reynolds numbers.}\label{fig:re-omega-zoom}
\end{figure}

The vorticity changes sign throughout the depth, so Figs.~\ref{fig:re-omega-full} and \ref{fig:re-omega-zoom} show the full cavity as well as enlargements of the strong near-lid gradients and the weak floor structure.

These trends agree with earlier CFD of deep cavities. Cheng and Hung \cite{cheng2006deep} considered depth-to-width ratios from 0.1 to 7 and Reynolds numbers from 0.01 to 5000; they found that the near-lid vortex changes size and position while the deeper vortices feel less inertia and are more nearly symmetric. Zhuo et al.\ \cite{zhuo2013deep} considered ratios from 1.5 to 7 and Reynolds numbers from 100 to 5000 and reported vortex positions, strengths and sizes. Different lid conditions and reference times prevent a quantitative comparison of vortex centers or circulation with those studies.

\FloatBarrier

\begin{figure}[!htbp]\centering
\includegraphics[width=\linewidth]{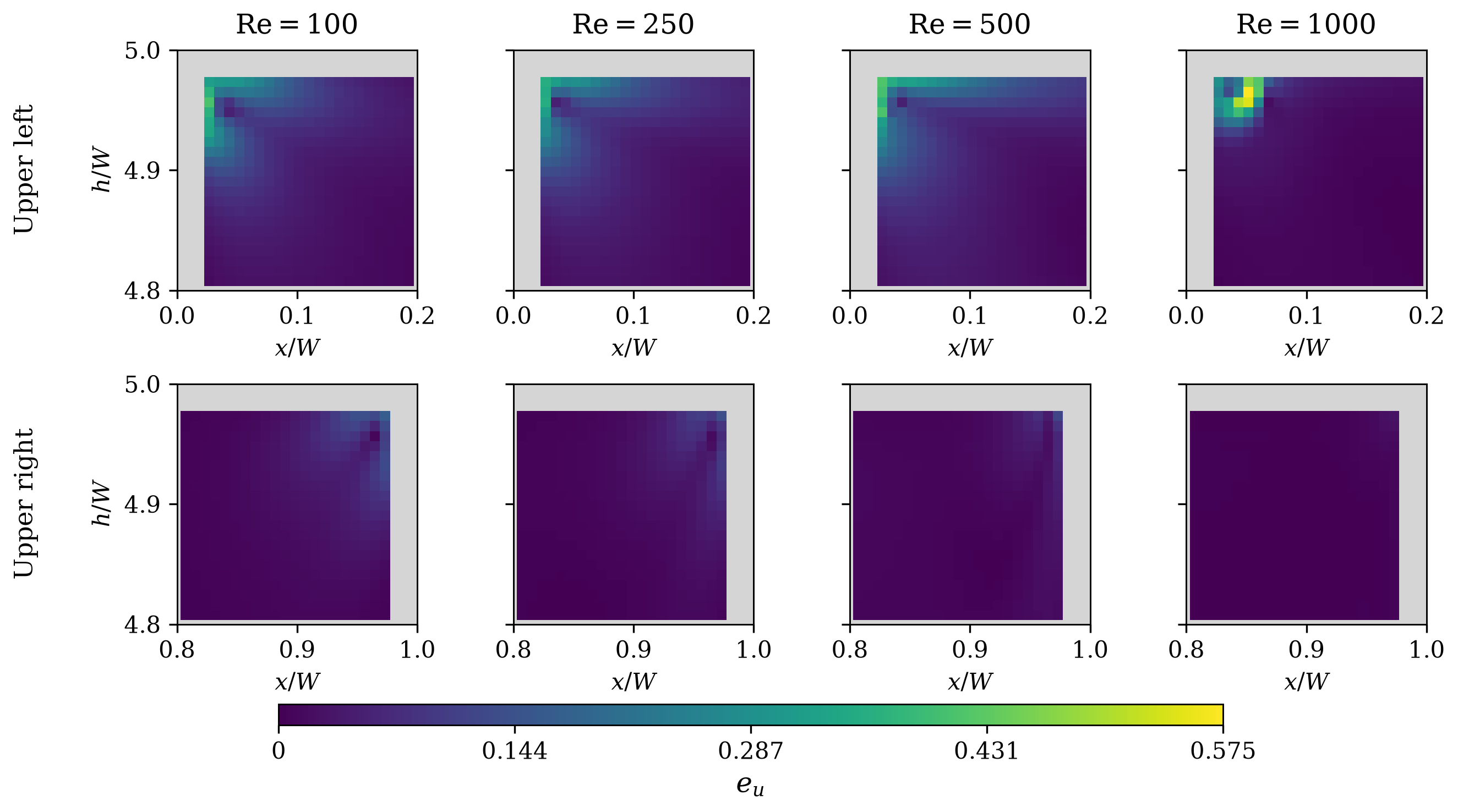}
\caption{Dimensionless velocity-vector discrepancy $e_u$ from Eq.~\eqref{eq:re-spatial-error-physical} at $D/W=5$, $\Rey=100$, 250, 500 and 1000 from left to right. Upper-left windows ($0\le x/W\le0.2$, top row) and upper-right windows ($0.8\le x/W\le1$, bottom row) both cover $4.8\le h/W\le5$ with equal spatial aspect ratios. The linear colour scale extends to the largest interior value; gray marks the excluded wall band.}\label{fig:re-velocity-error-physical}
\end{figure}

\begin{figure}[!htbp]\centering
\includegraphics[width=\linewidth]{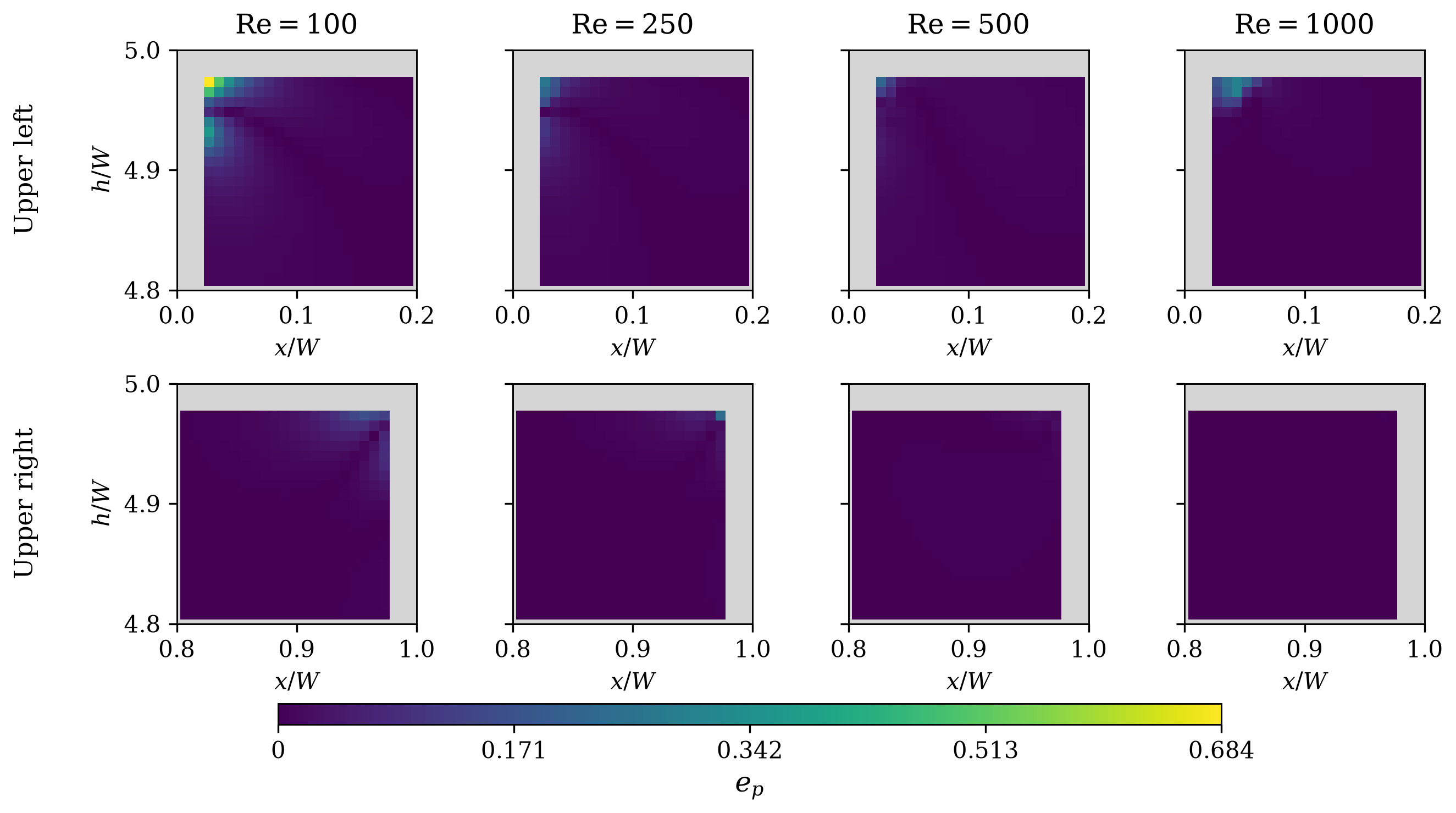}
\caption{Dimensionless gauge-aligned pressure discrepancy $e_p$, with the same upper-left (top) and upper-right (bottom) windows and Reynolds-number columns as Fig.~\ref{fig:re-velocity-error-physical}. Pressure retains the single full-interior mean alignment; no local recentering is applied. The linear colour scale extends to the largest interior value. The broader $\Rey=500$ pressure discrepancy also extends outside these windows.}\label{fig:re-pressure-error-physical}
\end{figure}

\begin{figure}[!htbp]\centering
\includegraphics[width=\linewidth]{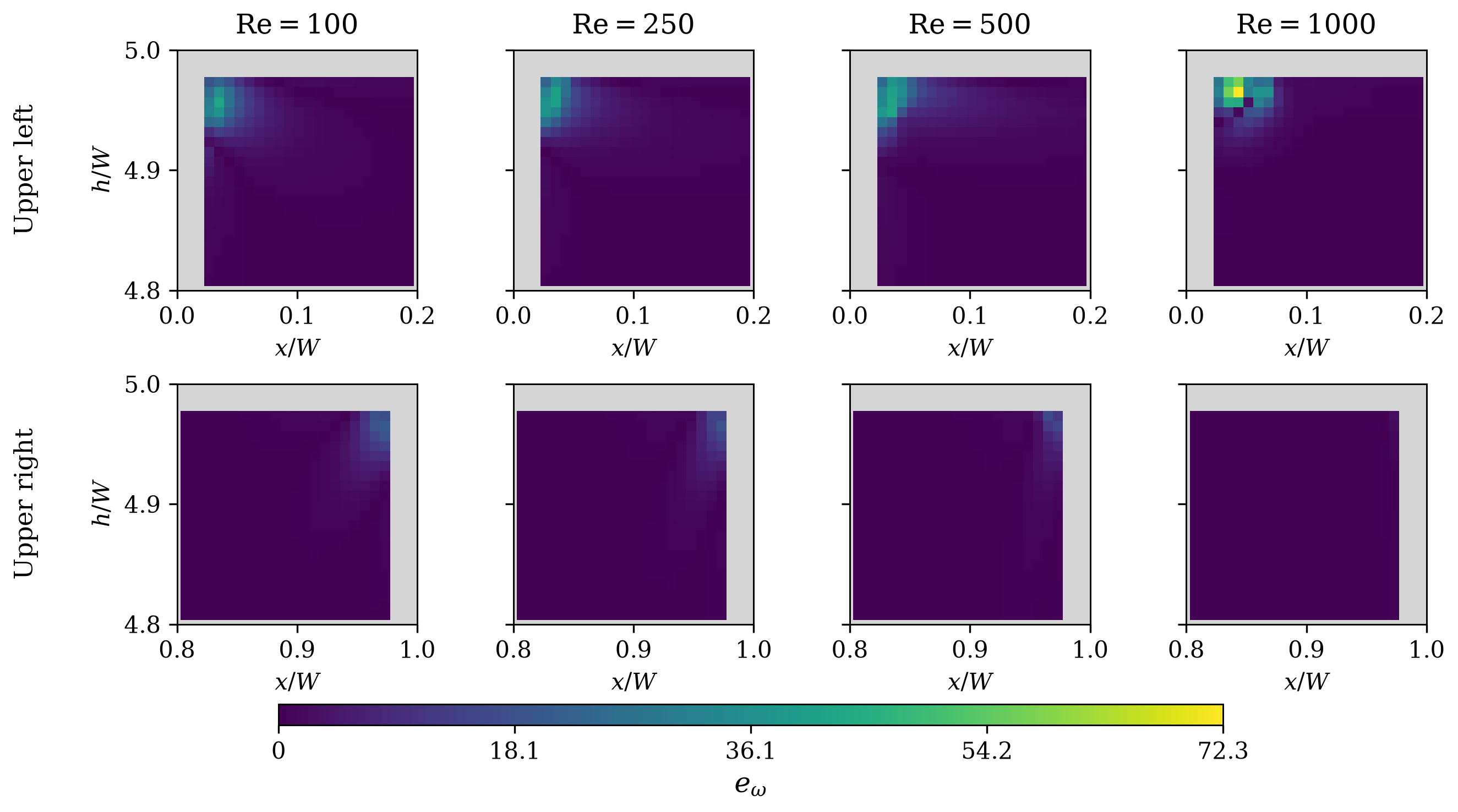}
\caption{Dimensionless sampled-vorticity discrepancy $e_\omega$, with the same upper-left (top) and upper-right (bottom) windows and Reynolds-number columns. Both sampled velocity fields retain the same second-order differentiation operator. The linear colour scale extends to the largest interior value; gray marks the excluded wall band. The samples are shown without smoothing.}\label{fig:re-omega-error-physical}
\end{figure}

Where is the error? Figures~\ref{fig:re-velocity-error-physical}--\ref{fig:re-omega-error-physical} show it near both upper corners on the common $121\times605$ grid. With the pressure gauge of Eq.~\eqref{eq:error}, the dimensionless pointwise discrepancies are
\begin{equation}
 e_u=\left\|\boldsymbol u^{\rm PINN}-\boldsymbol u^{\rm CFD}\right\|_2,\qquad
 e_p=\left|p_g^{\rm PINN}-p_g^{\rm CFD}\right|,\qquad
 e_\omega=\left|\omega^{\rm PINN}-\omega^{\rm CFD}\right|.
 \label{eq:re-spatial-error-physical}
\end{equation}
They are scaled by the fixed quantities $U$, $\rho U^2$ and $U/W$, not by local values, and are not percentages. Their RMS over a region gives Eq.~\eqref{eq:error}: $E_{\boldsymbol u}(\mathcal R)=U\langle e_u^2\rangle_{\mathcal R}^{1/2}$, $E_p(\mathcal R)=\rho U^2\langle e_p^2\rangle_{\mathcal R}^{1/2}$ and $E_\omega(\mathcal R)=(U/W)\langle e_\omega^2\rangle_{\mathcal R}^{1/2}$, where the brackets denote the mean over the sampled region.

Each figure shows windows at the upper-left and upper-right corners, $0\le x/W\le0.2$ and $0.8\le x/W\le1$, both with $4.8\le h/W\le5$. Each quantity has one linear color scale for all eight panels, from zero to its largest interior value over the four Reynolds numbers, without clipping or smoothing; gray marks the excluded wall band, not zero error.

The velocity discrepancy is concentrated near the lid (Fig.~\ref{fig:re-velocity-error-physical}): the strip $h/W>4.35$ holds 13\% of the interior samples but more than 92\% of the squared discrepancy.

The pressure discrepancy also lies in the upper flow, with more than 90\% of its square in the upper $0.65W$ (Fig.~\ref{fig:re-pressure-error-physical}). At $\Rey=500$ it is spread more widely: the strip $h/W>4.90$ holds less than half of it, against about 90\% or more at the other Reynolds numbers, and part of it lies outside the corner windows.

\FloatBarrier

The vorticity discrepancy is even more localized (Fig.~\ref{fig:re-omega-error-physical}). The strip $h/W>4.90$ holds 1.5\% of the interior samples and more than 99\% of the squared discrepancy, and the two discs of radius $0.1W$ around the lid corners hold less than 0.2\% of the samples and more than 95\% of the discrepancy. The global vorticity norm therefore measures a real error that is confined to the lid corners, not one spread through the cavity.

\FloatBarrier

Near the floor the velocity and vorticity discrepancies are many orders of magnitude smaller than in the interior (Table~\ref{tab:re-floor}), but so is the flow. Small values on the lid scale therefore do not by themselves show that the weak recirculation is accurate; the profiles below test its amplitude and shape. The floor pressure RMS includes the offset left by the interior gauge alignment, which we separate from the pressure shape below.

\begin{table}[htbp]\centering\small
\caption{Floor RMS discrepancies, with $0.02<h/W<0.65$ and $0.02<x/W<0.98$. Entries use the physical scales specified in Eq.~\eqref{eq:error}. Pressure retains the single interior mean alignment.}\label{tab:re-floor}
\begin{tabular}{rrrr}\toprule
$\Rey$ & $E_{\boldsymbol u}/U$ & $E_p/(\rho U^2)$ & $E_\omega W/U$\\\midrule
100 & $1.255\times10^{-10}$ & $1.033\times10^{-4}$ & $4.259\times10^{-10}$\\
250 & $9.133\times10^{-10}$ & $2.444\times10^{-4}$ & $2.915\times10^{-9}$\\
500 & $2.332\times10^{-9}$ & $5.348\times10^{-4}$ & $7.867\times10^{-9}$\\
1000 & $1.256\times10^{-9}$ & $7.074\times10^{-5}$ & $4.027\times10^{-9}$\\\bottomrule
\end{tabular}
\end{table}

\begin{figure}[H]\centering
\includegraphics[width=\linewidth]{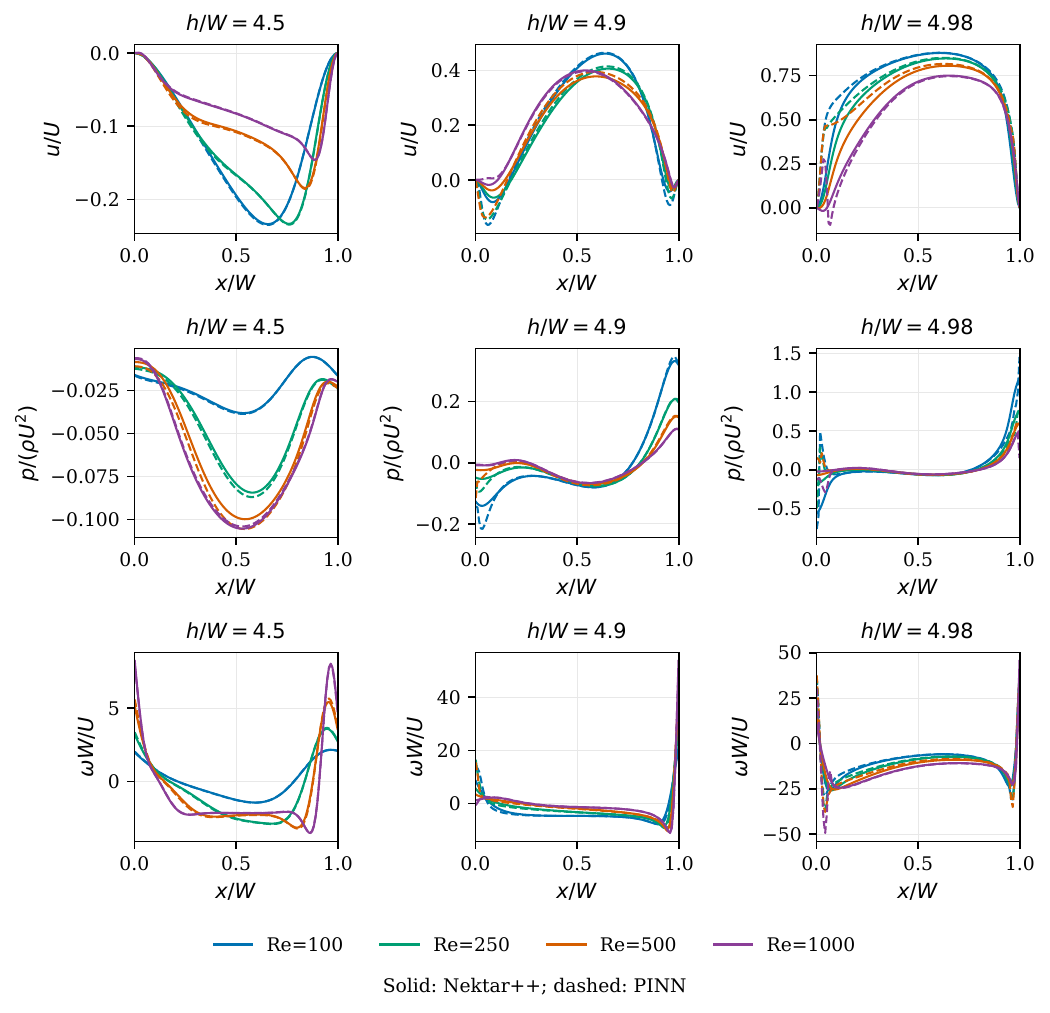}
\caption{Horizontal profiles at $h/W=4.5$, 4.9 and 4.98 at $D/W=5$, $\Rey=100$, 250, 500 and 1000. Rows show $u/U$, interior-aligned $p/(\rho U^2)$ and sampled $\omega W/U$. Color denotes Reynolds number, solid lines CFD and dashed lines PINN. Each panel has one shared amplitude axis for all curves.}\label{fig:re-upper-profiles}
\end{figure}

The horizontal profiles of Fig.~\ref{fig:re-upper-profiles} show the upper gradients directly. At $h/W=4.5$ the cut samples the return branch beneath the primary vortex, and its negative velocity trough shifts and changes shape as the Reynolds number increases. At $h/W=4.98$ it samples the lid-driven branch, where the four CFD fields have different velocity and shear profiles. Since the widths and extrema change, the vortex redistributes its motion rather than simply scaling up. The PINN follows most of this variation, with the clearest departures near the sidewalls at the highest cut. The pressure row tests the associated momentum balance, and the vorticity row, which depends on the profile slope, is the most sensitive test.

\begin{figure}[!htbp]\centering
\includegraphics[width=\linewidth]{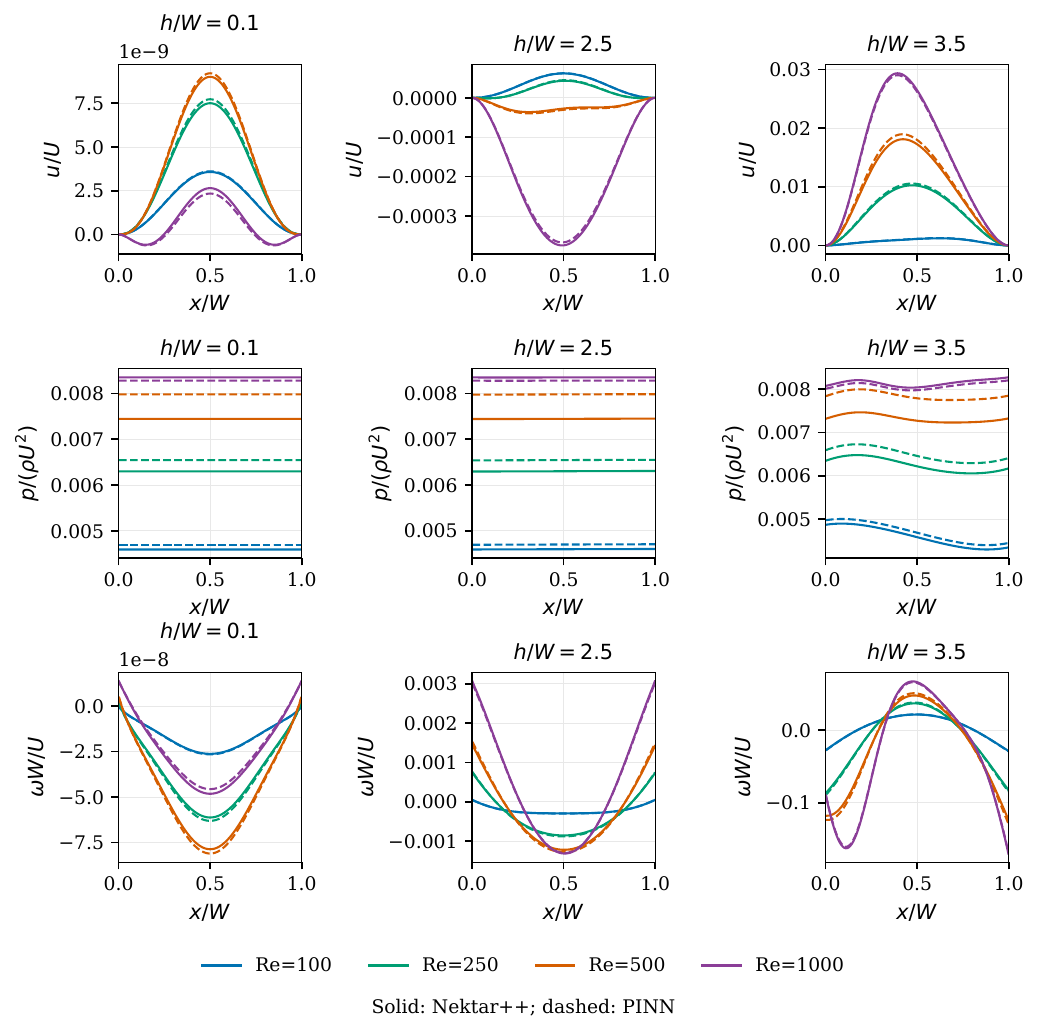}
\caption{Horizontal velocity, pressure and vorticity profiles at $D/W=5$, $\Rey=100$, 250, 500 and 1000 at $h/W=0.1$, 2.5 and 3.5. Colors and line styles match Fig.~\ref{fig:re-upper-profiles}. Scientific notation retains the weak floor amplitudes; pressure retains the global interior gauge rather than a separate gauge for each cut.}\label{fig:re-depth-profiles}
\end{figure}

The deeper cuts of Fig.~\ref{fig:re-depth-profiles} need more care. At $h/W=2.5$ and 3.5 a fixed horizontal line passes through different parts of an intermediate vortex as the stack shifts, so changes of sign or shape there must be read with Fig.~\ref{fig:re-streamlines}, not as a failure of the neural solution. The cut at $h/W=0.1$ tests the weak flow near the floor.

The sampled vorticity also depends on the evaluation grid. Recomputing it from every second grid point in each direction changes the interior vorticity by 1.5--7\% for CFD and by 8--18\% for the PINN, more at higher Reynolds number. The PINN fields are thus more sensitive to the sampling grid, and the vorticity discrepancies in Figs.~\ref{fig:re-velocity-error-physical}--\ref{fig:re-omega-error-physical} are meaningful only together with the grid and derivative used. This is a check of the postprocessing, not of the CFD mesh.

The CFD snapshots are at different times (Table~\ref{tab:re-sources}), so the series does not give a Reynolds-number trend at matched time or at steady state; that would need a sequence of late fields at each Reynolds number. It does show, at every Reynolds number, close agreement in the weak floor flow and a larger discrepancy in the velocity gradients near the lid.

To separate the floor pressure offset from the pressure shape, let $\Delta p=p_g^{\rm PINN}-p_g^{\rm CFD}$ in the floor region, where the superscripts denote the neural and reference pressure, and define $b_f=\langle\Delta p\rangle_f$ and $e_{p,f}^{\rm shape}=\langle(\Delta p-b_f)^2\rangle_f^{1/2}$. These quantities obey
\begin{equation}
 (e_{p,f}^{\rm global})^2=b_f^2+(e_{p,f}^{\rm shape})^2
\end{equation}
Here $\langle\cdot\rangle_f$ is the mean over the floor-region points, $b_f$ is the mean pressure offset, and $e_{p,f}^{\rm global}=\langle(\Delta p)^2\rangle_f^{1/2}$ is the RMS discrepancy under the interior gauge; the superscript shape denotes the discrepancy after the offset is removed. An independent check on the saved common arrays verifies the identity. Table~\ref{tab:floor-pressure-shape} shows that the floor pressure discrepancy is almost entirely an offset: the shape discrepancy is six to eight orders of magnitude smaller.
\begin{table}[htbp]\centering\small
\caption{Additional floor pressure decomposition on the same grid and region as Table~\ref{tab:re-floor}. Both quantities use units $\rho U^2$. The original globally gauge-aligned RMS remains in Table~\ref{tab:re-floor}.}\label{tab:floor-pressure-shape}
\begin{tabular}{rrr}\toprule
$\Rey$ & Floor mean offset $b_f$ & Local-gauge shape RMS\\\midrule
100 & $1.033\times10^{-4}$ & $1.152\times10^{-10}$\\
250 & $2.444\times10^{-4}$ & $8.709\times10^{-12}$\\
500 & $5.348\times10^{-4}$ & $8.174\times10^{-12}$\\
1000 & $-7.074\times10^{-5}$ & $3.965\times10^{-12}$\\\bottomrule
\end{tabular}
\end{table}

\FloatBarrier
\subsection{Deep-cavity agreement and regional trade-offs}
Table~\ref{tab:errors} compares the four workflows at $\Rey=1000$. The refined workflow, which starts from an earlier floor-refined $\Rey=1000$ model rather than from $\Rey=500$, more than halves the velocity discrepancy of the control in the interior and in the upper-left region, and also lowers it at the floor. Its gain therefore combines regional sampling with a different initialization (Section~\ref{sec:formulation}). In every workflow the upper-left discrepancy is eight to ten times the interior value, so the region near the lid corner limits the accuracy. The loss histories are given in Appendix~\ref{app:optimization-histories}; the rankings here come from the field comparisons, not from the losses.
\begin{table}[htbp]\centering\small
\caption{Regional velocity RMS discrepancy $E_{\boldsymbol u}/U$ against the same CFD field at $t\simeq317$, $\Rey=1000$, $D/W=5$, on the common $161\times801$ grid. The interior excludes the $0.02W$ wall band; upper-left and floor regions are subsets and do not include the geometric corners.}\label{tab:errors}
\begin{tabular}{p{2.7cm}p{3.7cm}rrr}\toprule
Workflow & Initialization / residual choice & Interior & Upper-left & Floor\\\midrule
Control & $\Rey=500$; $MR/E$ & $7.181\times10^{-3}$ & $6.868\times10^{-2}$ & $5.506\times10^{-10}$\\
Soft residual & $\Rey=500$; $MR/\sqrt E$ & $5.059\times10^{-3}$ & $4.143\times10^{-2}$ & $2.928\times10^{-9}$\\
Refined workflow & Prior floor-refined $\Rey=1000$; $MR/E$ & $3.394\times10^{-3}$ & $2.819\times10^{-2}$ & $3.612\times10^{-10}$\\
Half envelope & Extra $\Rey=500$ reconditioning; half $\kappa$ & $4.908\times10^{-3}$ & $3.854\times10^{-2}$ & $1.523\times10^{-9}$\\\bottomrule
\end{tabular}
\end{table}

\begin{table}[htbp]\centering\small
\caption{Matched-region velocity NRMSE on the detailed $161\times801$ grid, reported as dimensionless ratios. Each denominator is the CFD vector RMS on that same region with the original uniform sample weights. The original absolute RMS values remain in Table~\ref{tab:errors}. Each workflow was trained once, from a different starting field.}\label{tab:matched-velocity-nrmse}
\begin{tabular}{lrrr}\toprule
Workflow & Interior & Upper-left & Floor $h/W<0.4$\\\midrule
Control & 0.056 & 0.33 & 0.033\\
Soft residual & 0.040 & 0.20 & 0.18\\
Refined & 0.027 & 0.13 & 0.022\\
Half envelope & 0.038 & 0.18 & 0.092\\
\bottomrule\end{tabular}\end{table}

The two scale variants move accuracy between regions. Softening the residual scale or halving the envelope decay reduces the interior and upper-left velocity discrepancies of the control by 30--45\%, but raises the floor discrepancy by a factor of about three to five (Tables~\ref{tab:errors} and \ref{tab:matched-velocity-nrmse}). Weakening the emphasis on depth thus improves the upper flow at the expense of the floor. The regional balance is sensitive to these scales, but the comparison neither shows that the trade-off is unavoidable nor identifies an optimal scale.

\subsection{Full-depth structure and localized profiles}
\begin{figure}[H]\centering
\includegraphics[width=.90\linewidth]{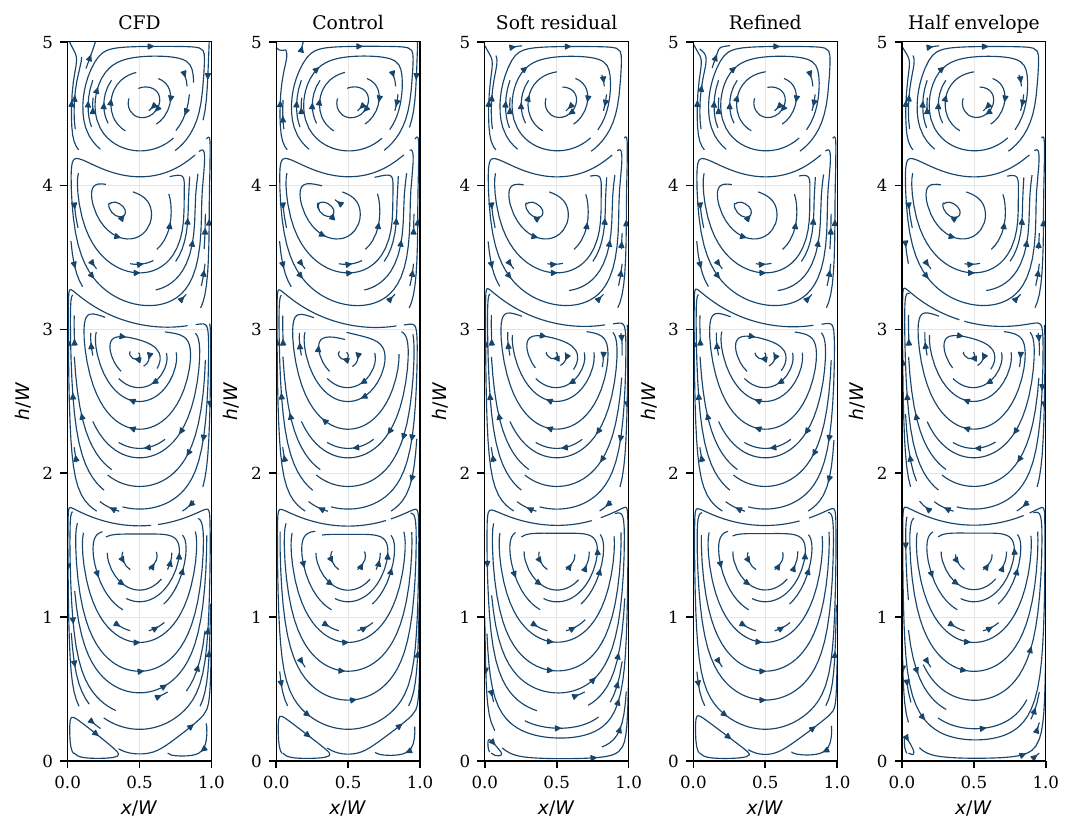}
\caption{Full-depth streamlines for the CFD field at $t\simeq317$ and four completed PINN workflows at $\Rey=1000$, $D/W=5$. The refined model combines transfer initialization and regional training.}\label{fig:pinn-full}
\end{figure}

\begin{figure}[H]\centering
\includegraphics[width=\linewidth]{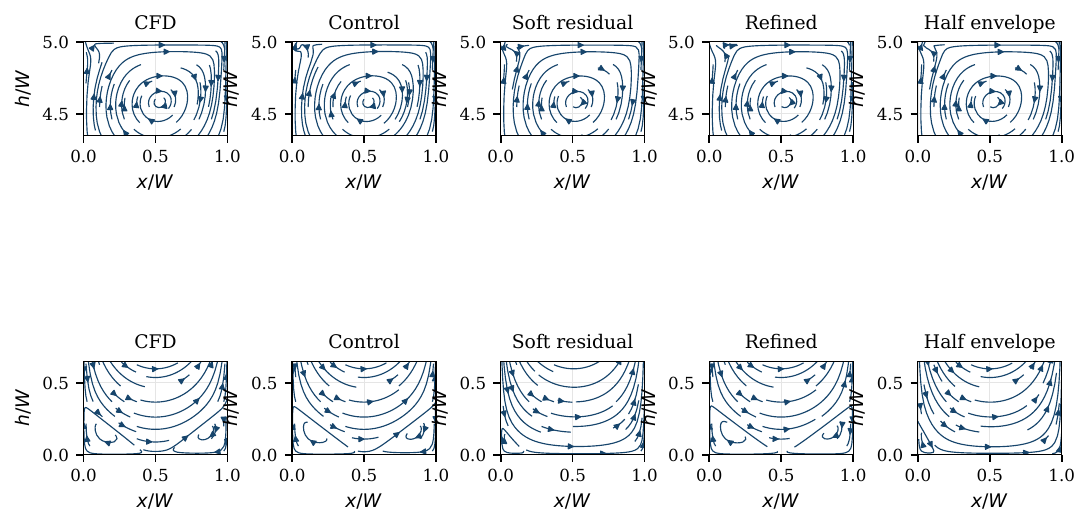}
\caption{Upper and lower zooms at $\Rey=1000$, $D/W=5$, of Fig.~\ref{fig:pinn-full}, with upper $4.35\leq h/W\leq5$ and lower $0\leq h/W\leq0.65$ views. Differences in streamlines complement, but do not replace, the amplitude errors in Table~\ref{tab:errors}.}\label{fig:pinn-zoom}
\end{figure}

Figures~\ref{fig:pinn-full} and \ref{fig:pinn-zoom} show the streamlines of the CFD and of the four workflows with the same plotting parameters; the zooms enlarge the flow near the lid and near the floor.

Figure~\ref{fig:pinn-full} shows how confinement turns a single moving wall into a stack of vortices. The rightward lid motion drives the primary clockwise circulation: fluid turns downward near the right wall and returns toward the left beneath the lid. Since the cavity is closed, this transport must be balanced by return flow. The lower flank of the primary vortex drags the fluid below it, where a counter-rotating vortex forms, and repetition of this coupling gives alternating rotation through the depth. The interfaces between vortices are dividing streamlines, not walls; viscous stresses transmit motion across them, and the induced circulation weakens away from the lid. This is why several lower vortices are visible in the streamlines although they carry little of the kinetic energy.

The two windows in Fig.~\ref{fig:pinn-zoom} show different regimes. In the upper window the flow turns from the lid into the stationary sidewalls, which produces strong velocity gradients even though the lid profile is smooth. The floor window shows the weak return flow above the stationary floor and, where visible, small corner eddies. These eddies are the local viscous response to two intersecting walls \cite{moffatt1964eddies}, not additional vortices of the stack.

All four workflows reproduce this organization, although the representation prescribes neither the number nor the position of the vortices. To compare them, the useful features are the shape of the primary circulation, the positions of the return branches and the spacing of the vortices. A small displacement of a dividing streamline moves motion between adjacent vortices and can produce a regional discrepancy while the full-depth pattern looks the same. Streamlines alone therefore cannot rank the workflows; Table~\ref{tab:errors} does.

\FloatBarrier

\begin{figure}[H]\centering
\includegraphics[width=\linewidth]{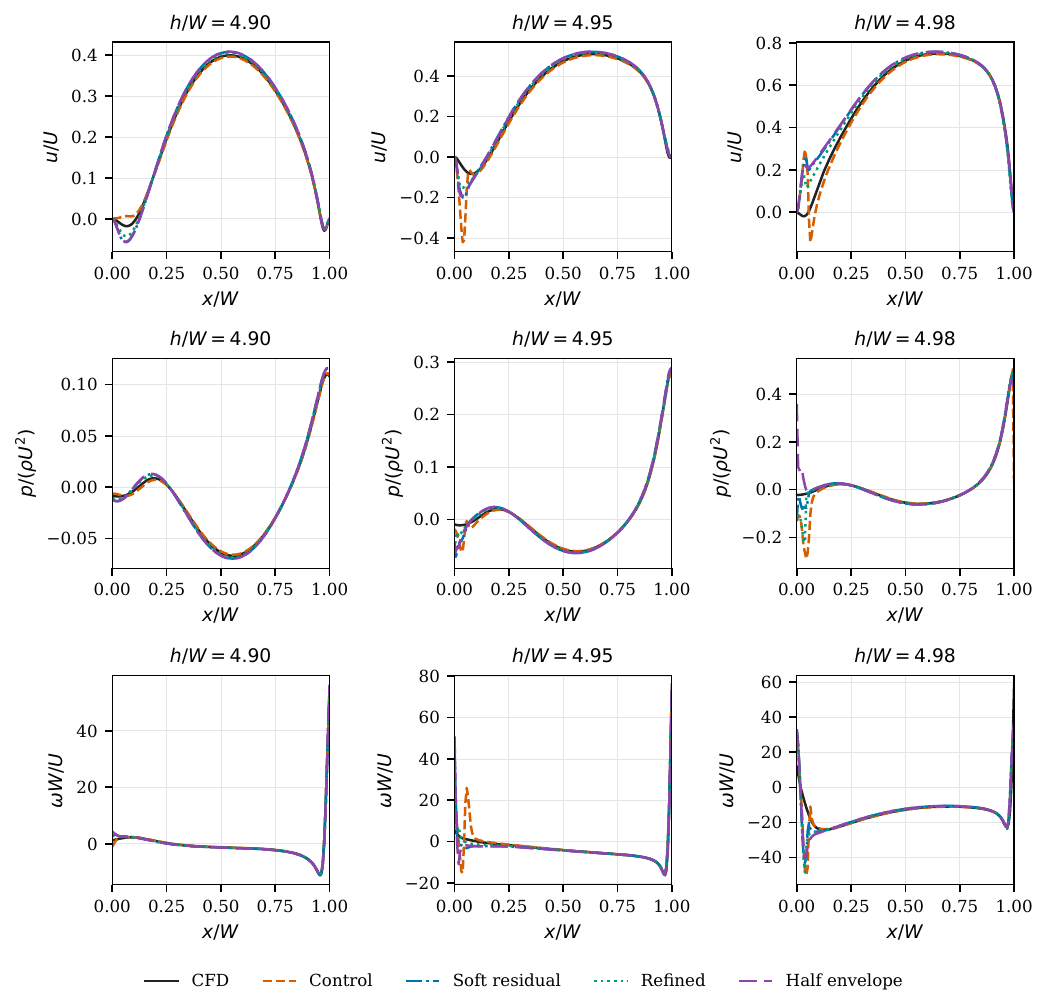}
\caption{Horizontal profiles at $\Rey=1000$, $D/W=5$, of velocity (top), pressure (middle) and vorticity (bottom) at three heights beneath the lid. Pressure gauges are aligned over the common interior. Vorticity uses the same second-order grid derivative for every field. Profiles are linearly sampled between grid heights.}\label{fig:profiles}
\end{figure}

The near-lid profiles of Fig.~\ref{fig:profiles} show where the upper-left discrepancy arises. The velocity row locates the positive lid-driven branch and the negative return branch, and a shift of their extrema reveals a displacement of the circulation as well as a change of amplitude. Pressure supplies the force that turns the flow, and vorticity measures the local rotation and shear (Eq.~\eqref{eq:vorticity_expanded}). Between $h/W=4.90$ and 4.98 all three change rapidly. A field can follow the broad velocity profile and still differ in the narrow regions of strong gradient, which show most clearly in the vorticity row. The refined curves improve on the control in all three rows, especially near the upper-left corner, but the near-lid region remains the main source of disagreement.

\FloatBarrier

\subsection{Pressure and vorticity in the upper and lower cavity}
Pressure and vorticity are more demanding tests than streamlines. Pressure is the second network output; vorticity is computed from the sampled velocities as before. In the contour plots the pressure mean is also removed within each displayed region, so that the very small variations near the floor remain visible.

\begin{figure}[H]\centering
\includegraphics[width=\linewidth]{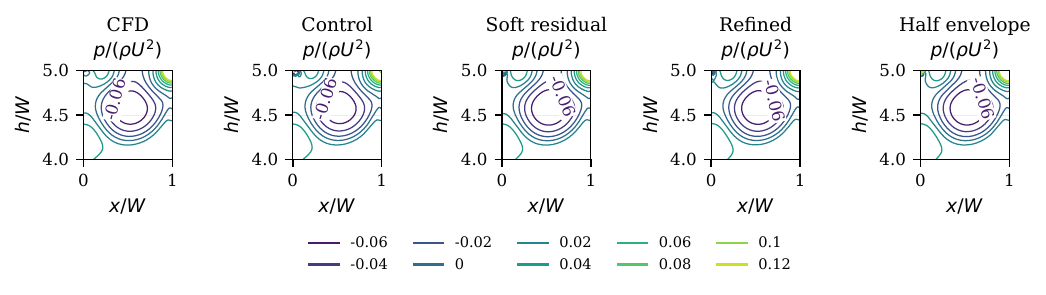}
\caption{Pressure contours in the upper cavity, $4\leq h/W\leq5$, for CFD, control, soft-residual, refined and half-envelope fields at $\Rey=1000$, $D/W=5$. Pressure is mean-centered over the displayed region, with common contour levels. The featureless lower part is omitted from this view.}\label{fig:pressure-full}\label{fig:pv-upper}
\end{figure}

\begin{figure}[H]\centering
\includegraphics[width=\linewidth]{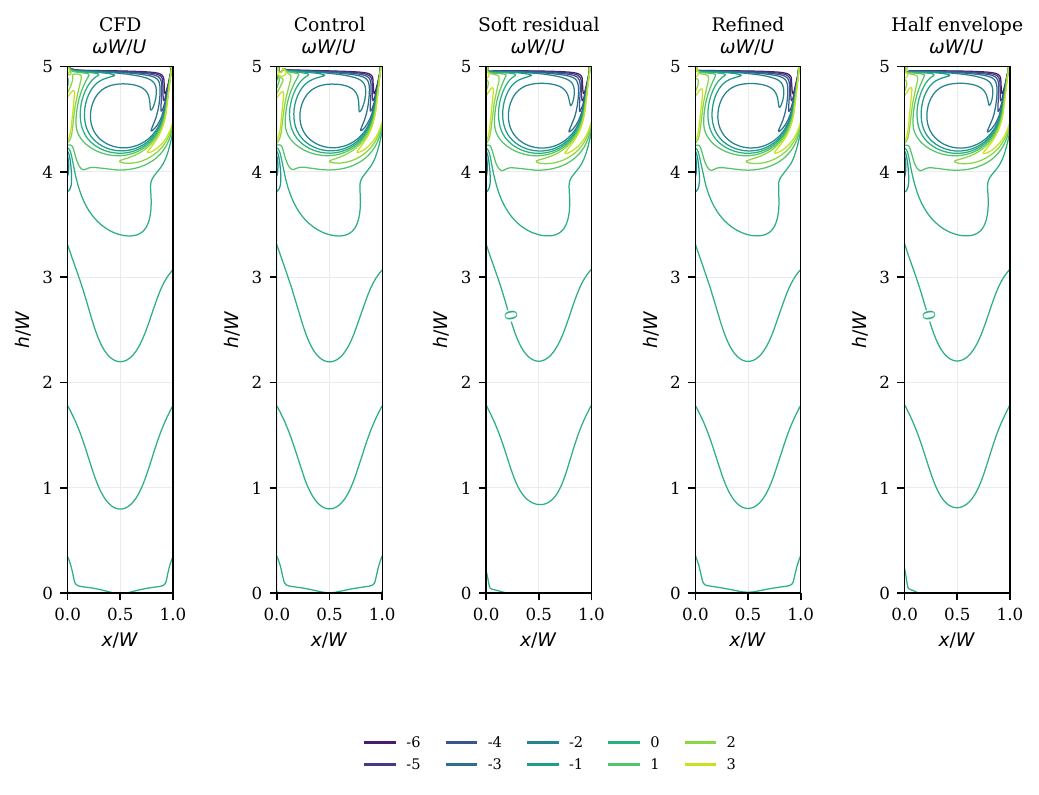}
\caption{Full-depth vorticity contours at $\Rey=1000$, $D/W=5$, in the same model order as Fig.~\ref{fig:pressure-full}. Values are in units of $U/W$, computed with the same sampled-grid derivative. Enlarged views below resolve structures that occupy little space on this full-depth scale.}\label{fig:vorticity-full}
\end{figure}

All models recover the pressure depression of the main circulation (Fig.~\ref{fig:pressure-full}) and the strong shear along the lid and right wall (Fig.~\ref{fig:vorticity-full}). Near the left lid corner, which is hard to judge from full-cavity plots, the profiles of Fig.~\ref{fig:profiles} show that the control departs most from CFD in velocity, pressure and vorticity. The soft-residual and refined fields reduce this departure, but not in the same order for every variable and region.

The pressure depression supplies the inward acceleration of the curved flow around the primary vortex. Pressure also turns the lid-driven branch into the sidewall flow, so its contours need not follow streamlines; Eq.~\eqref{eq:pressure_poisson_expanded} links its variation to products of velocity gradients. The strongest vorticity lies along the moving lid and the descending wall branch, and the deeper circulation is much weaker. A zero-vorticity contour marks a change in local rotation, not necessarily a streamline between two vortices, so a shear layer should not be mistaken for a separate vortex.

Table~\ref{tab:pv-errors} gives the pressure and vorticity discrepancies. Relative to the control, the refined workflow reduces both by a factor of 2.4 to 2.7 in the interior and in the upper-left region. Quantities with different units cannot be ranked against each other, but vorticity is the more sensitive test: it involves second derivatives of the streamfunction, so a small difference in the learned streamfunction produces a larger difference in curvature, especially near the lid. The best workflow depends on the quantity. The half envelope gives the smallest interior and upper-left pressure discrepancies, and the refined workflow the smallest interior and upper-left vorticity discrepancies. No single ranking based on velocity describes all the fields.

\FloatBarrier

\begin{table}[htbp]\centering\small
\caption{Pressure and sampled-vorticity RMS discrepancies at $\Rey=1000$, $D/W=5$, against CFD at $t\simeq317$. Entries use Eq.~\eqref{eq:error}; pressure retains the single interior gauge alignment in every region.}\label{tab:pv-errors}
\begin{tabular}{llrr}\toprule
Workflow & Region & $E_p/(\rho U^2)$ & $E_\omega W/U$\\\midrule
Control & Interior & $2.793\times10^{-3}$ & $8.009\times10^{-1}$\\
Soft residual & Interior & $1.195\times10^{-3}$ & $3.669\times10^{-1}$\\
Refined & Interior & $1.167\times10^{-3}$ & $3.142\times10^{-1}$\\
Half envelope & Interior & $1.064\times10^{-3}$ & $3.583\times10^{-1}$\\\midrule
Control & Upper-left & $2.684\times10^{-2}$ & $7.747\times10^{0}$\\
Soft residual & Upper-left & $7.798\times10^{-3}$ & $3.536\times10^{0}$\\
Refined & Upper-left & $9.974\times10^{-3}$ & $3.025\times10^{0}$\\
Half envelope & Upper-left & $4.688\times10^{-3}$ & $3.453\times10^{0}$\\\midrule
Control & Floor & $6.855\times10^{-5}$ & $2.056\times10^{-9}$\\
Soft residual & Floor & $2.963\times10^{-4}$ & $2.131\times10^{-8}$\\
Refined & Floor & $1.802\times10^{-4}$ & $3.025\times10^{-9}$\\
Half envelope & Floor & $3.131\times10^{-4}$ & $1.297\times10^{-8}$\\\bottomrule
\end{tabular}
\end{table}

\begin{table}[htbp]\centering\small
\caption{Detailed floor pressure decomposition, on the original $0.02<x/W<0.98$, $0.02<h/W<0.4$ region. All values use $\rho U^2$. The signed offset is PINN minus CFD after the single interior gauge alignment; the shape RMS subtracts only this floor difference mean. The identity $E_p^2=b_f^2+E_{p,\mathrm{shape}}^2$ preserves the original gauge-aligned RMS.}\label{tab:detailed-floor-pressure-decomposition}
\begin{tabular}{lrrr}\toprule
Workflow & Original $E_p/(\rho U^2)$ & $b_f/(\rho U^2)$ & Shape RMS$/ (\rho U^2)$\\\midrule
Control & $6.855\times10^{-5}$ & $-6.855\times10^{-5}$ & $3.260\times10^{-12}$\\
Soft residual & $2.963\times10^{-4}$ & $2.963\times10^{-4}$ & $4.129\times10^{-11}$\\
Refined & $1.802\times10^{-4}$ & $1.802\times10^{-4}$ & $4.016\times10^{-12}$\\
Half envelope & $3.131\times10^{-4}$ & $3.131\times10^{-4}$ & $3.293\times10^{-11}$\\
\bottomrule\end{tabular}\end{table}

Table~\ref{tab:detailed-floor-pressure-decomposition} gives the same decomposition for the four workflows. The reference floor pressure itself has a global-gauge RMS of $8.3\times10^{-3}\rho U^2$ but a shape RMS of only $1.7\times10^{-10}\rho U^2$: in the reference as in the discrepancy, almost all of the floor pressure is an offset from the interior mean, and the global RMS and the shape RMS answer different questions.

\begin{figure}[H]\centering
\includegraphics[width=\linewidth]{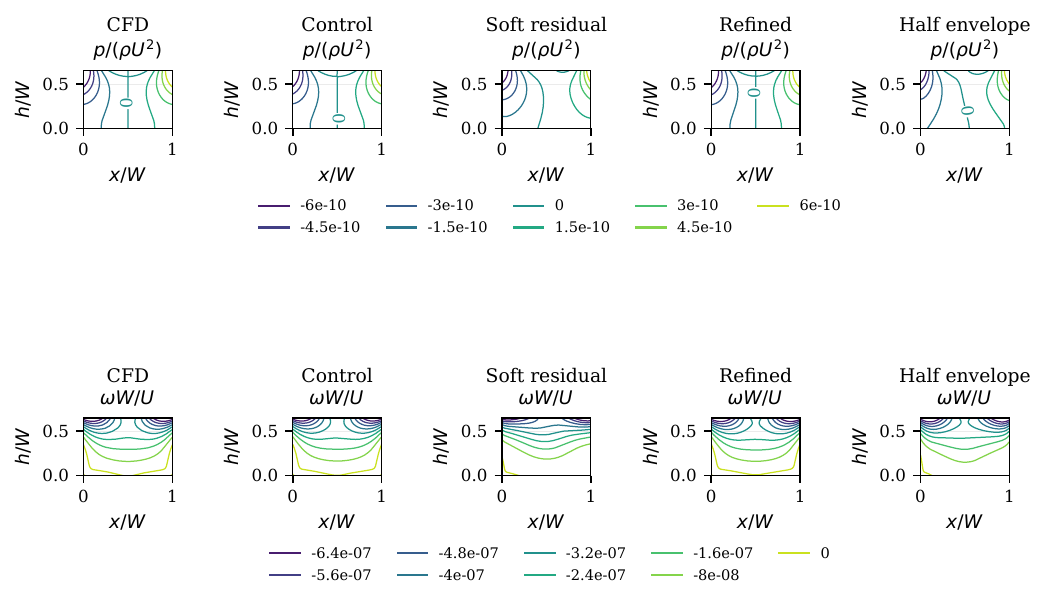}
\caption{Floor-region pressure (top) and vorticity (bottom) at $\Rey=1000$, $D/W=5$, in the same model order as Fig.~\ref{fig:pressure-full}. Pressure means are removed within the displayed region to expose its weak spatial variation; quantitative pressure errors retain the interior-aligned gauge. The small vorticity amplitudes are labeled in units of $U/W$.}\label{fig:pv-lower}
\end{figure}

\begin{figure}[H]\centering
\includegraphics[width=\linewidth]{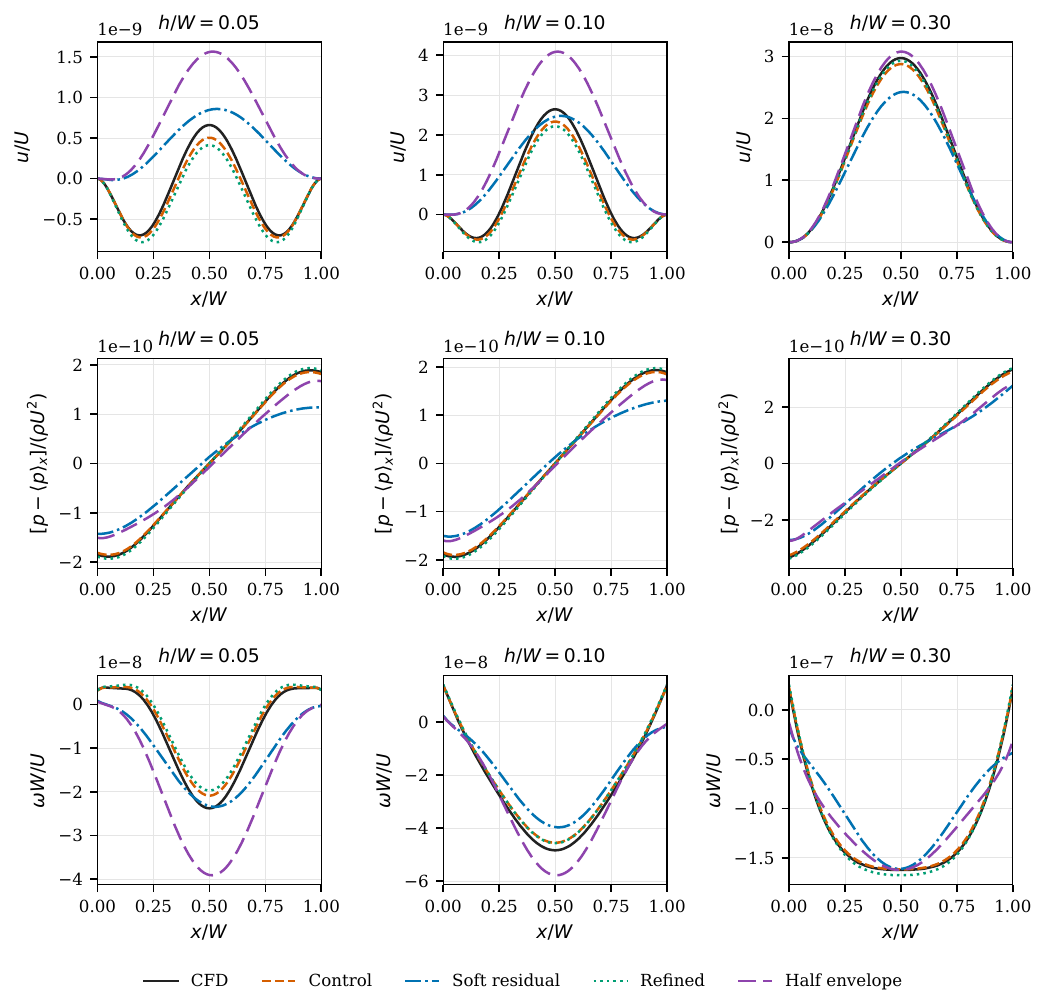}
\caption{Lower-cavity horizontal profiles at $\Rey=1000$, $D/W=5$, of velocity (top), pressure variation about each cut's mean (middle) and vorticity (bottom). Heights are $h/W=0.05$, 0.10 and 0.30. Scientific-notation axis scales retain the weak physical amplitudes instead of normalizing each curve.}\label{fig:lower-profiles}
\end{figure}

The lower cavity gives a different picture (Figs.~\ref{fig:pv-lower} and \ref{fig:lower-profiles}). Its velocities and vorticities are many orders of magnitude smaller than near the lid. The control and refined fields follow the lower flow more closely than the soft-residual field, which changes the shape of the horizontal velocity near the floor as well as the vorticity amplitude, consistent with its larger floor velocity discrepancy. Between control and refined the order depends on whether amplitudes or derivatives are examined: the refined field has the smaller floor velocity discrepancy, the control the smaller floor vorticity discrepancy (Tables~\ref{tab:errors} and \ref{tab:pv-errors}).

Pressure is nearly constant along the lowest cuts, so Fig.~\ref{fig:lower-profiles} shows it relative to the mean of each cut; the gauge-aligned values remain in Table~\ref{tab:pv-errors}. The control also has the smallest floor pressure-shape discrepancy, about 1.9\% of the CFD shape RMS, against 2.4\% for the refined field and larger values for the soft residual and half envelope (Table~\ref{tab:detailed-floor-pressure-decomposition}).

The floor contours of Fig.~\ref{fig:pv-lower} and the cuts of Fig.~\ref{fig:lower-profiles} test the shape of the weak flow as well as its magnitude. The velocity falls to zero at the floor, and the return flow above it produces shear and hence vorticity. Pressure varies little because both inertia and the transmitted motion are weak. The cuts at $h/W=0.05$, 0.10 and 0.30 follow the flow from the near-wall layer into the lower vortex. On these axes a visible difference can be large relative to the local circulation and still negligible on the lid scale.

\FloatBarrier
\subsection{Variation of physical amplitudes through the depth}
The functions $A_u(h)$, $A_p(h)$ and $A_\omega(h)$ below are horizontal RMS amplitudes of speed, pressure variation and vorticity at height $h$, and $\langle p\rangle_x(h)$ is the horizontal mean pressure at that height.
In Figs.~\ref{fig:pressure-full} and \ref{fig:vorticity-full} most strong contours lie near the lid, where the pressure gradient sustains the main circulation and wall shear generates large vorticity. One set of contour levels cannot show variations that span many orders of magnitude, so we also measure the amplitude of the flow at each height.

Why is a separate test of the weak flow needed? The upper $0.65W$ occupies 13\% of the cavity area but holds about 77\% of the kinetic energy, in the CFD and in all three workflows of Table~\ref{tab:depth-amplitudes}. This fraction agrees to within 0.2 percentage points although the floor discrepancies of the workflows differ several-fold, so an integrated measure cannot assess the weak lower flow.

\begin{figure}[H]\centering
\includegraphics[width=\linewidth]{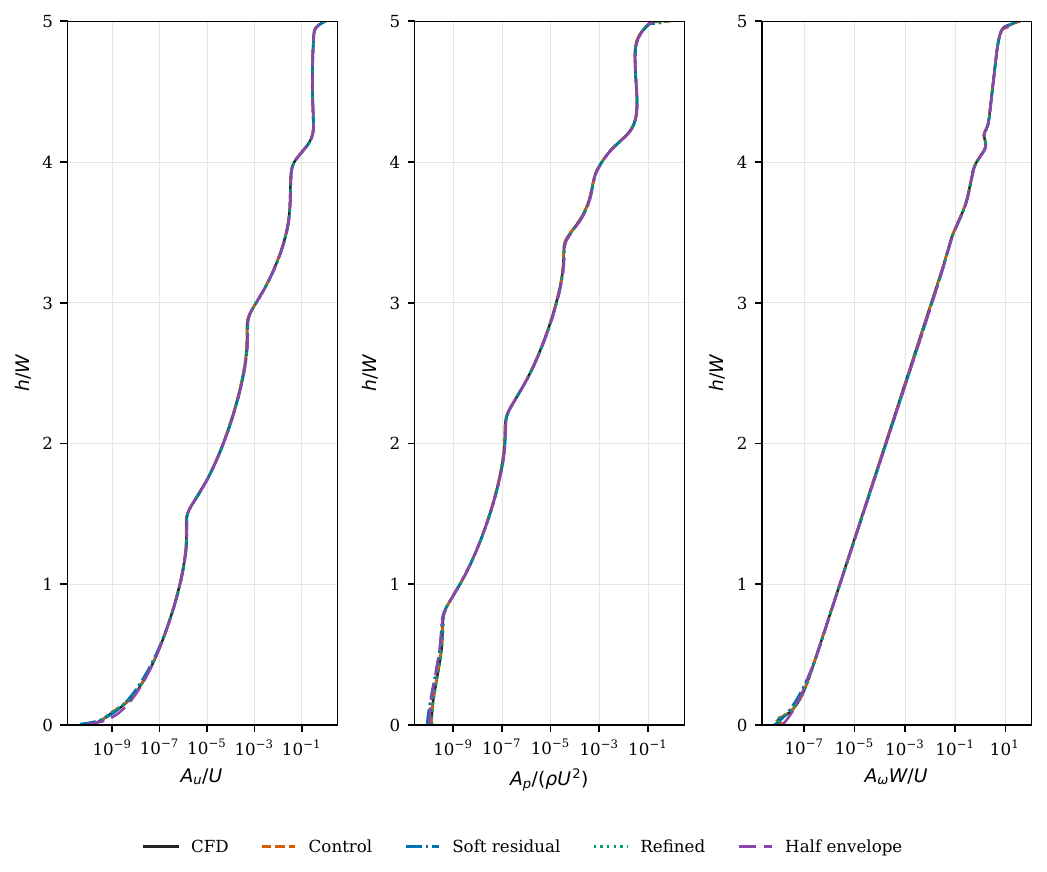}
\caption{Horizontal RMS velocity, pressure variation and vorticity at $\Rey=1000$, $D/W=5$, over the complete cavity depth. Logarithmic amplitude axes expose weak lower flow on its original physical scale. All models use the same quadrature grid; pressure is centered separately along each horizontal cut for this amplitude diagnostic.}\label{fig:depth-amplitudes}
\end{figure}

At each height, define the horizontal RMS amplitudes
\begin{align}
A_u(h)&=\left[\frac{1}{W}\int_0^W(u^2+v^2)\,dx\right]^{1/2},\\
A_p(h)&=\left[\frac{1}{W}\int_0^W[p-\langle p\rangle_x(h)]^2\,dx\right]^{1/2},\qquad \langle p\rangle_x(h)=\frac{1}{W}\int_0^W p\,dx,\\
A_\omega(h)&=\left[\frac{1}{W}\int_0^W\omega^2\,dx\right]^{1/2}.
\end{align}
Figure~\ref{fig:depth-amplitudes} evaluates these integrals by trapezoidal quadrature on the common grid, on logarithmic axes; only exactly zero amplitudes are omitted. $A_p$ is unaffected by a pressure constant, so it measures horizontal variation, not the offset between the upper and lower regions.

The curves connect the upper and floor views. Their local minima can accompany changes in the arrangement of the vortices, but a minimum alone does not identify a vortex or locate a separatrix; that requires the streamlines.

Why is the upper vortex sensitive to inertia while the lower stack is viscous? A local Reynolds number based on a vortex speed $U_c$ and size $L_c$ is $\Rey_c=\Rey(U_c/U)(L_c/W)$. As $U_c$ falls through the depth, $\Rey_c$ falls with it: at $\Rey=1000$ the second eddy already has $U_c/U$ of order $10^{-2}$ and the third of order $10^{-4}$, so from the third eddy down $\Rey_c$ is below one and diffusion dominates advection in Eq.~\eqref{eq:vort_transport_expanded}. The lower stack is then a Stokes flow, which Section~\ref{sec:lowRe} compares with Moffatt's prediction.
 The neural amplitude curves lie close to CFD in the upper cavity and further apart, in relative terms, at low amplitude.

The reference velocity amplitude falls by a factor of about $1.5\times10^7$ between $h/W=4.7$ and 0.3 (Table~\ref{tab:depth-amplitudes}). The neural fields match the upper amplitude to within 2\% but differ more at $h/W=0.3$, where the soft-residual field is 17\% low. A method can thus look right in a full-cavity image and still differ near the floor.

\begin{table}[htbp]\centering\small
\caption{Horizontal RMS velocity amplitudes and upper-region energy concentration from the common sampled fields. The energy fraction uses the integral of $u^2+v^2$ over $h/W\geq4.35$ divided by its full-depth integral.}\label{tab:depth-amplitudes}
\begin{tabular}{lrrr}\toprule
Field & $A_u(4.7W)/U$ & $A_u(0.3W)/U$ & Upper energy (\%)\\\midrule
CFD & 0.286 & $1.89\times10^{-8}$ & 77.4\\
Control & 0.284 & $1.82\times10^{-8}$ & 77.5\\
Soft residual & 0.290 & $1.56\times10^{-8}$ & 77.4\\
Refined & 0.288 & $1.85\times10^{-8}$ & 77.4\\\bottomrule
\end{tabular}
\end{table}

\FloatBarrier

\section{Results with wall-local boundary factors}\label{sec:local}

\subsection{The Reynolds-number series on the original comparison grid}\label{sec:reseries-local}
We first repeat the common assessment of Section~\ref{sec:reseries} with the wall-local formulation, on the same $121\times605$ grid, with the same reference snapshots, regions, pressure gauge and vorticity operator. Recomputing the discrepancies of the established formulation with this procedure reproduces the values reported above, so the two sets of numbers in Table~\ref{tab:reseries-local} are directly comparable.

\begin{table}[htbp]\centering\small
\caption{Common $121\times605$ assessment of Section~\ref{sec:reseries} for the established formulation (``earlier'') and the wall-local formulation (``local''), against the same Nektar++ snapshots. Entries use Eq.~\eqref{eq:error}; interior pressure retains the single interior gauge. Upper-left: $x/W<0.25$, $h/W>4.75$; floor: $0.02<h/W<0.65$.}
\label{tab:reseries-local}
\begin{tabular}{rlccccc}\toprule
 & & \multicolumn{3}{c}{Interior} & Upper-left & Floor\\\cmidrule(lr){3-5}\cmidrule(lr){6-6}\cmidrule(lr){7-7}
$\Rey$ & PINN & $E_{\boldsymbol u}/U$ & $E_p/(\rho U^2)$ & $E_\omega W/U$ & $E_\omega W/U$ & $E_{\boldsymbol u}/U$\\\midrule
100 & earlier & $8.05\times10^{-3}$ & $5.84\times10^{-3}$ & $0.520$ & $4.56$ & $1.26\times10^{-10}$\\
    & local   & $1.03\times10^{-4}$ & $2.39\times10^{-5}$ & $2.17\times10^{-3}$ & $1.38\times10^{-2}$ & $3.88\times10^{-10}$\\[2pt]
250 & earlier & $7.67\times10^{-3}$ & $2.50\times10^{-3}$ & $0.510$ & $4.65$ & $9.13\times10^{-10}$\\
    & local   & $7.41\times10^{-5}$ & $5.18\times10^{-6}$ & $3.68\times10^{-3}$ & $2.18\times10^{-2}$ & $6.99\times10^{-11}$\\[2pt]
500 & earlier & $9.54\times10^{-3}$ & $2.08\times10^{-3}$ & $0.562$ & $5.29$ & $2.33\times10^{-9}$\\
    & local   & $8.73\times10^{-5}$ & $3.94\times10^{-6}$ & $5.88\times10^{-3}$ & $3.26\times10^{-2}$ & $2.46\times10^{-10}$\\[2pt]
1000 & earlier & $6.54\times10^{-3}$ & $2.67\times10^{-3}$ & $0.665$ & $6.46$ & $1.26\times10^{-9}$\\
     & local   & $1.03\times10^{-4}$ & $3.80\times10^{-6}$ & $8.78\times10^{-3}$ & $4.81\times10^{-2}$ & $5.98\times10^{-10}$\\\bottomrule
\end{tabular}
\end{table}

\begin{figure}[!htbp]\centering
\includegraphics[width=0.92\linewidth]{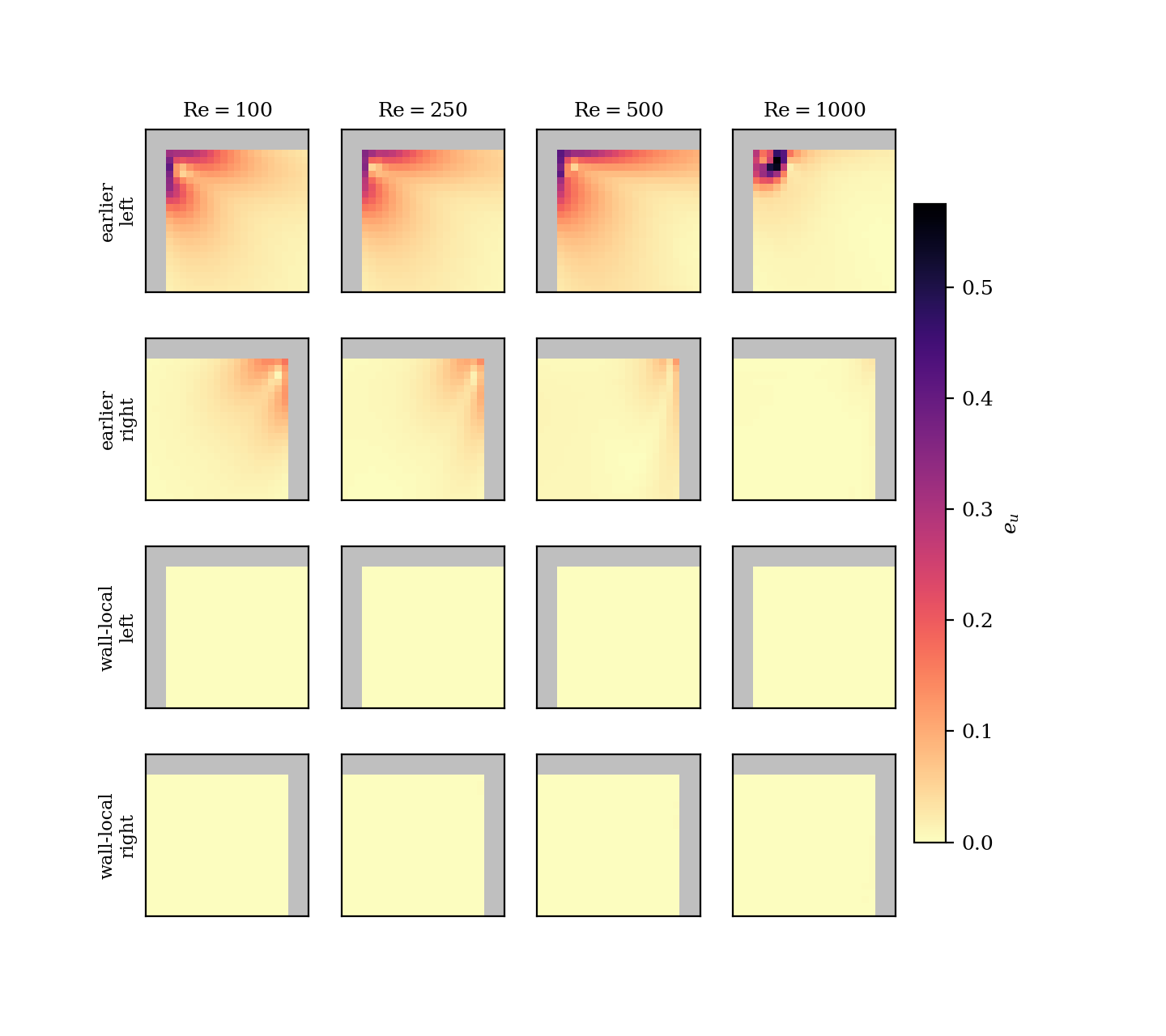}
\caption{Velocity discrepancy $e_u$ of Eq.~\eqref{eq:re-spatial-error-physical} in the upper-left and upper-right corner windows, $4.8\le h/W\le5$, for the established formulation (top two rows, as in Fig.~\ref{fig:re-velocity-error-physical}) and the wall-local formulation (bottom two rows), with the same linear scale. Gray: excluded wall band.}\label{fig:local-err-u}
\end{figure}

\begin{figure}[!htbp]\centering
\includegraphics[width=0.92\linewidth]{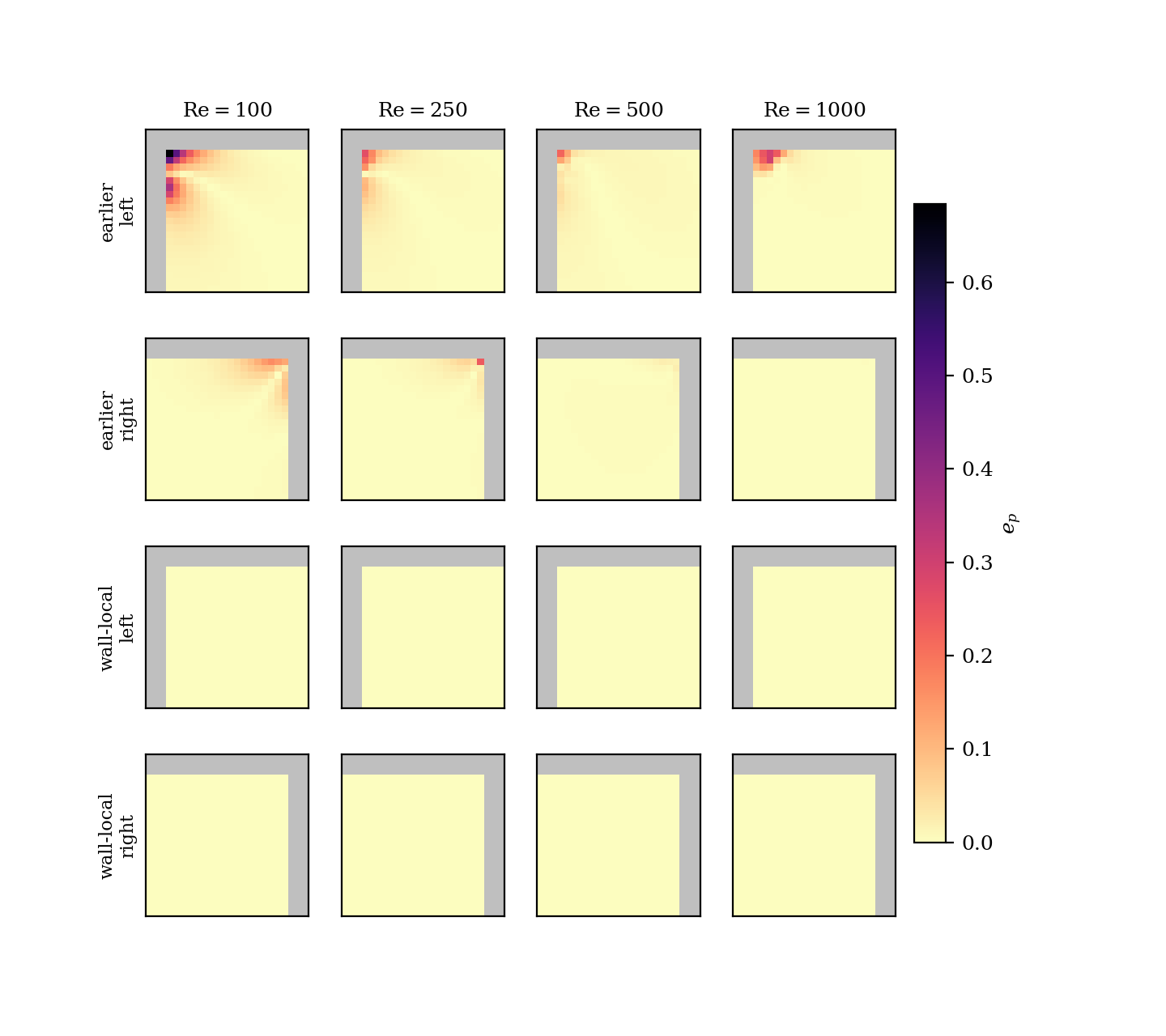}
\caption{Gauge-aligned pressure discrepancy $e_p$, arranged as Fig.~\ref{fig:local-err-u}, with the linear scale of Fig.~\ref{fig:re-pressure-error-physical}.}\label{fig:local-err-p}
\end{figure}

\begin{figure}[!htbp]\centering
\includegraphics[width=0.92\linewidth]{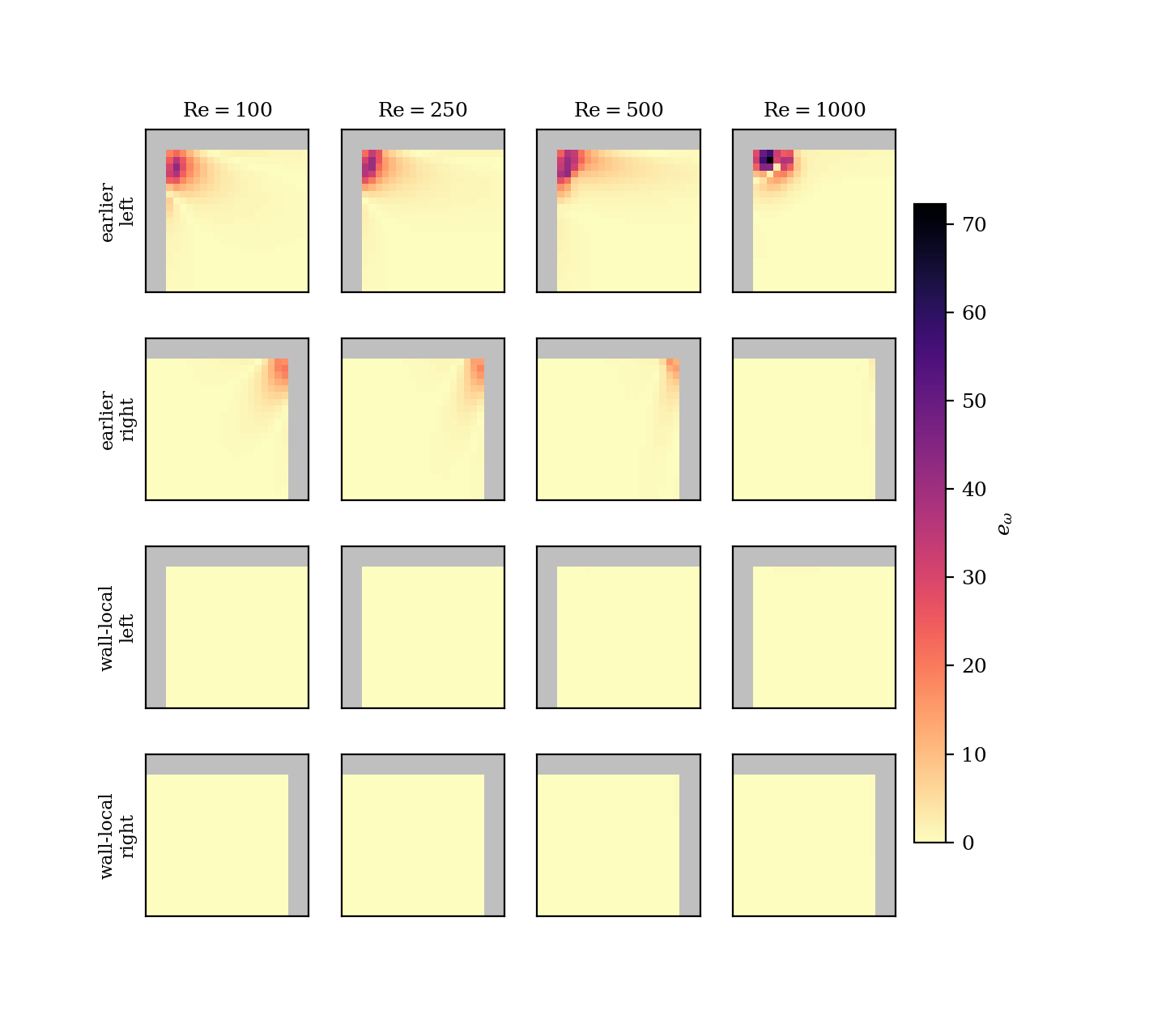}
\caption{Sampled-vorticity discrepancy $e_\omega$, arranged as Fig.~\ref{fig:local-err-u}, with the linear scale of Fig.~\ref{fig:re-omega-error-physical}.}\label{fig:local-err-w}
\end{figure}

In the interior the wall-local formulation reduces the velocity discrepancy by a factor of 60 to 110, the pressure discrepancy by one to three orders of magnitude and the vorticity discrepancy by a factor of 75 to 240; near the upper-left corner the vorticity discrepancy falls by more than two orders of magnitude at every Reynolds number. Figures~\ref{fig:local-err-u}--\ref{fig:local-err-w} show where this happens. They use the same corner windows and the same linear scales as Figs.~\ref{fig:re-velocity-error-physical}--\ref{fig:re-omega-error-physical}, and place the established and the wall-local formulations one above the other. The concentrated discrepancy at the left lid corner, which dominated every map of the established formulation, is no longer visible on these scales. The share of the squared interior vorticity discrepancy within $0.1W$ of the lid corners falls from more than 95\% to between a half and three quarters: the corners still hold most of what remains, but what remains is two orders of magnitude smaller.

At the floor the picture is different. The two formulations give floor discrepancies of the same order, and at $\Rey=100$ the wall-local one is larger. The floor eddy is also where the time-stepped reference is least converged (Section~\ref{sec:nektar}); against the steady reference at $\Rey=1000$ the floor discrepancy of the wall-local formulation is several times smaller (Table~\ref{tab:highRe}). We therefore do not read Table~\ref{tab:reseries-local} as a change at the floor in either direction.

\FloatBarrier

\begin{figure}[!htbp]\centering
\includegraphics[width=\linewidth]{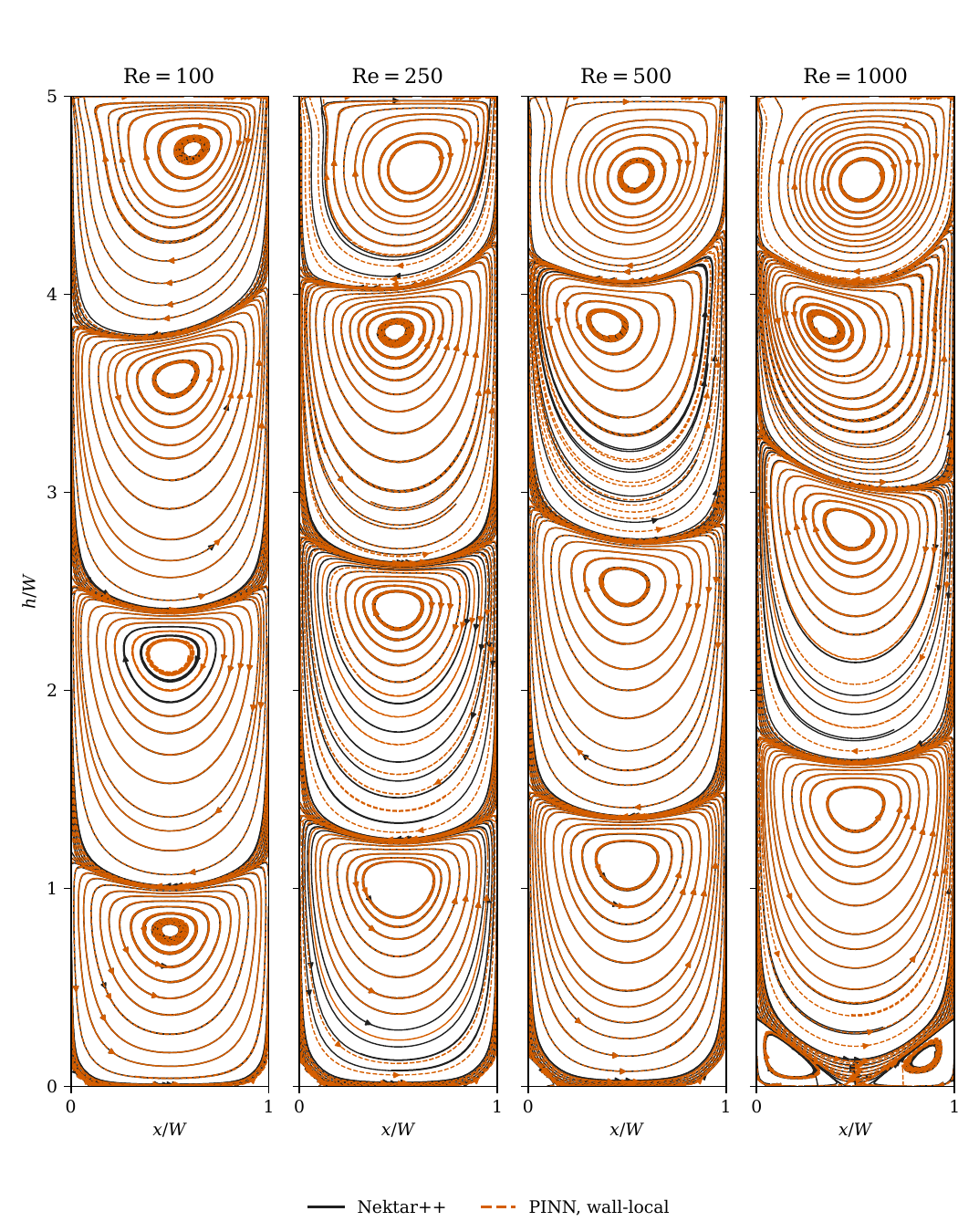}
\caption{Full-depth streamlines at $D/W=5$, $\Rey=100$, 250, 500 and 1000, as Fig.~\ref{fig:re-streamlines} but for the wall-local formulation. Black solid: Nektar++; orange dashed: PINN.}\label{fig:local-streamlines}
\end{figure}

\begin{figure}[!htbp]\centering
\includegraphics[width=\linewidth]{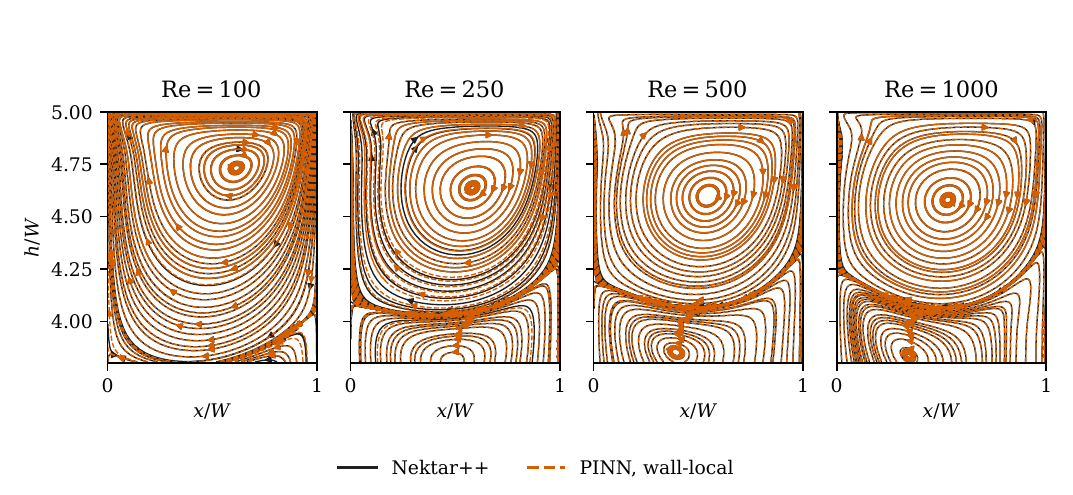}
\caption{Upper-cavity enlargement, $3.8\le h/W\le5$, of Fig.~\ref{fig:local-streamlines}.}\label{fig:local-streamlines-upper}
\end{figure}

\begin{figure}[!htbp]\centering
\includegraphics[width=\linewidth]{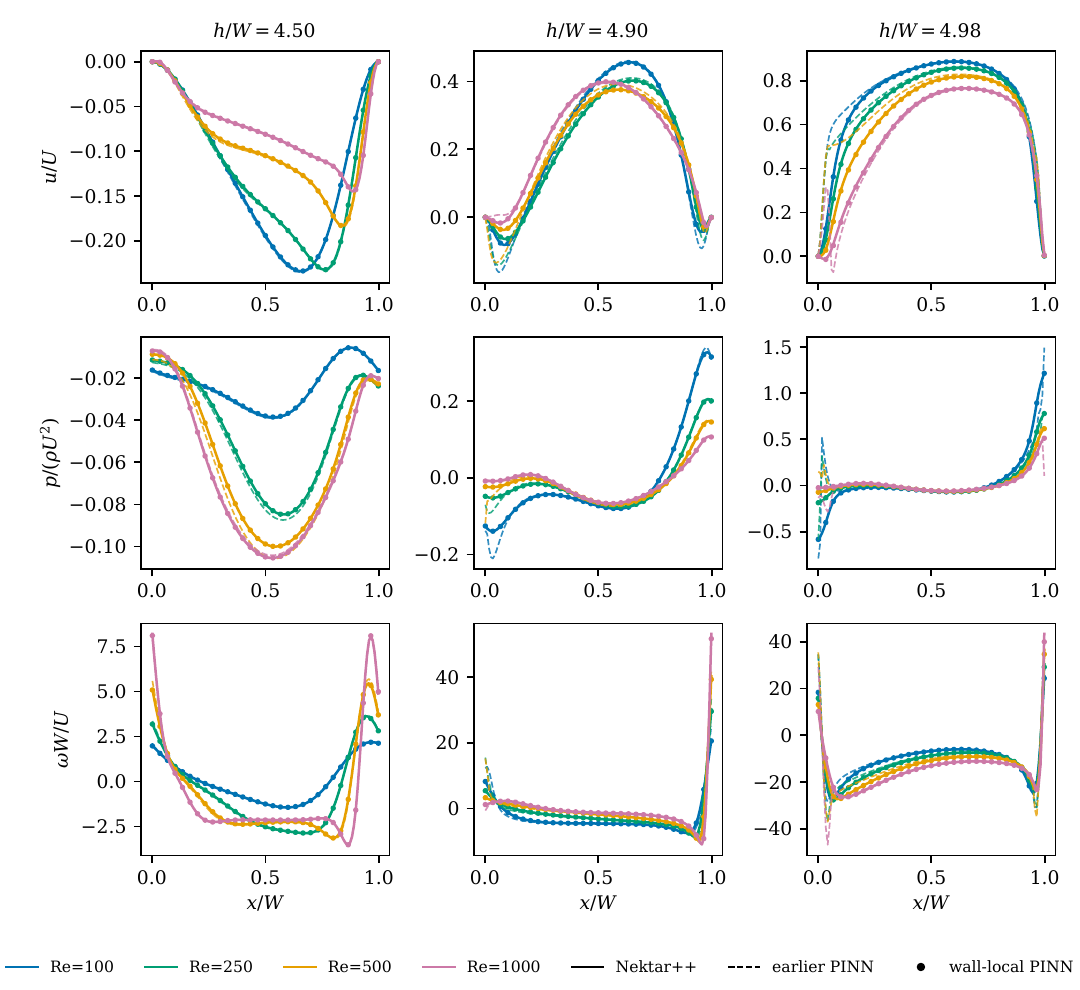}
\caption{Horizontal profiles of $u/U$, interior-aligned $p/(\rho U^2)$ and sampled $\omega W/U$ at $h/W=4.5$, 4.9 and 4.98, as Fig.~\ref{fig:re-upper-profiles}. Colour: Reynolds number; solid: Nektar++; dashed: established formulation; dots: wall-local formulation.}\label{fig:local-profiles-upper}
\end{figure}

\begin{figure}[!htbp]\centering
\includegraphics[width=\linewidth]{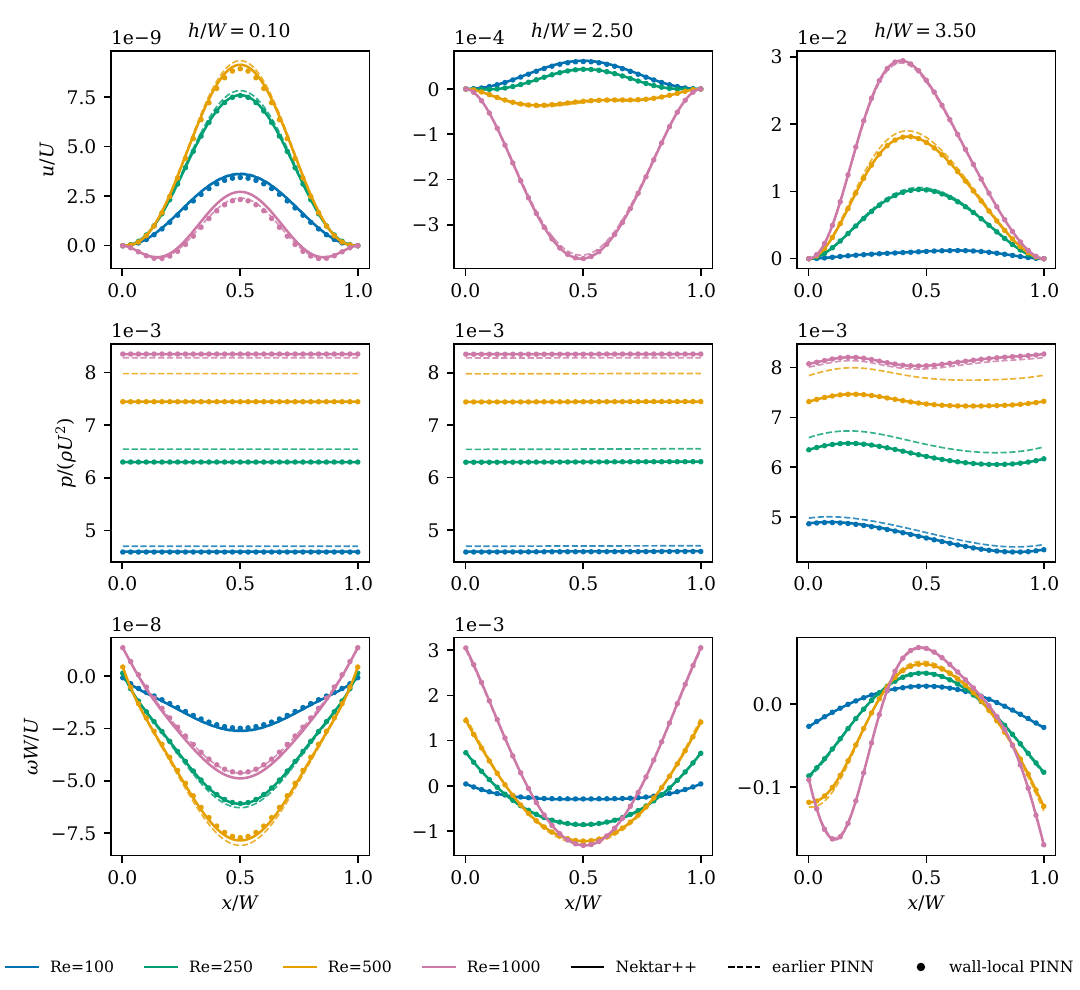}
\caption{As Fig.~\ref{fig:local-profiles-upper} at $h/W=0.1$, 2.5 and 3.5, as in Fig.~\ref{fig:re-depth-profiles}.}\label{fig:local-profiles-deep}
\end{figure}

The streamlines of the wall-local PINN coincide with those of the reference over the full depth and in the upper cavity (Figs.~\ref{fig:local-streamlines} and \ref{fig:local-streamlines-upper}), including the dividing streamlines and the cores where the established formulation departed locally. The horizontal profiles (Figs.~\ref{fig:local-profiles-upper} and \ref{fig:local-profiles-deep}) show the same improvement on physical scales. Near the lid the established formulation departs from the reference close to the side walls, most clearly in pressure and vorticity at the highest cut; the wall-local curves lie on the reference across the whole width. At depth both formulations follow the reference, consistent with the floor results above.

\FloatBarrier

\subsection{Matched comparison with the polynomial factor}\label{sec:matched}
The established formulation of Section~\ref{sec:results} differs from the wall-local one not only in its boundary factor but also in its training histories, which differed between Reynolds numbers. To separate the two, we retrained the polynomial factor (\ref{eq:b}) with raw inputs and the corner mask exactly like the wall-local formulation (Table~\ref{tab:setup}). Table~\ref{tab:main-local} compares the three PINNs with the spectral solution on the $121\times(120D/W+1)$ grid, in normalized RMS errors (NRMSE), the RMS discrepancy divided by the RMS of the reference on the same region; the corner region is $x<0.25W$ within $0.25W$ of the lid, the floor region $h<0.4W$.

\begin{table}[htbp]\centering\small
\caption{NRMSE (\%) against the spectral/$hp$ solution at $D/W=5$. ``Earlier'': the established formulation of Section~\ref{sec:results}. ``Polynomial'': polynomial factor trained with the same points and schedule as the wall-local formulation.}
\label{tab:main-local}
\begin{tabular}{rlccccc}\toprule
 & & \multicolumn{2}{c}{Interior} & \multicolumn{2}{c}{Corner} & Floor\\
\cmidrule(lr){3-4}\cmidrule(lr){5-6}\cmidrule(lr){7-7}
$\Rey$ & PINN & velocity & vorticity & velocity & vorticity & velocity\\\midrule
100 & earlier    & 7.6 & 43 & 28 & 84 & 1.2\\
    & polynomial & 0.07 & 0.24 & 0.20 & 0.45 & 1.9\\
    & local      & 0.10 & 0.18 & 0.16 & 0.25 & 4.5\\[2pt]
250 & earlier    & 6.7 & 41 & 32 & 96 & 3.0\\
    & polynomial & 0.56 & 0.55 & 0.46 & 0.82 & 13\\
    & local      & 0.065 & 0.30 & 0.20 & 0.45 & 0.25\\[2pt]
500 & earlier    & 7.9 & 45 & 39 & 128 & 3.8\\
    & polynomial & 2.6 & 2.3 & 1.3 & 1.2 & 75\\
    & local      & 0.072 & 0.47 & 0.24 & 0.79 & 1.2\\[2pt]
1000 & earlier   & 5.1 & 53 & 30 & 198 & 3.2\\
     & polynomial & \multicolumn{5}{c}{training diverged}\\
     & local     & 0.081 & 0.70 & 0.23 & 1.5 & 2.8\\\bottomrule
\end{tabular}
\end{table}

At $\Rey=100$ the polynomial factor, trained in the same way, is as accurate as the wall-local formulation. Most of the error of the earlier PINN at this Reynolds number therefore came from its training, not from its representation. As the Reynolds number increases the two factors part company. The error of the polynomial factor grows with every step, the floor eddy is lost by $\Rey=500$, and at $\Rey=1000$ the training diverges. The error of the wall-local formulation hardly changes in the interior and grows only slowly at the corners. This is the trend that Eq.~(\ref{eq:Bnear}) predicts: as the lid layer thins, the correction concentrates where the polynomial factor is smallest, until the network can no longer supply it.

\begin{figure}[!htbp]\centering
\includegraphics[width=\linewidth]{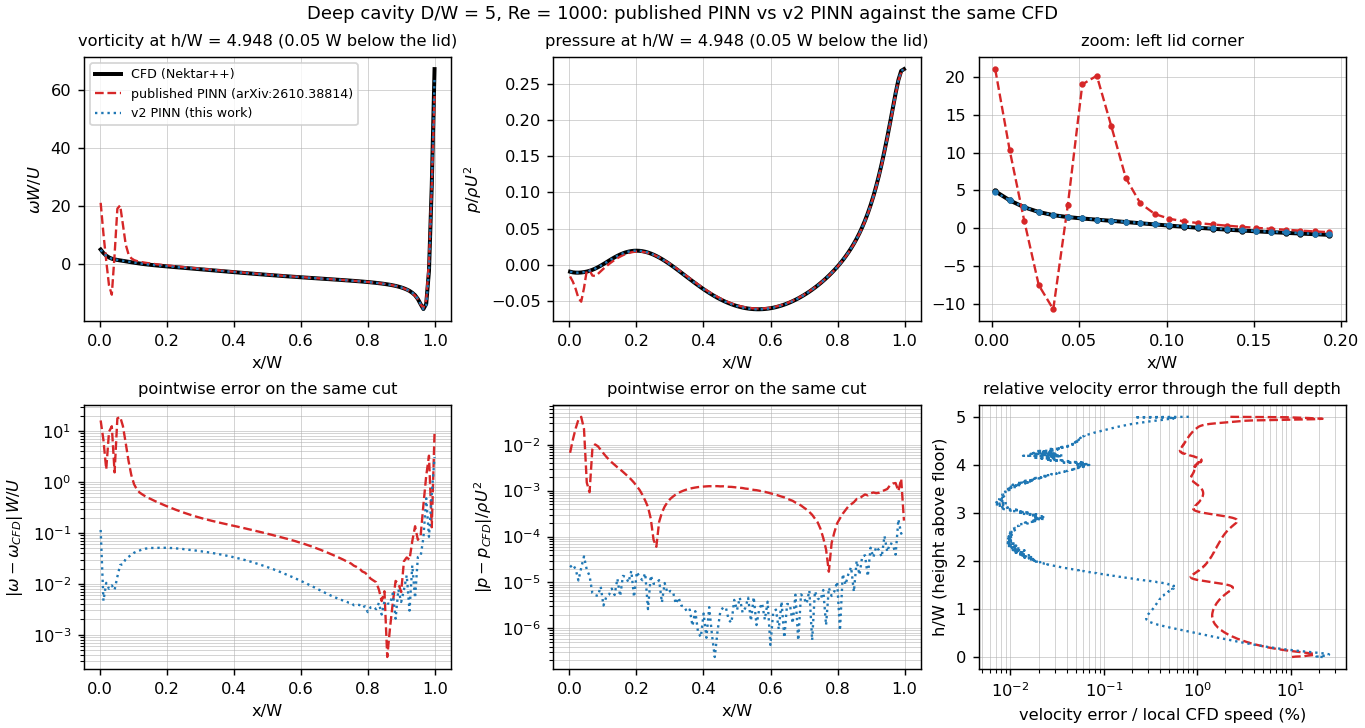}
\caption{$\Rey=1000$, $D/W=5$. Top: vorticity, pressure and a close-up of the vorticity near the left corner along $h=4.95W$, $0.05W$ below the lid. Bottom left and centre: absolute errors along the same line. Bottom right: horizontal RMS of the velocity error divided by the horizontal RMS of the reference speed, against height. Black: spectral/$hp$ reference; red dashed: earlier PINN; blue dotted: wall-local formulation.}
\label{fig:profiles-local}
\end{figure}

Figure~\ref{fig:profiles-local} shows what this means just below the lid at $\Rey=1000$. The vorticity of the earlier solution oscillates near the left corner, where the reference decreases smoothly, and overshoots near the right corner. The wall-local formulation follows the reference across the whole width, and its error becomes comparable with the vorticity itself only within a few hundredths of a width of the right wall, where the vorticity is largest.

\FloatBarrier

\subsection{Which ingredient matters}\label{sec:ablation}
\begin{figure}[!htbp]\centering
\includegraphics[width=\linewidth]{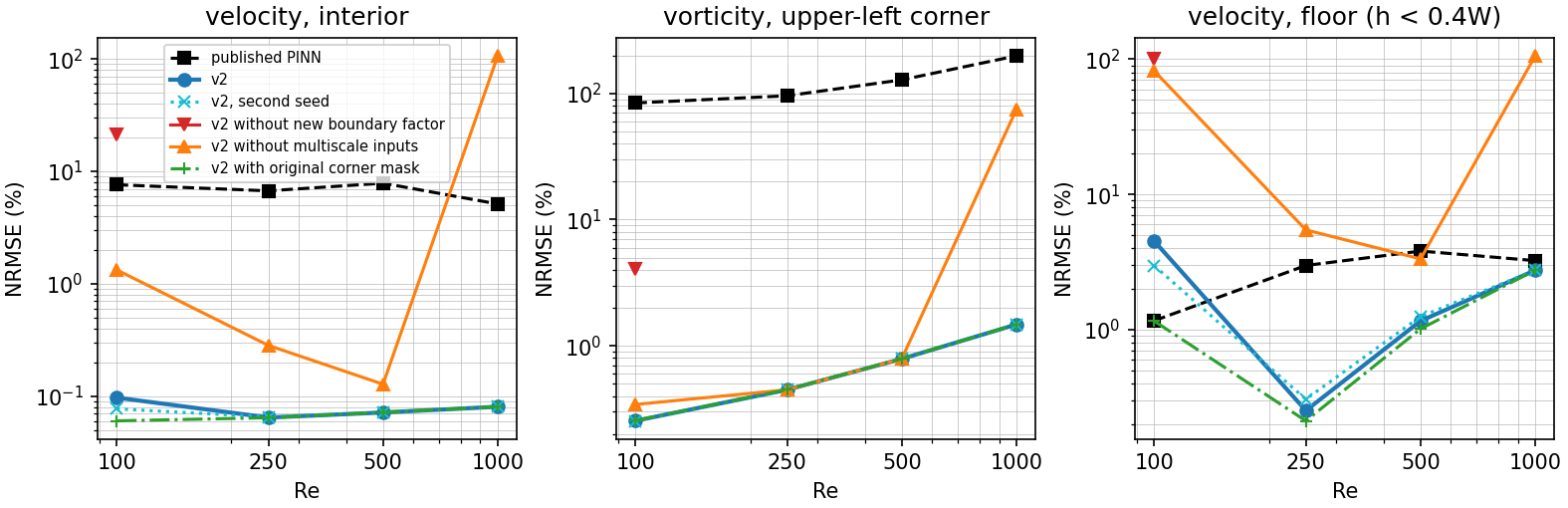}
\caption{NRMSE against Reynolds number for the earlier PINN, the wall-local formulation, a second seed and ablations (i)--(iii) at $D/W=5$. The ADF (case iv) coincides with the wall-local formulation at every Reynolds number and is not shown.}
\label{fig:ablation}
\end{figure}

To separate the ingredients of the wall-local formulation we repeated the sequence $\Rey=100\to1000$ with a second random seed and with one change at a time: (i) the polynomial factor with the wall-distance inputs, (ii) $B_{\tanh}$ with only $(x,y)$ as inputs, (iii) the original corner mask restored, and (iv) the ADF (\ref{eq:Badf}) instead of $B_{\tanh}$ (Table~\ref{tab:ablation}, Fig.~\ref{fig:ablation}).

\begin{table}[htbp]\centering\small
\caption{Ablations at $D/W=5$. NRMSE (\%) of the interior velocity, the corner vorticity and the floor velocity against the spectral/$hp$ solution.}
\label{tab:ablation}
\begin{tabular}{lcccccc}\toprule
 & \multicolumn{3}{c}{$\Rey=100$} & \multicolumn{3}{c}{$\Rey=1000$}\\
\cmidrule(lr){2-4}\cmidrule(lr){5-7}
Case & interior & corner $\omega$ & floor & interior & corner $\omega$ & floor\\\midrule
Matched original (polynomial) & 0.07 & 0.45 & 1.9 & \multicolumn{3}{c}{diverged}\\
Wall-local ($B_{\tanh}$, wall inputs) & 0.10 & 0.25 & 4.5 & 0.081 & 1.5 & 2.8\\
\quad second seed & 0.077 & 0.25 & 3.0 & 0.081 & 1.5 & 2.8\\
\quad (i) polynomial factor, wall inputs & 21 & 4.1 & 100 & \multicolumn{3}{c}{diverged at $\Rey=250$}\\
\quad (ii) raw inputs only & 1.3 & 0.34 & 82 & 106 & 75 & 106\\
\quad (iii) corner mask restored & 0.061 & 0.26 & 1.2 & 0.081 & 1.5 & 2.8\\
\quad (iv) ADF instead of $B_{\tanh}$ & 0.06 & 0.25 & 0.13 & 0.08 & 1.5 & 2.8\\\bottomrule
\end{tabular}
\end{table}

The second seed gives the same errors as the first from $\Rey=250$ on, so the result does not rest on a favourable initialization. The ADF gives the same interior and corner errors as $B_{\tanh}$ at every Reynolds number. We had expected it to be worse near the corners, because close to a corner it is several times smaller than $B_{\tanh}$; it is not. What matters, then, is that the factor is local to the walls and independent of the size of the domain, not its particular form, and the standard ADF is a good choice.

The wall-distance inputs become necessary as the boundary layers thin. Without them, $B_{\tanh}$ alone is adequate up to $\Rey=500$, but at $\Rey=1000$ the network converges to a wrong solution with a small training loss. The held-out test detected this: the residual between the collocation points was thousands of times the training loss. With only $(x,y)$ as inputs, the network apparently cannot represent a lid layer a few hundredths of a width thick with the available parameters and iterations.

The polynomial factor combined with the wall-distance inputs is worse than either alone, and its training diverges already at $\Rey=250$. We have not identified the mechanism. One possibility is that the inputs make it easy for the network to produce the very large near-wall amplitudes that the polynomial factor demands, and that optimizing such amplitudes is unstable. Restoring the corner mask makes no difference from $\Rey=250$ on: once the representation can express the corner flow, it does not matter whether the residual is enforced in the innermost corner region. This also explains the limited effect of the workflow changes of Section~\ref{sec:results}, which all kept the polynomial factor.
\FloatBarrier

\subsection{Dependence on depth}\label{sec:depth}
Equation~(\ref{eq:Bnear}) predicts that the polynomial factor becomes worse as the cavity deepens. We repeated the sequence $\Rey=100\to1000$ at $D/W=1$, 2 and 10 with the matched polynomial and the wall-local formulations (Table~\ref{tab:depth}). With the polynomial factor the training reaches $\Rey=1000$ at $D/W=1$ and 2, fails at $\Rey=1000$ at $D/W=5$, and fails at $\Rey=250$ at $D/W=10$. The wall-local formulation completes every case.

\begin{table}[htbp]\centering\small
\caption{Dependence on depth: NRMSE (\%) of the interior velocity and the corner vorticity. At $D/W=10$ the errors are given against both references; $\Rey=250$ and 500 are intermediate steps of the continuation, trained for half as many Broyden iterations as $\Rey=1000$ (Section~\ref{sec:lowRe}).}
\label{tab:depth}
\begin{tabular}{rrlcccc}\toprule
 & & & \multicolumn{2}{c}{Polynomial} & \multicolumn{2}{c}{Wall-local}\\
\cmidrule(lr){4-5}\cmidrule(lr){6-7}
$D/W$ & $\Rey$ & reference & interior $u$ & corner $\omega$ & interior $u$ & corner $\omega$\\\midrule
1 & 100 & finite volume & 0.035 & 0.09 & 0.034 & 0.09\\
1 & 1000 & finite volume & 0.13 & 0.20 & 0.13 & 0.21\\
2 & 100 & spectral/$hp$ & 0.003 & 0.03 & 0.004 & 0.03\\
2 & 1000 & spectral/$hp$ & 0.14 & 2.8 & 0.019 & 0.08\\
5 & 100 & spectral/$hp$ & 0.07 & 0.45 & 0.10 & 0.25\\
5 & 1000 & spectral/$hp$ & \multicolumn{2}{c}{diverged} & 0.08 & 1.5\\
10 & 100 & spectral/$hp$ & 6.1 & 3.1 & 0.40 & 0.07\\
10 & 100 & finite volume & 6.1 & 3.1 & 0.39 & 0.10\\
10 & 250 & finite volume & \multicolumn{2}{c}{failed} & 1.2 & 0.30\\
10 & 500 & finite volume & \multicolumn{2}{c}{---} & 0.91 & 0.31\\
10 & 1000 & spectral/$hp$ & \multicolumn{2}{c}{diverged at $\Rey=250$} & 0.35 & 0.21\\
10 & 1000 & finite volume & \multicolumn{2}{c}{diverged at $\Rey=250$} & 0.36 & 0.24\\\bottomrule
\end{tabular}
\end{table}

In the square cavity the two factors give the same errors at both Reynolds numbers, as expected when the lid is never far from the floor. At $\Rey=100$ they are also equally accurate for $D/W=2$ and 5, but at $D/W=10$ the polynomial factor is an order of magnitude worse. At $\Rey=1000$ it is already several times worse at $D/W=2$ and fails at the larger depths. The polynomial factor thus degrades in both directions that Eq.~(\ref{eq:Bnear}) predicts, with depth at fixed Reynolds number and with Reynolds number at fixed depth. At $D/W=10$ the two references give the same errors for the wall-local formulation, which shows that these interior errors belong to the PINN and not to either reference.

\begin{figure}[!htbp]\centering
\includegraphics[width=0.82\linewidth]{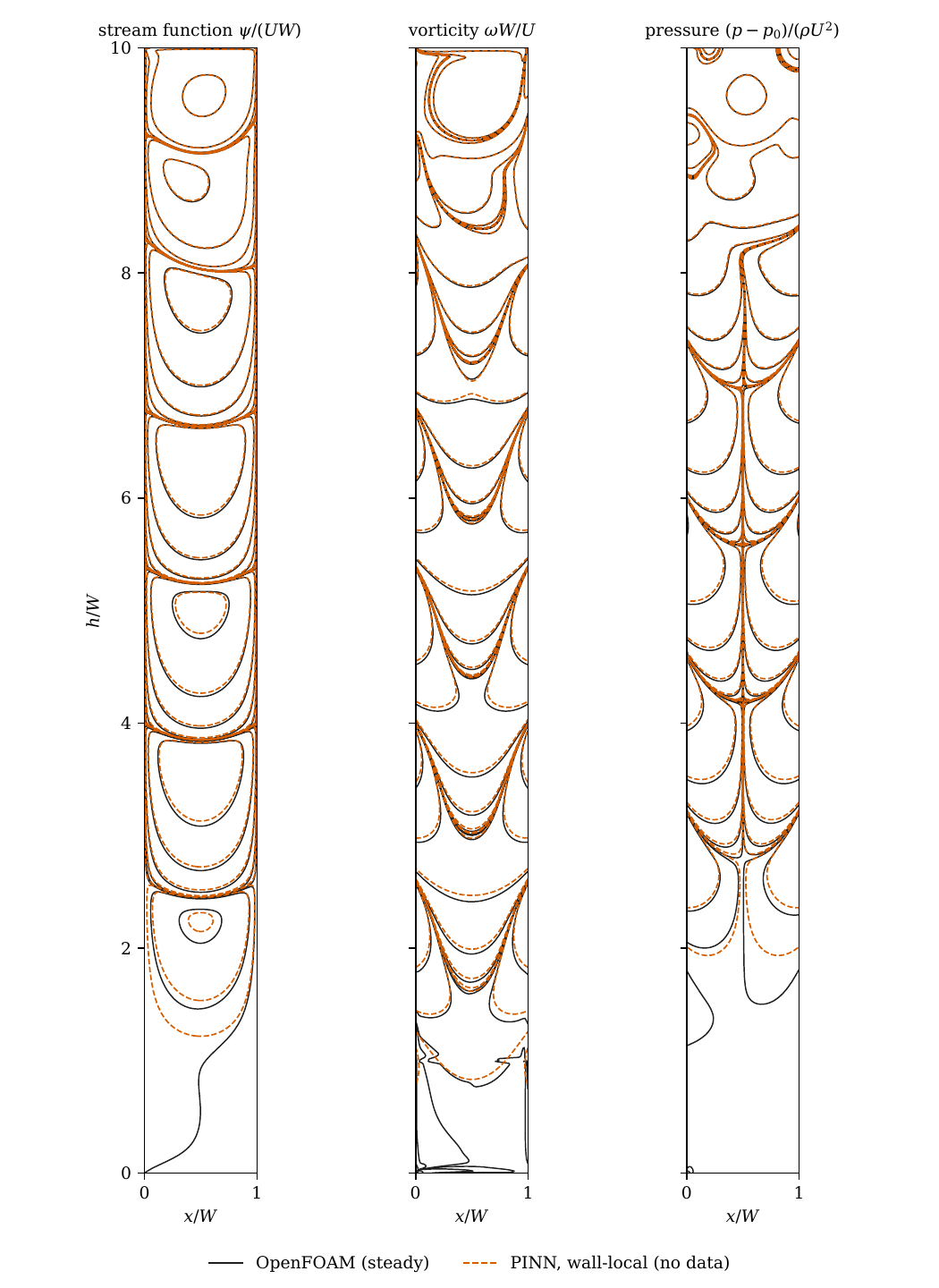}
\caption{$D/W=10$, $\Rey=1000$: contours of the stream function $\psi/(UW)$, the sampled vorticity $\omega W/U$ and the pressure $(p-p_0)/(\rho U^2)$, with $p_0$ the pressure at the centre of the floor. Levels $\pm10^{-k}$ for integer $k$. Black solid: steady finite-volume solution; orange dashed: wall-local PINN. Below $h\approx W$ the finite-volume contours are affected by round-off.}\label{fig:D10-full}
\end{figure}

\begin{figure}[!htbp]\centering
\includegraphics[width=\linewidth]{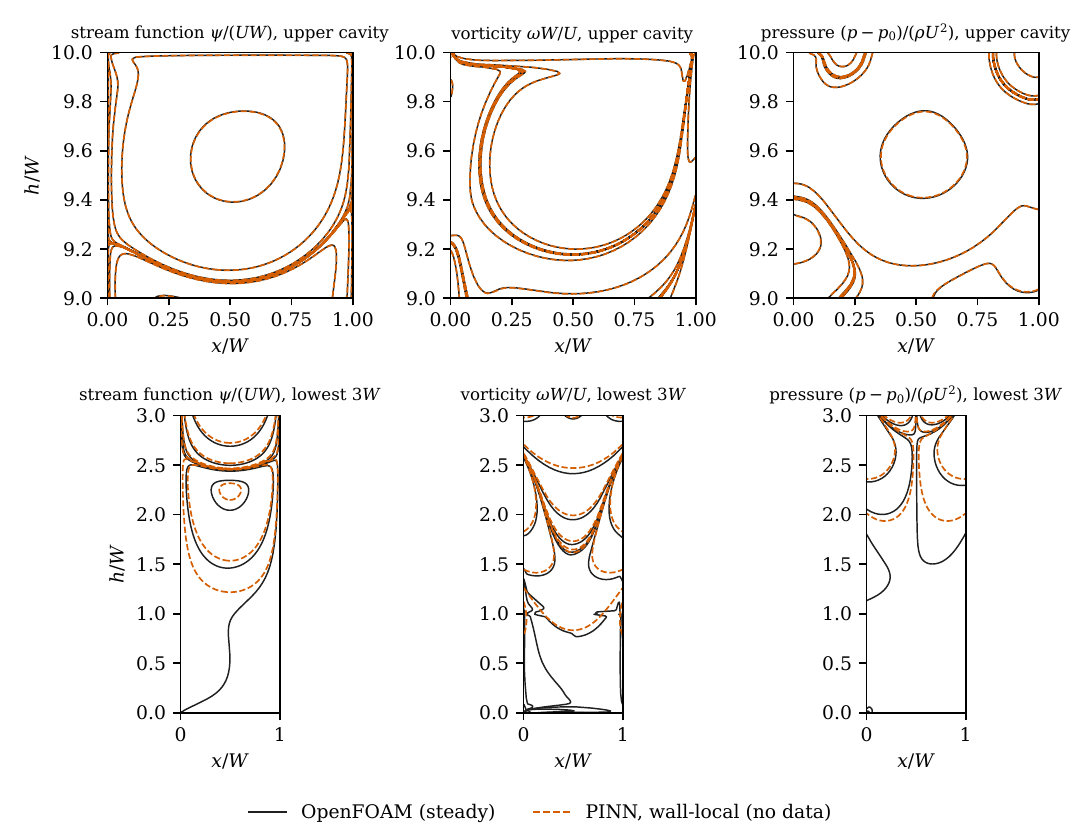}
\caption{Enlargements of Fig.~\ref{fig:D10-full}: the upper width (top) and the lowest three widths (bottom).}\label{fig:D10-zoom}
\end{figure}

\begin{figure}[!htbp]\centering
\includegraphics[width=0.55\linewidth]{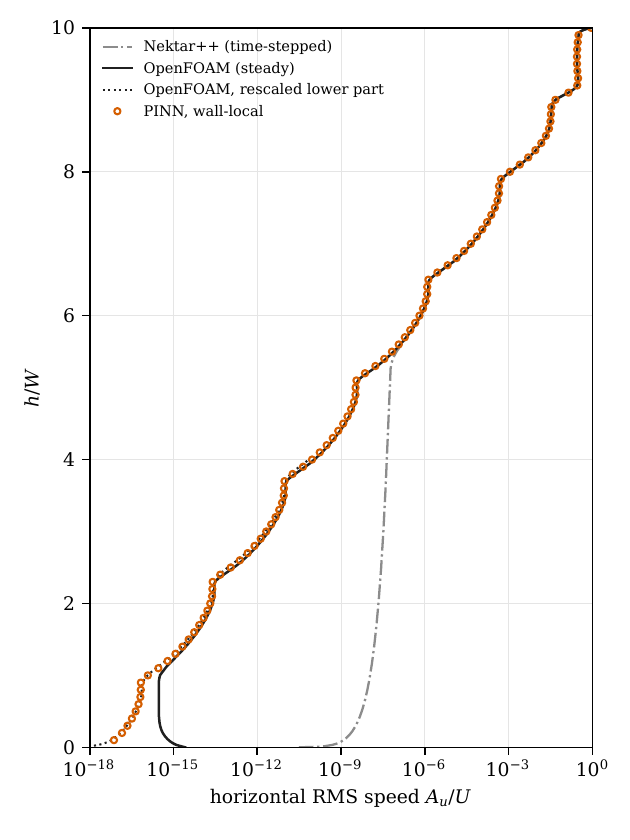}
\caption{$D/W=10$, $\Rey=1000$: horizontal RMS speed against height for the time-stepped spectral/$hp$ solution at $t=480\,W/U$, the full steady finite-volume solution, the rescaled finite-volume solution of the lowest four widths, and the wall-local PINN.}\label{fig:D10amp}
\end{figure}

Figures~\ref{fig:D10-full} and \ref{fig:D10-zoom} show the cavity ten widths deep at $\Rey=1000$, in the line-contour style of the earlier figures. The stream function, vorticity and pressure contours of the steady reference and the PINN have the same eight counter-rotating eddies and coincide in the upper part of the cavity. Figure~\ref{fig:D10amp} shows that the speed falls by about sixteen orders of magnitude from the lid to the floor. The time-stepped spectral solution follows the other two down to mid-depth and then levels off, as discussed in Section~\ref{sec:nektar}; below about one width above the floor the full steady solution reaches round-off, which also explains the irregular reference contours there in Figs.~\ref{fig:D10-full} and \ref{fig:D10-zoom}.

\begin{table}[htbp]\centering\small
\caption{$D/W=10$, $\Rey=1000$: horizontal RMS speed $A_u/U$ in the lower part of the cavity from the full finite-volume solution, the rescaled sub-domain solution (divided by $S=10^8$) and the PINN.}
\label{tab:floor10}
\begin{tabular}{rcccc}\toprule
$h/W$ & full FV & sub-domain FV & PINN & PINN / sub-domain\\\midrule
3.85 & $2.28\times10^{-11}$ & $2.28\times10^{-11}$ & $1.86\times10^{-11}$ & 0.82\\
3.10 & $2.93\times10^{-12}$ & $2.93\times10^{-12}$ & $2.41\times10^{-12}$ & 0.82\\
2.10 & $2.52\times10^{-14}$ & $2.52\times10^{-14}$ & $2.10\times10^{-14}$ & 0.83\\
1.34 & $1.42\times10^{-15}$ & $1.40\times10^{-15}$ & $1.05\times10^{-15}$ & 0.75\\
0.85 & $2.54\times10^{-16}$ & $7.28\times10^{-17}$ & $6.05\times10^{-17}$ & 0.83\\
0.35 & $2.61\times10^{-16}$ & $2.61\times10^{-17}$ & $2.03\times10^{-17}$ & 0.78\\
0.10 & $5.38\times10^{-16}$ & $6.00\times10^{-18}$ & $4.55\times10^{-18}$ & 0.76\\\bottomrule
\end{tabular}
\end{table}

Table~\ref{tab:floor10} tests the lowest eddies against the rescaled solution. Where the full steady solution is above round-off, the two finite-volume solutions agree to three or four digits; below that, the full solution stops decaying while the rescaled one continues to the floor. The PINN decays at the same rate as the rescaled solution, but its speed is about a fifth lower throughout the lowest four widths, and the ratio does not drift with depth. The same deficit appears higher up and grows gradually from the second eddy downwards, in vorticity and pressure as well as in velocity. We read this as follows. Below mid-depth the flow is close to Stokes flow, and the rate at which the eddies weaken is set locally; the PINN reproduces it. The amplitude of each eddy is inherited from the one above, so a small error in the intermediate eddies is passed down unchanged. The interior NRMSE does not reveal this deficit, because it is dominated by the strong upper flow; as in Section~\ref{sec:results}, the amplitudes through the depth have to be compared on their own scales.
\FloatBarrier

\subsection{The eddy stack at lower Reynolds numbers}\label{sec:lowRe}
\begin{figure}[!htbp]\centering
\includegraphics[width=\linewidth]{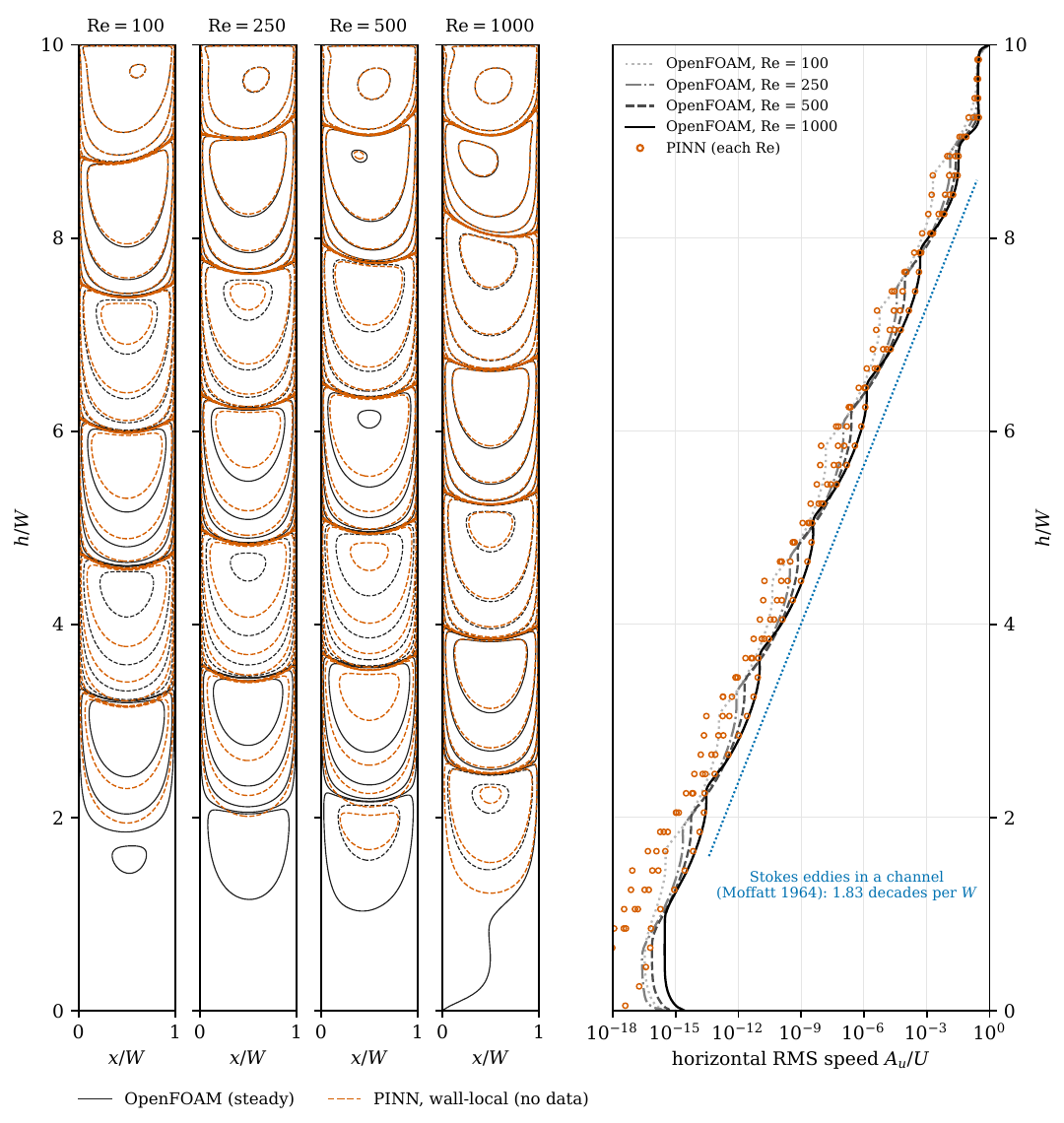}
\caption{$D/W=10$ at $\Rey=100$, 250, 500 and 1000. Left: stream-function contours, levels $\pm10^{-k}$ for $k=1,\dots,16$; black solid: steady finite-volume solution, orange dashed: wall-local PINN. Right: horizontal RMS speed against height, finite-volume solutions as lines and the PINN as circles; the dotted line has the slope of Moffatt's Stokes eddies in a channel. Below about one width above the floor the finite-volume solutions are at round-off.}\label{fig:D10-Re}
\end{figure}

At $D/W=10$ we also computed steady finite-volume solutions at $\Rey=100$, 250 and 500, the earlier steps of the continuation, and compared them with the PINN at each step (Fig.~\ref{fig:D10-Re}). Table~\ref{tab:eddies} lists what we measured along the vertical centreline: where the stream function changes sign, which gives the dividing streamlines, and the largest $|\psi|$ between two of them, which gives the strength of each eddy.

\begin{table}[htbp]\centering\small
\caption{The eddy stack at $D/W=10$ from the steady finite-volume solutions, measured along $x=W/2$. Length and strength ratio of the lower eddies are the same for every eddy from the third down to the last one above round-off. The last column is the ratio of the horizontal RMS speeds of the PINN and the finite-volume solution at $h=3W$.}
\label{tab:eddies}
\begin{tabular}{rcccccc}\toprule
 & Primary eddy & \multicolumn{3}{c}{Lower eddies} & Strength ratio & PINN/FV\\
\cmidrule(lr){3-5}
$\Rey$ & depth & length & strength ratio & decay rate & first/second & at $h=3W$\\\midrule
100 & $1.20W$ & $1.397W$ & 359 & 1.83 decades/$W$ & 112 & 0.20\\
250 & $0.97W$ & $1.397W$ & 359 & 1.83 decades/$W$ & 18 & 0.24\\
500 & $0.93W$ & $1.397W$ & 359 & 1.84 decades/$W$ & 12 & 0.25\\
1000 & $0.93W$ & $1.397W$ & 359 & 1.84 decades/$W$ & 10 & 0.82\\\midrule
Stokes theory \cite{moffatt1964eddies} & -- & $1.396W$ & 357 & 1.83 decades/$W$ & -- & --\\\bottomrule
\end{tabular}
\end{table}

The Reynolds number changes the top of the stack. The primary eddy is longest at $\Rey=100$ and becomes shorter and rounder as inertia grows, as it does at $D/W=5$ (Section~\ref{sec:reseries}). The second eddy changes most. It is driven by the primary eddy across their dividing streamline, and its strength relative to the primary rises from about one hundredth at $\Rey=100$ to one tenth at $\Rey=1000$; at $\Rey=1000$ it is also shorter, about one width instead of 1.4. Its local Reynolds number based on its strength, $\Rey\,\psi_{\max}/(UW)$, grows from about 0.1 at $\Rey=100$ to about ten at $\Rey=1000$, while that of the third eddy stays below one; the second eddy is the only one besides the primary in which inertia comes to matter.

Below it the Reynolds number does not matter. From the third eddy down, every eddy in the finite-volume solutions is $1.397W$ long and 359 times weaker than the eddy above it, the same to three digits at all four Reynolds numbers, and the speed falls by 1.83 decades per width. These are Moffatt's values for Stokes eddies between parallel walls, $1.396W$ and 357 (Section~\ref{sec:problem}). The lower stack is therefore a Stokes flow whose shape is fixed by the width of the cavity alone. The upper eddies set only where it begins and how strong it is. Because the second eddy shortens as the Reynolds number increases, the whole stack moves up by about $0.6W$ between $\Rey=100$ and 1000, and one more eddy fits above the floor, eight instead of seven. The strength of the stack at a given height changes much less: at mid-depth the speed at $\Rey=1000$ is only about twice that at $\Rey=100$. A deeper cavity adds eddies to the bottom of the stack without changing those above, and each additional width lowers the speed at the floor by almost two orders of magnitude. The same coefficient, $4.212/W$, sets the depth envelope (\ref{eq:e}) of the PINN; it was taken from the Stokes limit before these solutions existed, and the solutions now show that it is the actual decay rate of the cavity below the upper eddies at every Reynolds number studied.

The envelope does not, however, give the PINN the decay rate for free. At $\Rey=1000$, the final step of the continuation, the PINN places every dividing streamline within $0.02W$ of the reference and its eddies are $1.39W$ long, but each is about 367 times weaker than the one above, 2.5\% more than in the reference. Over the lower stack this small excess accumulates to the deficit of about a fifth found in Table~\ref{tab:floor10}. At the intermediate steps, which were trained with 10{,}000 instead of 20{,}000 Broyden iterations and stopped at a ten times larger loss, the dividing streamlines are still within $0.06W$ of the reference down to $h=3W$, but below mid-depth each eddy is between 1.2 and 2 times too weak relative to the one above, and at $h=3W$ the PINN speed is only a fifth to a quarter of the reference. The interior NRMSE of about one per cent (Table~\ref{tab:depth}) does not show this. The decay through the weakest part of the stack is the last thing the network learns, and it is learned only when the training is run to convergence.
\FloatBarrier

\subsection{Higher Reynolds numbers}\label{sec:highRe}
We continued the wall-local formulation at $D/W=5$ to $\Rey=1500$, 2000 and 2250 (Table~\ref{tab:highRe}).

\begin{table}[htbp]\centering\small
\caption{Wall-local formulation at $D/W=5$ and higher Reynolds numbers: NRMSE (\%) against the spectral/$hp$ (S) and the steady finite-volume (FV) solutions, and the held-out residual ratio. The spectral/$hp$ solution at $\Rey=2000$ is not converged near the floor (Section~\ref{sec:nektar}).}
\label{tab:highRe}
\begin{tabular}{rlcccccc}\toprule
 & & \multicolumn{2}{c}{Interior} & \multicolumn{2}{c}{Corner} & Floor & held-out\\
\cmidrule(lr){3-4}\cmidrule(lr){5-6}\cmidrule(lr){7-7}
$\Rey$ & ref. & velocity & vorticity & velocity & vorticity & velocity & ratio\\\midrule
1000 & S  & 0.081 & 0.70 & 0.23 & 1.5 & 2.8 & 0.36\\
     & FV & 0.23 & 0.38 & 0.09 & 0.12 & 0.47 & \\[2pt]
1500 & -- & & & & & & 0.17\\[2pt]
2000 & S  & 0.04 & 0.09 & -- & 0.28 & ($\approx$100) & 0.15\\
     & FV & 0.58 & 1.2 & 0.25 & 0.36 & 0.35 & \\[2pt]
2250 & FV & 0.66 & 1.5 & 0.29 & 0.42 & 0.29 & 0.44\\\bottomrule
\end{tabular}
\end{table}

\begin{figure}[!htbp]\centering
\includegraphics[width=\linewidth]{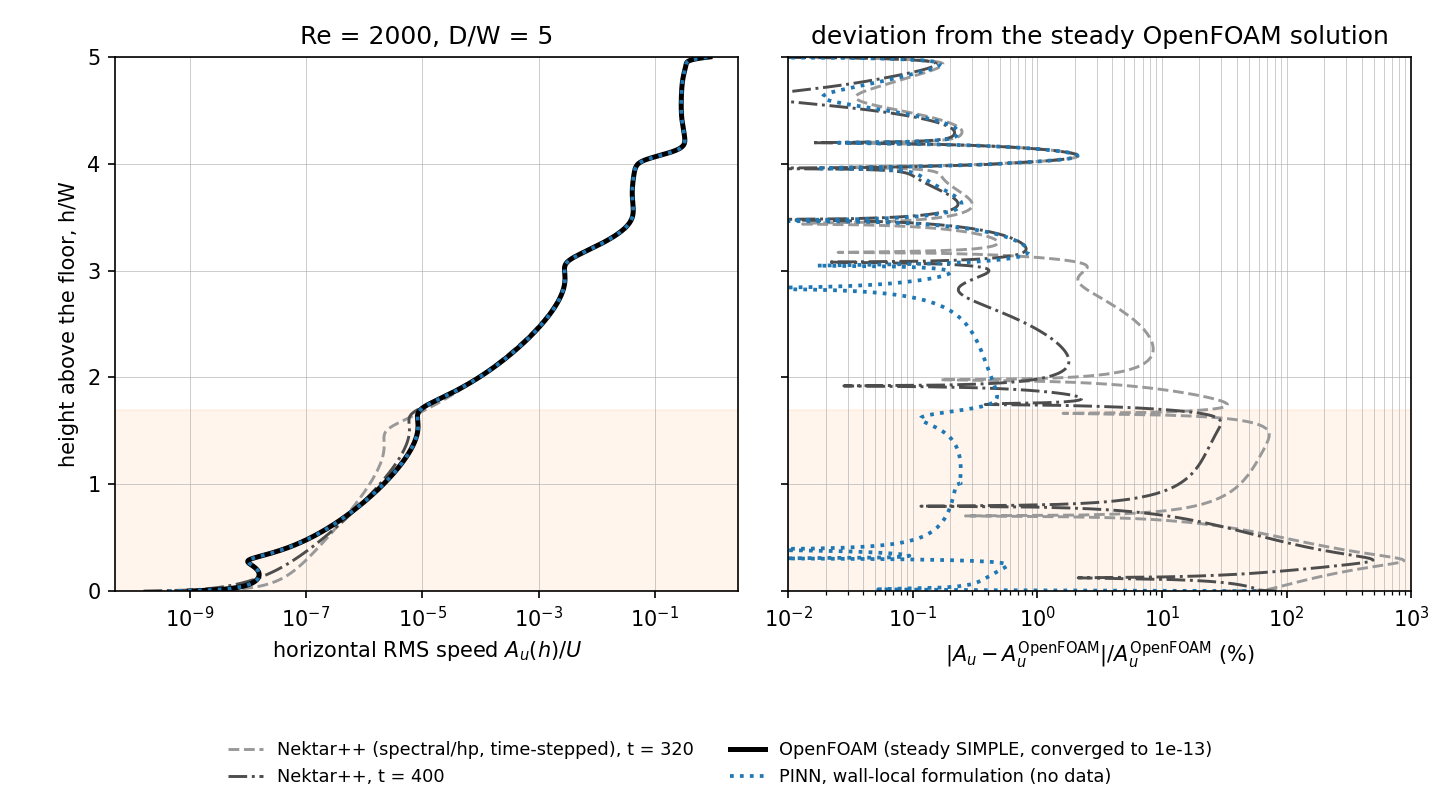}
\caption{$\Rey=2000$, $D/W=5$. Left: horizontal RMS speed against height. Right: its deviation from the steady finite-volume solution. Grey dashed and dark dash-dotted: time-stepped spectral/$hp$ solution at $t=320$ and $400\,W/U$; black: steady finite-volume solution; blue dotted: wall-local PINN. Shaded: the lowest $1.7W$.}
\label{fig:floor2000}
\end{figure}

\begin{figure}[!htbp]\centering
\includegraphics[width=0.62\linewidth]{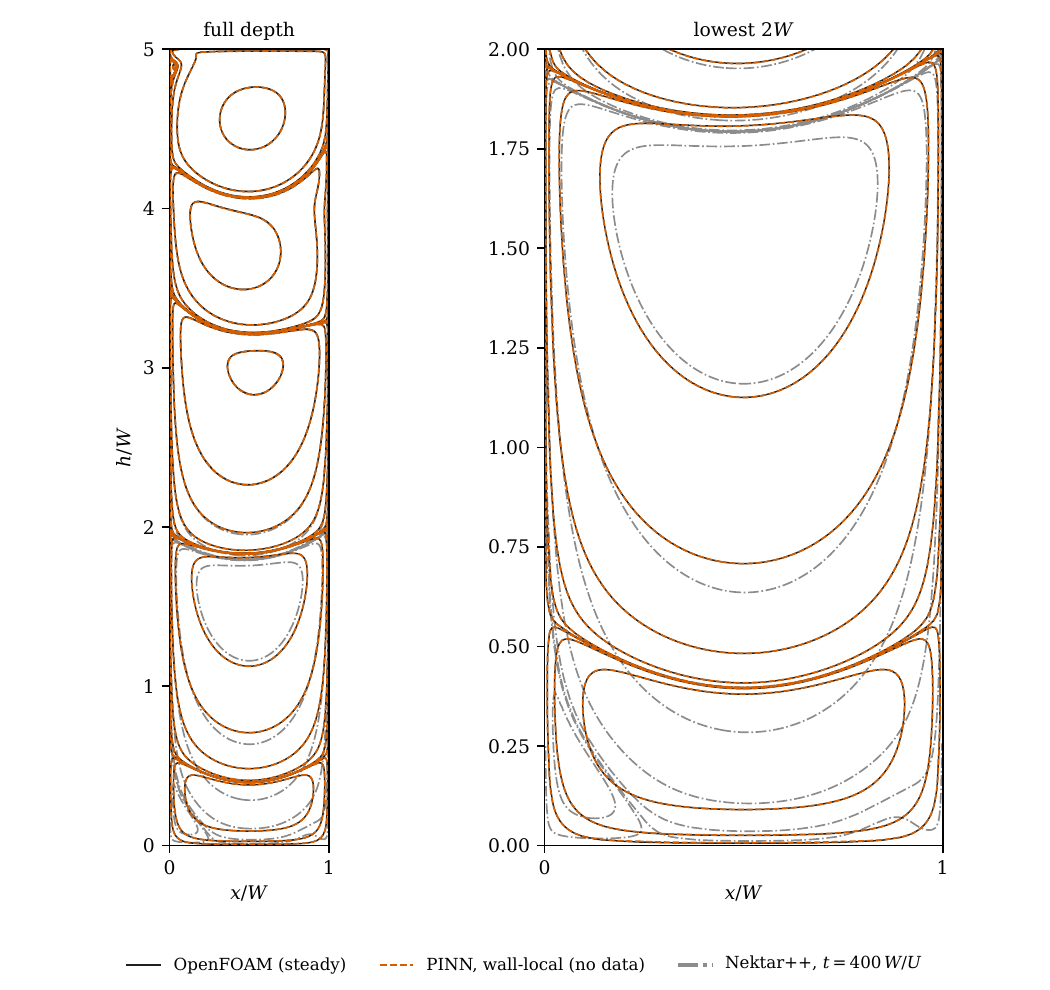}
\caption{$\Rey=2000$, $D/W=5$: stream-function contours ($\pm10^{-k}$) of the time-stepped spectral/$hp$ solution at $t=400\,W/U$ (grey dash-dotted), the steady finite-volume solution (black solid) and the wall-local PINN (orange dashed), over the full depth and in the lowest two widths.}
\label{fig:stream2000}
\end{figure}

At $\Rey=2000$ the PINN and the spectral solution agree closely everywhere except at the floor, where they differ completely. Three PINNs with different boundary factors and envelopes gave the same floor eddy to three significant figures, which made an error of the PINN unlikely, and the steady finite-volume solution confirms the PINN: in the lowest third of the cavity the two agree to a fraction of a per cent (Fig.~\ref{fig:floor2000}). The spectral solution is still moving towards them, and its floor eddy grew substantially between $t=320$ and $400\,W/U$. The streamlines (Fig.~\ref{fig:stream2000}) show the same thing: the steady solution and the PINN both have a separate counter-rotating eddy just above the floor, which the time-stepped solution has not yet formed.

In the upper flow the order of the two references is reversed. The PINN is closer to the spectral solution than to the finite-volume solution, at $\Rey=1000$ and at $\Rey=2000$. Since the PINN and the spectral solution are independent, most of the difference from the finite-volume solution must be the discretization error of the second-order solver, which grows as the shear layers thin. We therefore take each reference where it is reliable, the spectral solution for the upper flow and the corners, and the steady solution near the floor. On that basis the PINN at $\Rey=2250$ is as accurate as at $\Rey=2000$: its floor profile agrees with the steady solution to three digits, and in its two lowest eddies the ratio of strengths and the spacing are close to Moffatt's values \cite{moffatt1964eddies}.

The continuation stopped at $\Rey=2500$. The steady finite-volume iteration converges at $\Rey=3000$, so the difficulty is not a loss of a steady solution. In three attempts, with the tanh factor and with the ADF and from two different starting points, the training loss became very small while the held-out residual ratio rose by seven to eight orders of magnitude, and the two factors disagreed on the strength of the second eddy. The networks fitted their collocation points with a flow that is wrong between them. We report these runs as failures; the held-out test identified each of them without a reference solution. Two further attempts, one with half the step in Reynolds number and one with twice as many collocation points, failed in the same way. The limit is therefore more likely set by the size of the network or by the optimizer than by the sampling or the continuation, and we leave it for further work.

\FloatBarrier

\section{Discussion}
\subsection{Why the training changes could not remove the corner error}
The two parts of this study answer each other. In the first, regional collocation, transferred initialization and changes to the residual scaling and the depth envelope moved the error of the established formulation by at most a factor of two, and the remaining error stayed at the lid corners in every workflow and at every Reynolds number. In the second, a single change to the boundary factor, made with the training held fixed, reduced the same discrepancies by about two orders of magnitude. The comparison shows that the corner error was a property of the representation, not of the training. Hard boundary constraints are usually described as a way to remove boundary terms from the loss; they also set the scale of the network output everywhere. The polynomial factor is small wherever one wall is close and the opposite wall is far, and in a deep cavity this happens exactly at the lid corners, where the flow is most demanding. The consequence is not a uniform loss of accuracy but a limit that moves to lower Reynolds numbers as the cavity deepens.

The matched retraining also shows how much a training history can cost: at $\Rey=100$ the same polynomial factor gave 7.6\% interior error in the first part and 0.07\% with matched training. Representations should be compared only under matched training, and the workflow comparisons that follow should be read with that in mind.

\subsection{What the workflow comparison establishes}
No workflow is best everywhere. The refined workflow has the smallest velocity discrepancies and the smallest interior vorticity discrepancy; the half envelope has the smallest interior and upper-left pressure discrepancies; the control has the smallest floor pressure and floor vorticity discrepancies. Each workflow was trained once, so these rankings say nothing about the variation between runs from different seeds.

The hard-constrained representation, depth envelope, reciprocal residual scaling and staged SOAP--Broyden optimizer come from the established formulation; what was tested here is regional collocation, transferred initialization and two scale variants. Because the workflows start from different fields and have different total training (Section~\ref{sec:formulation}), only control and soft residual form a clean comparison, and the advantage of the refined workflow includes its earlier training at $\Rey=1000$. We draw no conclusion about the cost of the PINN relative to CFD.

\subsection{Velocity agreement and sampled rotational structure}
Velocity, pressure and vorticity test different properties of the solution. The refined field matches the CFD velocity to a few thousandths of the lid speed in the interior, yet its vorticity discrepancy near the upper-left corner is about $3U/W$ (Tables~\ref{tab:errors} and \ref{tab:pv-errors}), because vorticity depends on second derivatives of the streamfunction. The Reynolds-number series, which uses the control rather than the refined field at $\Rey=1000$, shows the same contrast (Figs.~\ref{fig:re-velocity-error-physical} and \ref{fig:re-omega-error-physical}).

The sampled vorticity applies one finite-difference operator to both fields; it does not compare the native spectral derivatives with the neural automatic derivatives, and it does not remove interpolation error. The PINN fields respond more strongly to coarsening of the sampling grid, but this sensitivity cannot be subtracted from the discrepancies to give a corrected error. Because the wall band is excluded, the regional norms also do not assess the wall shear or the corners themselves.

Locally centered pressure displays show shape only; the error statistics use one interior gauge, and the floor decomposition separates offset and shape. Streamlines locate the vortex stack but do not measure circulation, separatrix area or stability, and the changes in vortex geometry agree with known deep-cavity flow. Below the upper eddies, however, the stack at $D/W=10$ follows Moffatt's Stokes eddies quantitatively at every Reynolds number studied (Section~\ref{sec:lowRe}).

The main error lies near the lid, at its junctions with the sidewalls (Figs.~\ref{fig:re-velocity-error-physical}--\ref{fig:re-upper-profiles}). The omitted collocation points and the mask weaken the residual in the innermost corner neighborhoods, but the mask is already close to unity at the evaluation points, so the location of the error alone does not show that the mask causes it. Separating the effects of point omission, residual weighting and finite training would need a controlled comparison.

\subsection{The Stokes stack as a test that needs no reference}
The lower eddies of a deep cavity have a property that a solver can be checked against without any reference solution. Below the second eddy their length and their strength ratio do not depend on the Reynolds number or on the depth, only on the width, and they must take Moffatt's values. A solution whose lower eddies have the wrong length, or weaken too fast or too slowly, is wrong there, however small its global error. This is how the deficit of the less-trained PINNs at $\Rey=100$--500 shows itself (Section~\ref{sec:lowRe}): their dividing streamlines are nearly in the right place, but each eddy is too weak by a factor that grows with depth, which no interior norm reveals. The test works in the other direction too. A time-stepped solution that has not yet spun up its lower eddies, such as the spectral solution at $D/W=10$, shows a speed that stops decaying instead of falling by 1.83 decades per width, and a steady solution near round-off shows the same. Together with the held-out residual, which tests the equations between the collocation points, the eddy ratio tests the weakest part of the flow, where neither the loss nor a global error measure has any weight.

\subsection{Reference sensitivity and scope}
The Nektar++ solution is a numerical reference, not an exact one. Its sensitivity was measured at $t\simeq170$, earlier than the detailed comparison at $t\simeq317$, so it gives no error bound for the floor velocity, pressure or vorticity of the reference, and the floor discrepancies must be read with this in mind.

The Reynolds-number series covers one aspect ratio, with CFD fields at different times and networks with different training. It shows no monotonic error trend, but does not show that the attainable accuracy is independent of Reynolds number. Some losses were still decreasing when training stopped (Appendix~\ref{app:optimization-histories}), so the comparisons are between the fields reached after a fixed number of steps, not between converged solutions.

Finally, a steady solution says nothing about its stability or uniqueness. Hopf thresholds in deep cavities depend on geometry and lid conditions \cite{zhuo2013bifurcation,lin2013transition}, so thresholds for the conventional lid do not carry over to the regularized one, and the present calculations neither locate an instability nor describe turbulence.

\section{Conclusions}
We assessed data-free PINNs for the steady flow in deep lid-driven cavities, region by region and for velocity, pressure and vorticity, against independent references that use the same smoothed lid.

With the established stream-function formulation, the PINNs reproduce the stack of counter-rotating eddies of a cavity five widths deep and its weak floor flow at Reynolds numbers from 100 to 1000. Their error, however, stays at several per cent and sits almost entirely at the two lid corners. Regional collocation, transferred initialization, softer residual scaling and a slower depth envelope each change the regional ranking of the solutions, but none of them removes this error, and no workflow is best for every observable and region.

The cause is the polynomial boundary factor. Near the lid of a deep cavity it is proportional to the distance from the lid divided by the depth, so it is very small exactly where the flow is most demanding. With identical training it is as good as any other factor in a square cavity, but it degrades as the Reynolds number grows and as the cavity deepens. Factors that are local to the walls remove this limit, and their particular form is unimportant: a product of hyperbolic tangents and the standard R-function distance function behave identically. As the boundary layers thin, the network also needs the wall distances as inputs. With these changes the same network and optimizer reduce the velocity, pressure and vorticity discrepancies of the established formulation by about two orders of magnitude, and remain accurate at every depth from one to ten widths and up to $\Rey=2250$.

The weakest eddies needed more care than the PINN. A time-stepped solver integrated from rest had not formed them, because their spin-up time grows with both the Reynolds number and the depth, and a steady solver loses them to round-off in the deepest cavity unless the equations are rescaled. With such steady references the cavity ten widths deep shows a simple structure. Only the primary eddy and, at the higher Reynolds numbers, the second eddy feel inertia. Below them the eddies are Stokes eddies with Moffatt's length and strength ratio, independent of the Reynolds number, so the Reynolds number only shifts the stack and sets its strength. The PINN reproduces this structure; fully trained, it places the dividing streamlines within a fiftieth of a width, but its eddies weaken slightly too fast, and the small excess accumulates to an amplitude about a fifth too low near the floor. Because the Stokes values do not depend on the flow above, they give a check of the weakest part of any computed deep-cavity flow that needs no reference solution.

The work has clear limits. All cases are steady and two-dimensional, the lid velocity is smoothed at the corners so the singularity of the classical problem is not represented, and the continuation stopped at $\Rey=2500$, where the held-out residual test, which also needs no reference solution, identified the failures.

\section*{Acknowledgments}
The calculations used the Unity research computing environment at the University of Massachusetts Amherst and University of Florida Research Computing resources. C.J.M. acknowledges support from the U.S. Department of Energy, Office of Fusion Energy Sciences, under Award No. DE-SC0024634.

\section*{Declaration of AI Use}

The authors used OpenAI ChatGPT (Sol 6.1 and GPT-6 Astra Ultra) and Anthropic
Claude Opus 5.5 to assist with code improvement and debugging,
and with editing and improving the manuscript text.
The authors take full responsibility for the final manuscript,
code, results, and conclusions.

\appendix
\section{Reproducibility and data availability}
Training uses seed 12345 and double precision; the held-out test points use seed 20264. For each case we record the Reynolds number, initialization, residual mask, depth scale and sampling distribution. The evaluation array labeled ``Re500'' is named after its initialization; it is compared at $\Rey=1000$.

The public project repository is \url{https://github.com/Ehsan-Roohi/FlowMLLab}; it contains related teaching and validation code, not the training states of this study. A separate private repository, \texttt{Cavity-PINN-CFD-Gallery}, holds derived case figures and comparison summaries. The manuscript package records the source fields and figure settings in \texttt{figure\_manifest.json} and \texttt{multiphysics\_analysis.json}; \texttt{build\_figures.py} and \texttt{build\_multiphysics\_figures.py} reproduce the figures and \texttt{build\_manuscript.py} assembles the LaTeX project. The training code is \texttt{cavity.py}, \texttt{train.py} and \texttt{optimizer.py}, with the regional distribution defined in \texttt{pinn\_d5\_balanced\_experiment.py} and the phases run by \texttt{pinn\_d5\_balanced\_run.py}. The Nektar++ case generator \texttt{prepare\_nektar.py} records the geometry, polynomial order, lid and time-integration settings.

The comparison arrays use axes $(x,h)$; the plotting routines transpose them to $(h,x)$. Pressure is the network output, not reconstructed from velocity. Vorticity is $\partial_xv-\partial_hu$, computed on the common grid with second-order central differences in the interior and second-order one-sided differences at the boundaries. Contour levels are chosen between the first and ninety-ninth percentiles of the reference in each displayed region and shared across models; profiles and error calculations use all values.

\texttt{build\_multiphysics\_figures.py} reconstructs the pressure and vorticity comparisons, and \texttt{multiphysics\_analysis.json} records source hashes, plotting settings and discrepancies; the scripts do not modify the raw fields or model states.

For the Reynolds-number series, \texttt{analyze\_reynolds\_series.py} records the Secure Hash Algorithm 256-bit (SHA-256) hashes of the source fields, the array orientation, sampling, gauge alignment and derivative settings in \texttt{reynolds\_analysis.json}. The $\Rey=250$ and 500 weights are used only for inference, without loading an optimizer or changing a checkpoint. The common-grid arrays, \texttt{reynolds\_common\_fields.npz}, accompany the analysis; raw weights and third-party code are not in the manuscript archive. The source records of the GitHub cases were checked against the downloaded CFD fields. The archive contains the figures in Portable Document Format (PDF) with editable TikZ sources. The floor pressure decomposition is recorded in \texttt{floor\_pressure\_shape\_check.json} and verified on the common arrays, without changing them, by \texttt{check\_floor\_pressure\_shape.py}.

Supplementary Movies S1 and S2 show the $\Rey=1000$, $D/W=5$ case. Movie S1 shows saved iterates of the refined PINN during self-scaled Broyden optimization, from iteration 25 to 20,000, in full-cavity, floor and upper-left views; its axis is the iteration count, not physical time. Movie S2 shows the transient Nektar++ solution for $1\leq tU/W\leq188$, before the reference time, in full-cavity and floor views. Frames are held longer for readability but not interpolated in time. In the PINN viewing copy a monotonic contrast adjustment makes faint lines readable without moving contours or changing levels. Hashes and presentation settings accompany the movies.

A private bundle held by the authors contains the four final $\Rey=1000$ weight files with their final-phase histories and configuration files, the common-grid and detailed fields, the $t\simeq317$ CFD snapshot, and a postprocessing script with reference metrics and hash manifests. It allows the sampled results to be checked, but not the whole study to be rerun. It includes a $\Rey=100$ CFD session file as an example, but not the $\Rey=1000$ session and restart files, the full optimizer archives or the two missing initialization histories; inherited training and evaluation code whose redistribution rights are unresolved is also excluded. The bundle has not been released and has no public DOI; any release is subject to the authors' review of rights and scope. A CSV file documents the $t\simeq170$ sensitivity diagnostics; the $t\simeq160$ source field is missing.

The public viewing copies are \href{https://youtu.be/pgQ-09eDHII}{Movie S1 (PINN optimization)} and \href{https://youtu.be/xDTzcYgPHrY}{Movie S2 (transient CFD)}.

\section{Matched-reference supplemental diagnostics}\label{app:matched-diagnostics}
Table~\ref{tab:matched-omega-pressure} uses the detailed $161\times801$ fields and exactly the masks of Tables~\ref{tab:errors} and \ref{tab:pv-errors}. Sampled vorticity uses the common finite-difference operator, not native automatic or spectral derivatives. The CFD vorticity RMS denominators are $1.27$, $3.41$ and $1.06\times10^{-7}$ in units $U/W$ for the interior, upper-left and floor regions. Pressure retains the global interior gauge except in the explicitly labeled regional-shape column. The floor shape denominator is $1.7\times10^{-10}\rho U^2$; its ratios should be read together with the absolute decomposition, not as a pressure-gradient accuracy claim.
\begin{table}[htbp]\centering\small
\caption{Matched-region NRMSE ratios for sampled vorticity, globally gauge-aligned pressure and regional pressure shape. Uniform weights and finite nonzero reference denominators are used throughout.}\label{tab:matched-omega-pressure}
\begin{tabular}{llrrr}\toprule
Workflow & Region & $\mathrm{NRMSE}_\omega$ & Pressure & Pressure shape\\\midrule
Control & Interior & 0.629 & 0.112 & 0.112\\
Control & Upper-left & 2.27 & 2.05 & 2.2\\
Control & Floor & 0.0193 & 0.00822 & 0.0191\\
Soft residual & Interior & 0.288 & 0.0478 & 0.0478\\
Soft residual & Upper-left & 1.04 & 0.597 & 0.648\\
Soft residual & Floor & 0.2 & 0.0355 & 0.242\\
Refined & Interior & 0.247 & 0.0467 & 0.0467\\
Refined & Upper-left & 0.888 & 0.763 & 0.829\\
Refined & Floor & 0.0284 & 0.0216 & 0.0236\\
Half envelope & Interior & 0.281 & 0.0426 & 0.0426\\
Half envelope & Upper-left & 1.01 & 0.359 & 0.384\\
Half envelope & Floor & 0.122 & 0.0376 & 0.193\\
\bottomrule\end{tabular}\end{table}
\FloatBarrier

\begin{table}[htbp]\centering\small
\caption{Matched-reference velocity and sampled-vorticity NRMSE in the common $121\times605$ Reynolds-number series. The original interior and upper-left masks are retained; here the floor uses $0.02<h/W<0.65$, unlike the detailed-workflow floor mask. The $\Rey=1000$ field is the control. Each denominator is the CFD RMS on the identical sampled region.}\label{tab:common-nrmse}
\begin{tabular}{rlrr}\toprule
$\Rey$ & Region & $\mathrm{NRMSE}_{\boldsymbol u}$ & $\mathrm{NRMSE}_{\omega}$\\\midrule
100 & Interior & 0.0761 & 0.428\\
100 & Upper-left & 0.282 & 0.84\\
100 & Floor & 0.0139 & 0.0111\\
250 & Interior & 0.0671 & 0.411\\
250 & Upper-left & 0.323 & 0.957\\
250 & Floor & 0.0297 & 0.0297\\
500 & Interior & 0.0785 & 0.449\\
500 & Upper-left & 0.391 & 1.28\\
500 & Floor & 0.0491 & 0.053\\
1000 & Interior & 0.0513 & 0.531\\
1000 & Upper-left & 0.297 & 1.98\\
1000 & Floor & 0.0187 & 0.0157\\
\bottomrule\end{tabular}\end{table}
\FloatBarrier

\section{Recorded optimization histories}\label{app:optimization-histories}
All four detailed final histories record SOAP 10,000 and SSB 20,000 with seed 12345. The earlier training of the refined initializer and the reconditioning phase of the half envelope are not in these histories (Section~\ref{sec:formulation}), so the total training differs between the workflows. The recorded wall-clock time covers the Broyden phase only and is not used to compare cost with CFD.

\begin{figure}[H]\centering
\includegraphics[width=\linewidth]{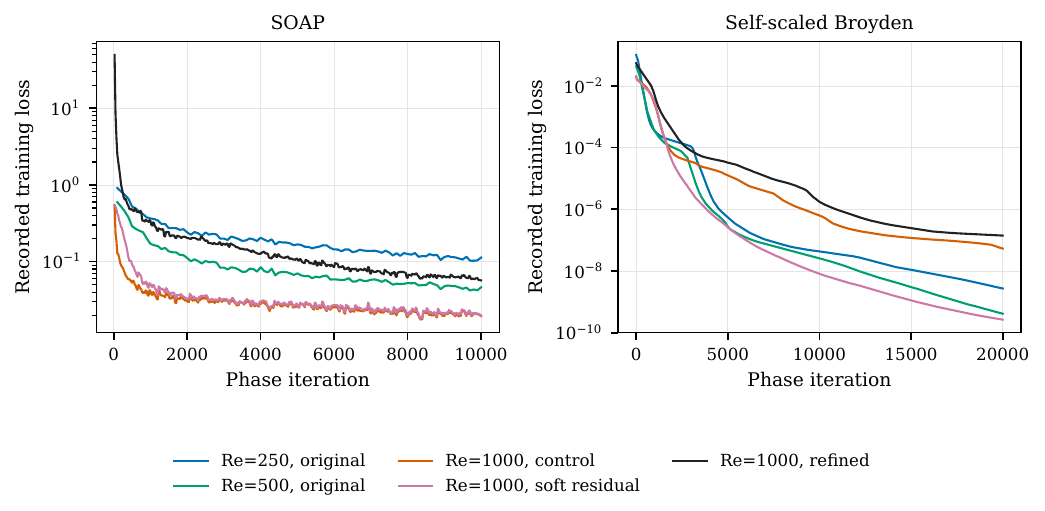}
\caption{Recorded weighted momentum-loss histories at $D/W=5$, $\Rey=250$, 500 and 1000, including the detailed $\Rey=1000$ variants: changing-batch SOAP (left) and fixed-batch self-scaled Broyden (right). Curves use the retained completed histories without smoothing. The horizontal coordinate is the iteration within each phase; vertical axes are logarithmic. Different objectives and initialization histories prevent interpreting the loss ordering as a physical-accuracy ranking.}\label{fig:training-loss}
\end{figure}

Figure~\ref{fig:training-loss} separates the SOAP and self-scaled Broyden histories rather than joining phases with different point distributions. It uses the saved training records for the original $\Rey=250$ and 500 cases and for the $\Rey=1000$ control, soft-residual and refined workflows. During SOAP, periodic resampling produces visible fluctuations; the fixed-point Broyden phase decreases more smoothly. The plots retain the recorded losses without smoothing. The different residual scales, sampling distributions and initial fields belong to each workflow, so vertical separation between curves is not a comparison of physical field error. The smaller soft-residual loss, for example, accompanies a larger floor discrepancy than the refined workflow in Table~\ref{tab:errors}.

Several histories are still decreasing when training stops, so the reported fields are those reached after a fixed number of steps, not converged solutions. Longer training might improve them, but the loss curves cannot say by how much or where; only the field comparisons measure accuracy.

\FloatBarrier

\setlength{\bibsep}{2pt}

\end{document}